\documentclass[a4paper,fleqn,usenatbib,useAMS]{mn2e}
\usepackage[applemac]{inputenc}
\usepackage[draft=false, colorlinks=true, allcolors=blue]{hyperref}
\usepackage{graphicx}	% Including figure files
\usepackage{amsmath}	% Advanced maths commands
\usepackage{dsfont}
\usepackage{bm}
\usepackage{mathrsfs}
\usepackage[normalem]{ulem}
\newcommand{\percent}{\ensuremath{\mathrm{per\, cent}}}
\renewcommand{\arcmin}{\ensuremath{\mathrm{arcmin}}}
\newcommand{\msun}{\ensuremath{\mathrm{M_\odot}}}
\newcommand{\mh}{\ensuremath{h^{-1} \, \mathrm{M_\odot}}}

\newcommand{\cmpc}{\ensuremath{\mathrm{cMpc}}}

\newcommand{\gpc}{\ensuremath{\mathrm{Gpc}}}

\newcommand{\kpc}{\ensuremath{\mathrm{kpc}}}

\newcommand{\mean}[1]{\bar{#1}}
\newcommand{\amean}[1]{\langle#1\rangle}
\newcommand{\diff}{\mathrm{d}}
\newcommand{\vect}[1]{\ensuremath{\mathbf{#1}}}

\newcommand{\Z}{\ensuremath{\mathsf{Z}}}
\newcommand{\X}{\ensuremath{\mathsf{X}}}
\newcommand{\Xs}{\ensuremath{\mathsf{X}^*}}
\newcommand{\K}{\ensuremath{\mathsf{K}}}

\newcommand{\KXX}{\ensuremath{\K_{\X\X}}}
\newcommand{\KXXs}{\ensuremath{\K_{\X\Xs}}}
\newcommand{\KXsX}{\ensuremath{\K_{\Xs\X}}}
\newcommand{\KXsXs}{\ensuremath{\K_{\Xs\Xs}}}

\newcommand{\Om}{\ensuremath{\Omega_\mathrm{m}}}
\newcommand{\Ob}{\ensuremath{\Omega_\mathrm{b}}}

\newcommand{\On}{\ensuremath{\Omega_\nu}}
\newcommand{\Ok}{\ensuremath{\Omega_\mathrm{k}}}

\newcommand{\zc}{\ensuremath{\zeta_\mathrm{c}}}
\newcommand{\Sc}{\ensuremath{\Sigma_\mathrm{c}}}
\newcommand{\tin}{\ensuremath{\theta_1}}
\newcommand{\tout}{\ensuremath{\theta_2}}
\newcommand{\tmax}{\ensuremath{\theta_\mathrm{m}}}
\newcommand{\Rin}{\ensuremath{R_1}}
\newcommand{\Rout}{\ensuremath{R_2}}
\newcommand{\Rmax}{\ensuremath{R_\mathrm{m}}}
\newcommand{\gt}{\ensuremath{g_\mathrm{T}}}
\newcommand{\gammat}{\ensuremath{\gamma_\mathrm{T}}}
\newcommand{\Dl}{\ensuremath{D_\mathrm{l}}}
\newcommand{\Ds}{\ensuremath{D_\mathrm{s}}}
\newcommand{\Dls}{\ensuremath{D_\mathrm{ls}}}

\usepackage[T1]{fontenc}
\usepackage{ae,aecompl}

\usepackage{newtxtext,newtxmath}
\title[Halo aperture mass function]{Why are we still using 3D masses
  for cluster cosmology?}

\author[S.N.B. Debackere et al.]{Stijn N.B. Debackere\thanks{Contact e-mail:
    \href{mailto:debackere@strw.leidenuniv.nl}{debackere@strw.leidenuniv.nl}
  }$^{1}$,
  Henk Hoekstra$^{1}$,
  Joop Schaye$^{1}$,
  Katrin Heitmann$^{2}$,
  \newauthor
  Salman Habib$^{2,3}$
  \\
  $^{1}$Leiden Observatory, Leiden University, PO Box 9513, NL-2300 RA
  Leiden, The Netherlands
  \\
 $^{2}$High Energy Physics Division, Argonne National Laboratory, Lemont, IL 60439, USA\\
  $^{3}$Computational Science Division, Argonne National Laboratory, Lemont, IL 60439, USA\\}

\date{Last updated --; in original form --}

\pubyear{2021}

\begin{document}\label{firstpage}
\pagerange{\pageref{firstpage}--\pageref{lastpage}}
\maketitle

% Abstract of the paper
\begin{abstract}
  The abundance of clusters of galaxies is highly sensitive to the
  late-time evolution of the matter distribution, since clusters form
  at the highest density peaks. However, the 3D cluster mass cannot be
  inferred without deprojecting the observations, introducing
  model-dependent biases and uncertainties due to the mismatch between
  the assumed and the true cluster density profile and the neglected
  matter along the sightline. Since projected aperture masses
  \emph{can} be measured directly in simulations and observationally
  through weak lensing, we argue that they are better suited for
  cluster cosmology. Using the Mira--Titan suite of gravity-only
  simulations, we show that aperture masses correlate strongly with 3D
  halo masses, albeit with large intrinsic scatter due to the varying
  matter distribution along the sightline. Nonetheless, aperture
  masses can be measured $\approx 2-3$ times more precisely from
  observations, since they do not require assumptions about the
  density profile and are only affected by the shape noise in the weak
  lensing measurements. We emulate the cosmology dependence of the
  aperture mass function directly with a Gaussian process. Comparing
  the cosmology sensitivity of the aperture mass function and the 3D
  halo mass function for a fixed survey solid angle and redshift
  interval, we find the aperture mass sensitivity is higher for $\Om$
  and $w_a$, similar for $\sigma_8$, $n_\mathrm{s}$, and $w_0$, and
  slightly lower for $h$. With a carefully calibrated aperture mass
  function emulator, cluster cosmology analyses can use cluster
  aperture masses directly, reducing the sensitivity to
  model-dependent mass calibration biases and uncertainties.
\end{abstract}

% Select between one and six entries from the list of approved keywords.
% Don't make up new ones.

%%%%%%%%%%%%%%%%%%%%%%%%%%%%%%%%%%%%%%%%%%%%%%%%%%

%%%%%%%%%%%%%%%%% BODY OF PAPER %%%%%%%%%%%%%%%%%%

\section{Introduction}\label{sec:introduction}
The next decade of cosmological galaxy surveys such as
\emph{Euclid}\footnote{\url{https://www.euclid-ec.org}} and the Rubin
Observatory Legacy Survey of Space and Time (LSST)
\footnote{\url{https://www.lsst.org/}} will elucidate the late-time
evolution of the Universe by measuring the large-scale distribution of
galaxies out to a redshift of $z \approx 2$. The sheer volume of these
surveys will result in the detection of over a billion galaxies that
can be used to trace the underlying dark matter distribution. The main
focus of these surveys is on measuring the matter distribution through
the clustering of galaxies and through the lensing-induced distortion
of galaxy shapes due to the intervening large-scale structure, the
cosmic shear.

Galaxy clusters, located at the most significant peaks of the density
field, will be another particularly powerful probe. Due to the
hierarchical growth of structure, the abundance of clusters as a
function of mass and time depends sensitively on the amount of matter,
$\Om$, how clustered it is, $\sigma_8$, and also on the late-time
expansion due to dark energy, quantified by its equation-of-state
parameter $w_0$ and its time derivative $w_a$
\citep[e.g.][]{Haiman2001, allen2011, pratt2019}. More than $10^5$
galaxy clusters will be detected in the coming decade
\citep[e.g.][]{sartoris2016}, transforming galaxy cluster cosmology
into a cosmological probe limited only by our understanding of its
systematic uncertainties \citep[e.g.][]{kohlinger2015}.

Observationally, clusters are identified as highly significant peaks
in maps of some observed signal, $\mathcal{O}$, that traces the total
mass distribution, such as the galaxy overdensity, the weak lensing
shear, the X-ray emission, or the Sunyaev-Zel'dovich (SZ) effect
signal. Next, after some quality cuts on the cluster candidates, we
are left with a cluster catalogue for the surveyed volume. To derive
cosmological constraints from this catalogue, we need a theoretical
prediction for the cosmology-dependent cluster abundance, and a way to
link the theoretical predictions to the observed clusters. In
principle, any halo property that depends on cosmology can be used,
but the halo mass, $\mathcal{M}$, is the most obvious candidate. This
then requires knowledge of the dependence of the halo mass function,
$n(\mathcal{M}|\mathbf{\Omega})$, on the cosmological parameters,
$\mathbf{\Omega}$, the mass--observable relation,
$P(\mathcal{O}|\mathcal{M})$, and the cluster selection function,
$\mathcal{S}$. Any systematic error in these quantities will degrade
the cosmological constraints from cluster cosmology.

To calibrate the mass--observable relation, we need observational
measurements of the halo mass, $\mathcal{M}$, for a subsample of the
detected clusters. We will denote the halo mass inferred from
observations as $\mathcal{M}_\mathrm{obs}$. There are multiple ways in
which halo masses can be defined, since haloes do not have clear
boundaries. Weak lensing observations have become the de facto
standard to calibrate cluster masses as they provide the only way to
directly probe both baryonic and dark matter \citep[for a review,
see][]{hoekstra2013}. Masses can be obtained from weak lensing
observations either by fitting a density profile to the observed shear
and inferring the mass within some radius, or by directly adding up
the surface mass density---which can be obtained from the
shear---within some aperture. Since we are only able to securely
identify clusters above some threshold in the observed signal,
$\mathcal{O}_\mathrm{lim}$, a correct calibration of the
mass--observable relation also requires the abundance of clusters to
be taken into account. After all, the number of haloes around the
detection limit will depend not only on the uncertainty in the
mass--observable relation, but also on the expected number of haloes
at that given mass (see \citealp{mantz2019} for a clear discussion of
this effect).

A full cluster cosmology analysis then calibrates the cosmology- and
redshift-dependent relations
$P(\mathcal{O}, z|\mathcal{M}_\mathrm{obs}, \mathbf{\Omega},
\mathcal{S})$ and
$P(\mathcal{M}_\mathrm{obs}, z|\mathcal{M}, \mathbf{\Omega},
\mathcal{S})$, by fitting them jointly with the theoretical halo
abundance, $n(\mathcal{M}, z|\mathbf{\Omega}, \mathcal{S})$, to the
observed cluster number counts within bins $\mathcal{O}_i$ and $z_j$,
$N(\mathcal{O}_i, z_j)$. The halo abundance possibly depends on the
selection function for quality cuts based on the halo environment, for
example to exclude chance alignments or mergers. We write out the
forward model as
\begin{equation}
  \label{eq:N_obs}
  N(\mathcal{O}_i, z_j| \mathbf{\Omega}, \mathcal{S}) = \Omega_\mathrm{sky} 
  \begin{aligned}[t]
    & \int\limits_{\mathcal{O}_i}^{\mathcal{O}_{i+1}} \diff
      \mathcal{O} \int\limits_{z_j}^{z_{j+1}} \diff z \frac{\diff V(z,
      \mathbf{\Omega})}{\diff \Omega \diff z} \int
      \diff \mathcal{M}\, \diff \mathcal{M}_\mathrm{obs} \\
    & \times P(\mathcal{O}, z|\mathcal{M}_\mathrm{obs}, \mathbf{\Omega},
      \mathcal{S})
      P(\mathcal{M}_\mathrm{obs}, z|\mathcal{M}, \mathbf{\Omega},
      \mathcal{S}) \\
    & \times n(\mathcal{M}, z|\mathbf{\Omega}, \mathcal{S}) \, ,
  \end{aligned}
\end{equation}
where $\mathcal{O}$ and $z$ are integrated over their respective bins,
and $\mathcal{M}_\mathrm{obs}$ and $\mathcal{M}$ over all possible
values. We convert the halo number density to the number counts taking
into account the cosmology-dependent comoving volume at redshift $z$,
$V(z, \mathbf{\Omega})$, probed by a survey covering a solid angle
$\Omega_\mathrm{sky}$. Correctly modelling the cluster selection is of
vital importance in any attempt to derive cosmological constraints
from galaxy clusters. Ideally, we would detect clusters through an
observable that has a straightforward selection function. Since the
selection function depends on the survey under consideration, we will
assume here that the selection has been modelled correctly. This
simplifies the derivation of the main points we want to make.

Currently, cluster analyses infer 3D halo masses from weak lensing
observations to determine the mass--observable relation \citep[see
e.g.][]{bocquet2020, descollaboration2020}. The appeal of 3D halo
masses stems from analytic arguments such as the (extended)
Press-Schechter theory \citep{press1974, bond1991}, that predict that
the 3D halo mass function has a universal shape set only by the
significance of the seed perturbation of a halo in the initial
Gaussian density field. In recent years, however, ever larger suites
of cosmological dark matter-only (DMO) simulations have shown that the
assumed \emph{universality} of the 3D halo mass function does not hold
in detail. Simulated abundances can deviate from the universal
prediction by $> 10 \, \percent$ depending on the redshift and the
exact cosmology \citep[see e.g.][]{tinker2008, bhattacharya2011,
  despali2016, diemer2020a}. Hence, suites of large-volume
cosmological simulations run on a grid of different cosmological
parameter values are vital to capture the cosmology dependence of the
halo mass function through either analytic fitting functions
\citep{tinker2008, bhattacharya2011} or emulators
\citep{mcclintock2019a, nishimichi2019, bocquet2020}.

Problematically, 3D halo masses cannot be measured directly from
observations, which first need to be deprojected. Generically,
deprojection requires the assumption of a spherically symmetric
density profile, which will be affected by baryons and scatter
introduced by halo triaxiality, substructures, and correlated
structures \citep[see e.g.][]{becker2011, oguri2011a, bahe2012b,
  henson2017, debackere2021}. This introduces model-dependent biases
and increases the uncertainty in the inferred 3D halo masses,
degrading the cosmological constraints from cluster cosmology. Note
that this step is only required to transform the observations to
theory predictions. As we argue in this paper, such a procedure is not
necessary.

Since modern theoretical predictions for the halo abundance already
rely on large simulation suites, it is possible to perform the cluster
mass calibrations with halo properties that \emph{can} be measured
directly in both observations and simulations. This has the additional
advantage that dark matter-only simulations can optionally be replaced
by hydrodynamical simulations in order to account for baryonic effects
on the halo mass function \citep[e.g.][]{velliscig2014} or to directly
predict a baryonic observable. We focus on weak lensing observations
because they probe the total matter content and are thus less
sensitive to uncertainties in how baryonic matter traces the dark
matter. From the weak lensing shear signal we can directly measure
projected aperture masses within apertures of a fixed angular or
physical size, without the need to assume any density profile
\citep[see e.g.][]{schneider1996, bartelmann2001a}. Importantly, these
aperture masses can also be measured directly in simulations.

Aperture masses have been studied before in the context of cluster
cosmology with purely shear-selected samples in order to bypass
uncertainties due to the selection based on some baryonic observable
such as the X-ray luminosity, the SZ signal or the galaxy overdensity
\citep[e.g.][]{reblinsky1999a}. \citet{marian2010} argued that future
surveys would no longer need to convert shear peaks to 3D halo masses,
if predictions for the halo abundance as a function of their aperture
mass were available. However, \citet{hennawi2005} showed that while
almost all massive clusters produce significant aperture mass peaks,
there is a large population of significant peaks that cannot be
ascribed to a single cluster but rather is the result of chance
superpositions along the line-of-sight due to the broad lensing
kernel. Hence, to decrease the number of false-positive cluster
detections, baryonic observables are still required for confirmation.
More recently, \citet{hamana2015}, \citet{shan2018} and
\citet{martinet2018} have used peaks identified from weak lensing
observations to constrain the matter density and clustering of the
Universe.

With the availability of large-volume simulation suites run for many
different cosmological models, it is now possible to calibrate the
cosmology dependence of the halo aperture mass function. Importantly,
with aperture mass measurements the theoretical model assumptions
separate cleanly from the purely observational data in
Eq.~\eqref{eq:N_obs}. That is, Eq.~\eqref{eq:N_obs} splits into an
observational scaling relation,
$P(\mathcal{O}, z|\mathcal{M}_\mathrm{obs}, \mathbf{\Omega},
\mathcal{S})$, independent of the cluster density profile, and a
calibration between the observed and the simulated aperture mass
measurement,
$P(\mathcal{M}_\mathrm{obs}, z|\mathcal{M}, \mathbf{\Omega},
\mathcal{S})$. The uncertainty in the observational scaling relation
will depend on how accurately $\mathcal{O}$ can be measured in the
survey, and how strongly it correlates with the aperture mass. The
theoretical calibration, on the other hand, will have a fixed
uncertainty set by the shape noise of the observations, since the
aperture mass measured from the weak lensing shear is an unbiased
measure of the true aperture mass \citep{schneider1996}. Moreover, as
shown by \citet{debackere2021}, halo aperture masses are expected to
be less sensitive to baryonic effects, especially when measured within
larger apertures that are able to capture more of the ejected halo
baryons. We study how baryons modify aperture mass measurements in
\citet{debackere2022}.

Here, we investigate the behaviour of the different components that
enter the model for the cluster number counts in Eq.~\eqref{eq:N_obs},
that is, the uncertainty in the mass--observable relation and the halo
mass function for halo aperture masses. We will show that the
mass--observable relation can be calibrated more precisely with
aperture masses than with the standard deprojected 3D halo masses.
Additionally, we will use an emulator calibrated on the Mira--Titan
suite of large-volume cosmological N-body simulations to show that the
halo aperture mass function is also highly sensitive to variations in
the cosmological parameters, in agreement with \citet{marian2010}.
This study serves as a proof-of-concept that can be applied in future
cosmological analyses when carefully calibrated emulators for the halo
aperture mass function are available.

The paper is structured as follows: first, we introduce the
large-volume simulation suite that we use for our analysis in
Section~\ref{sec:sims}. Then, in Section~\ref{sec:aperture_masses}, we
study the dependence of the aperture mass on both the 3D halo mass and
the aperture size, and use the clean separation between the
theoretical and observational uncertainties in aperture mass
measurements to study the behaviour of the mass--observable relation.
In Section~\ref{sec:aperture_mass_function}, we build an emulator to
investigate the sensitivity of the aperture mass function to changes
in the cosmological parameters, comparing it to the 3D halo mass
function. We compare our analysis with the wider literature, discuss
advantages and possible difficulties, and provide future applications
in Section~\ref{sec:discussion}. Finally, we conclude in
Section~\ref{sec:conclusions}.

\section{Simulations}\label{sec:sims}
We use the Mira--Titan suite of cosmological, gravity-only
simulations, run with the HACC N-body code (Hardware/Hybrid
Accelerated Cosmology Code, \citealt{habib2016}). This simulation
suite is well-suited to our purpose: it contains large-volume
simulations with cosmological parameters sampled using a nested
space-filling design that is ideal for interpolating the simulation
predictions. The simulations include dynamical dark energy and massive
neutrinos. The publically available data products of the simulation
suite are described in more detail in \citet{heitmann2019}. So far,
Mira--Titan has been used to construct emulators for the matter power
spectrum \citep{heitmann2016,lawrence2017} and the 3D halo mass
function \citep{bocquet2020}.

\begin{table}
  \caption{Cosmological parameter values for the Mira--Titan suite of
    large-volume, cosmological N-body simulations.}
  \begin{tabular}{lrr}
    Parameter & Min & Max \\
    \hline
    $\Om h^2$ & $0.12$ & $0.155$ \\
    $\Ob h^2$ & $0.0215$ & $0.235$ \\
    $\On h^2$ & $0.0$ & $0.01$ \\
    $\sigma_8$ & $0.7$ & $0.9$ \\
    $h$ & $0.55$ & $0.85$ \\
    $n_\mathrm{s}$ & $0.85$ & $1.05$ \\
    $w_0$ & $-1.3$ & $ -0.7$ \\
    $w_\mathrm{b} \equiv (-w_0 - w_a)^{1/4}$$^\dagger$ & $0.3$ & $1.3$ \\
    $w_a$ & $-1.56$ & $1.29$
  \end{tabular}
  \\
  \footnotesize{$^\dagger$ \citet{heitmann2016} show that this
    rescaling improves the prediction accuracy of cosmological models
    with $w_0 + w_a \approx
    0$ by putting slightly more points near the $w_0+w_a=0$ boundary.}
  \label{tab:sim_prms}
\end{table}

The simulation suite consists of a grid of 111 simulations that vary 8
different cosmological parameters. The cosmological parameters are
chosen according to a nested lattice design that enforces
space-filling properties at multiple design steps~\citep[see Section 3
of][]{heitmann2016}. This design works well with Gaussian process
emulators and has an important global convergence property that allows
systematic improvement of emulation accuracy as more design points are
added. All cosmologies are spatially flat with
$\Ok=0$. The models vary the cosmological parameters within the ranges
shown in Table~\ref{tab:sim_prms}. The full grid of cosmological
parameters is shown in figure 1 of \citet{bocquet2020}.

The Mira--Titan suite consists of 3 nested tessellations that refine
the higher level grids (M011-M036, M037-M065, and M066-M111,
respectively). These models all include massive neutrinos. To enable
accurate predictions for the Standard Model of cosmology with massless
neutrinos, the simulation suite includes an additional 10 simulations
with $m_\nu=0$ with the remaining 7 cosmological parameters sampled on
a symmetric Latin hypercube (M001-M010). All simulations have box
sizes of $2.1 \, \gpc$ (except for M006, M023, and M046 with $2.091$,
$2.085$, and $1.865 \, \gpc$, respectively) and include $3200^3$
particles with masses
$m_\mathrm{dm}=7.23 \times 10^9 - 1.22 \times 10^{10} \, \msun$
depending on the cosmology. Hence, groups and clusters with
$m>10^{13} \msun$ are generally resolved with $>1000$ particles. All
simulations use a force softening length of $\epsilon=6.6 \, \kpc$.
For our analysis, we focus on the 100 simulations with massive
neutrinos (M011-M110, for the distribution of the cosmological
parameters, see fig. 1 of \citealt{bocquet2020}).

We now briefly describe how dynamical dark energy and massive
neutrinos are included in the simulations, referring to
\citet{upadhye2014} and \citet{heitmann2016} for the full details.
Both massive neutrinos and dynamical dark energy are included at the
level of the background evolution, $H(z)$, and the initial conditions.
Particularly, the linear $z=0$ transfer function includes dark matter,
baryons, and massive neutrinos and is normalized to the correct
$\sigma_8$. Then, the matter component including dark matter and
baryons is evolved back to the initial redshift assuming a
scale-independent growth factor including all species in the
homogeneous background and used to determine the initial particle
positions and velocities. This ensures that the $z=0$ linear power
spectrum of the simulation is correct on large scales. For power
spectrum calculations, the neutrino contribution needs to be included
by hand. Hence, the simulations do not account for neutrino
clustering, which is no cause for concern, since this effect is much
smaller than the suppression of the halo mass function due to neutrino
free-streaming for the neutrino mass range considered.

The saved simulation data products had to be chosen carefully due to
the large volume of the simulations and the size of the cosmological
parameter hypercube. For each simulation output, the full particle
data is downsampled by a factor $100$ before saving. Simulation haloes
are identified on the fly, i.e. from the full particle data,
using a friends-of-friends (FoF) algorithm with linking length
$b=0.168$. Subsequently, spherical overdensity masses, defined as
$m_{\Delta\mathrm{c}} = 4/3\pi \Delta \rho_\mathrm{crit}(z) r_{\Delta
  \mathrm{c}}^3$, with overdensity $\Delta=200$ are determined around
the potential minimum of the FoF halo. For all haloes with $>1000$
particles (corresponding to
$m_\mathrm{FoF} \gtrsim 10^{13} \, \msun$), all the particles
belonging to the FoF halo are also saved separately. We will use the
downsampled particle catalogues to compute the projected aperture
masses around the identified FoF haloes with spherical overdensity
masses $m_\mathrm{200c} > m_\mathrm{200c,lim} = 10^{13.5} \, \msun$.
In Fig.~\ref{fig:sigma_delta_m_m000} and
Section~\ref{sec:aperture_calc}, we show that the Poisson noise due to
the downsampling introduces an uncertainty of $> 15 \, \percent$ in
the measured aperture masses of haloes with
$m_\mathrm{200c} < 10^{14} \, \msun$. Hence, we will mainly focus on
haloes with $m_\mathrm{200c} > 10^{14} \, \msun$ in the rest of this
paper.

\section{Aperture mass--observable relation}\label{sec:aperture_masses}
To quantify the uncertainties in the aperture mass--observable
relation, we first need to measure the halo aperture masses. In
Section~\ref{sec:aperture_calc}, we describe how we extract the halo
aperture masses from the Mira--Titan suite. We show how halo aperture
masses depend on the 3D halo mass and the aperture size in
Section~\ref{sec:aperture_behaviour}. Finally, we investigate the
possible theoretical and observational uncertainties in the aperture
mass--observable relation and compare our results to 3D halo masses in
Section~\ref{sec:aperture_uncertainty}.

\begin{figure*}
  \centering
  \includegraphics[width=\textwidth]{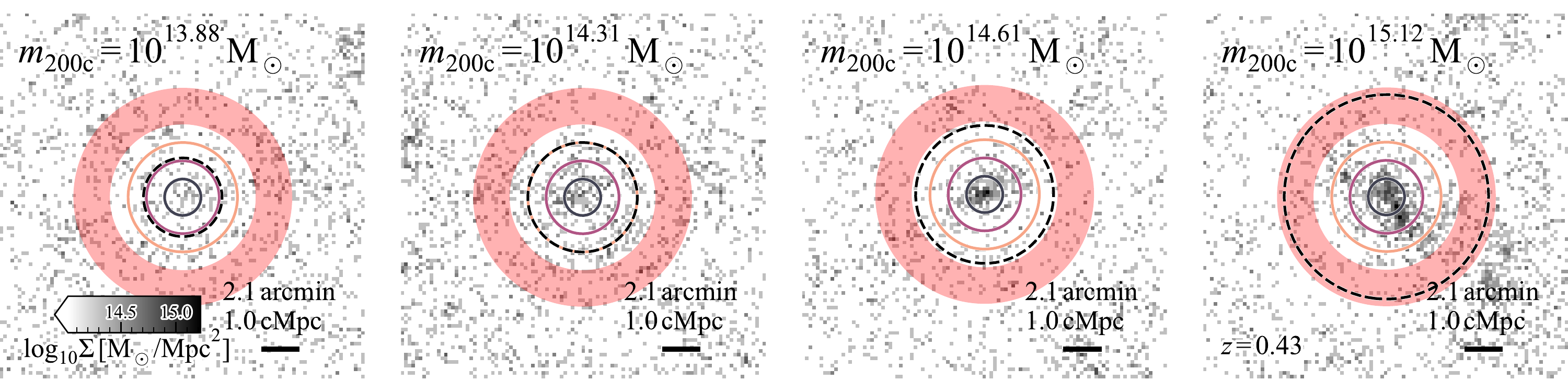}
  \caption{Surface mass density maps for randomly selected haloes in
    mass bins
    $\log_{10}m_\mathrm{200c}/\msun \in [13.5, 14.0, 14.5, 15, 15.5]$
    for simulation M000 at $z=0.43$. Each cutout has size
    $10\,\cmpc \times 10\,\cmpc$ and is plotted on the same colour
    scale. The dashed circles indicate the spherical overdensity
    radius $r_\mathrm{200c}$ for each halo. With the coloured lines,
    we show the inner apertures $\Rin=[0.5, 1.0, 1.5] \, \cmpc$ which
    we use throughout this work. The red shaded region shows the outer
    control annulus between $\Rout=2.0\, \cmpc$ and
    $\Rmax=3.0 \, \cmpc$ for the background subtraction.}
  \label{fig:map_sigma_m000}
\end{figure*}
\subsection{Extraction from the simulations}\label{sec:aperture_calc}
We will use the term aperture mass, in accordance with the literature,
to refer to the projected mass difference
\begin{align}
  \label{eq:delta_m}
  \Delta M(<\Rin|\Rout, \Rmax) & = \pi \Rin^2 (\mean{\Sigma}(\leq \Rin) - \mean{\Sigma}(\Rout < R \leq \Rmax)) \\
  \nonumber
                               & = M(\leq\Rin) - M_\mathrm{bg}(\leq \Rin)\, ,
\end{align}
where we have introduced the mean enclosed surface mass density,
$\mean{\Sigma}$, which is defined as
\begin{equation}
  \label{eq:mean_kappa}
  \mean{\Sigma}(\Rout < R \leq \Rmax) = \frac{2}{\Rmax^2 - \Rout^2} \int_{\Rout < R < \Rmax}  \diff R \, R\Sigma(R) \, .
\end{equation}
The second term in Eq.~\eqref{eq:delta_m} corrects the mass within the
aperture $\Rin$ for the average surface mass density within the
control annulus bounded by $\Rout$ and $\Rmax$, which acts as a local
background subtraction, $M_\mathrm{bg}$. Both terms get the same
contribution from the mean cosmological background density along the
line-of-sight, which cancels out in the difference. The background
subtraction makes the aperture mass independent of the line-of-sight
integration length, provided it is large compared with the clustering
length \citep[as also noted by][]{marian2010}. We verify this below.

The power of the aperture mass defined in Eq.~\eqref{eq:delta_m} is
that it can be obtained directly from weak lensing observations, as
shown in Eq.~\eqref{eq:delta_m_from_zeta_c} in
Appendix~\ref{app:weak_lensing}. Moreover, choosing fixed physical or
angular aperture sizes removes the need to assume a cluster density
profile, in contrast to spherical overdensity radii. We will measure
aperture masses within three different but fixed apertures of
$\Rin=[0.5, 1.0, 1.5] \, \cmpc$, with $\Rout=2.0 \, \cmpc$ and
$\Rmax=3.0\, \cmpc$. These apertures are similar to the typical
aperture sizes used in weak lensing cluster mass calibrations
\citep[e.g.][]{hoekstra2015, applegate2014}. Moreover, they also
roughly correspond to the halo radii for haloes with
$m_\mathrm{200c} > 10^{13} \, \msun$. Smaller apertures will give
better signal-to-noise ratios (SNRs) for lower-mass haloes since they
are better matched to their sizes \citep{schneider1996}. To compare
these results with aperture masses inferred from observations, the
distances in the simulations need to be converted into angular
positions, $\theta$, using the angular diameter distance to the lens
for the simulated cosmology.

Since the aperture mass from weak lensing observations is inferred
from the shear signal within the annulus between $\Rin$ and $\Rmax$,
the optimal choice of the aperture sizes balances the increased signal
from decreasing $\Rin$ and increasing $\Rmax$, respectively, against
the increased modelling uncertainty due to contamination from cluster
member galaxies and miscentring errors, and the contribution of cosmic
noise in the cluster outskirts \citep[e.g.][]{mandelbaum2010a}. We
stress that the aperture mass in Eq.~\eqref{eq:delta_m} will be
computed directly from the simulation data without any assumptions
about the weak lensing observations. Any observational uncertainty in
converting the weak lensing signal to the surface mass density will
thus be included in the $P(\Delta M_\mathrm{obs}|\Delta M, z)$ term in
Eq.~\eqref{eq:N_obs}, leaving the aperture mass function unaffected.
We discuss such observational uncertainties in
Section~\ref{sec:aperture_uncertainty}. In practice, the observed weak
lensing aperture mass includes the contribution of mass along the
line-of-sight, weighted by the lensing kernel. However, as we will
show in Fig.~\ref{fig:delta_m_los}, the total aperture mass is
dominated by the correlated structure within $\approx 30 \, \cmpc$ of
the cluster, which justifies neglecting the lensing kernel weighting
in our analysis.

Given the downsampled particle catalogue, calculating halo aperture
masses is relatively straightforward. First, we correct the particle
catalogues for the downsampling (see Section~\ref{sec:sims}) by
increasing the particle masses by a factor $100$. We investigate the
effect of this downsampling on the accuracy of the derived halo masses
below. We generate projected maps of the surface mass density,
$\Sigma$, along the three principal axes of the simulation volume on a
grid of $21000\times 21000$ pixels, corresponding to a pixel size of
$(L/21000)^2 = (0.1 \, \cmpc)^2$ (except for the simulations with
smaller box sizes). Subsequently, we can directly obtain halo aperture
masses from the surface mass density maps by calculating
Eq.~\eqref{eq:delta_m} centred on the identified halo centres.

\begin{figure}
  \centering
  \includegraphics[width=\columnwidth]{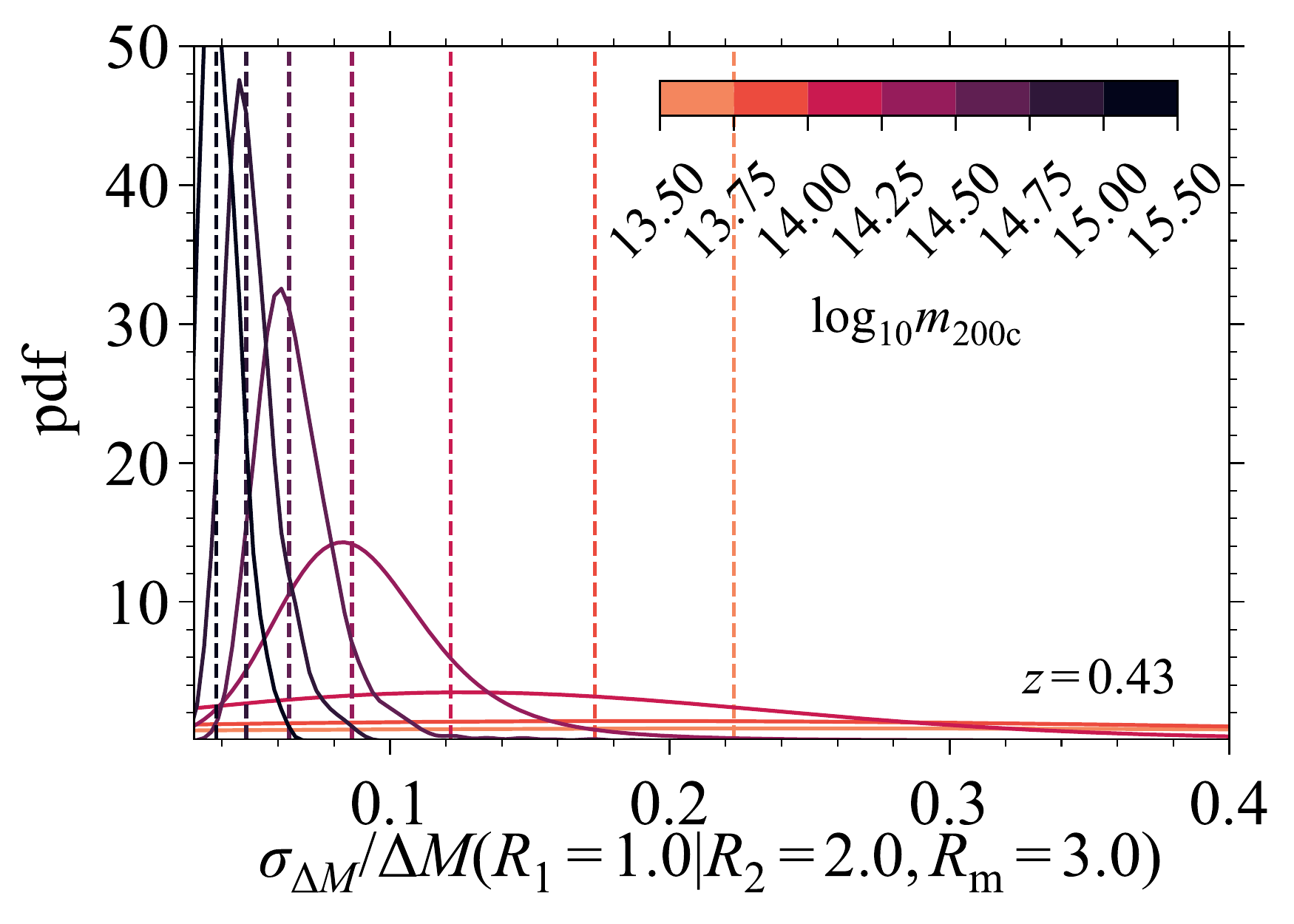}
  \caption{Distribution of the uncertainty in the aperture mass, given
    in Eq.~\eqref{eq:delta_m}, for different 3D halo mass bins due to
    the factor of $100$ downsampling of the saved simulation particle
    catalogues. We add the Poisson uncertainties of the downsampled
    number of simulation particles within $R<\Rin = 1.0 \, \cmpc$ and
    $\Rout \leq R < \Rmax$ in quadrature for all haloes with
    $m_\mathrm{200c}>10^{13.5} \, \msun$ in M000 at $z=0.43$.
    Different coloured lines correspond to different 3D halo mass bins
    and the dashed lines indicate the median uncertainty. The
    downsampling results in a significant uncertainty in the derived
    aperture masses for haloes with
    $m_\mathrm{200c} < 10^{14.25} \, \msun$.}
  \label{fig:sigma_delta_m_m000}
\end{figure}
In Fig.~\ref{fig:map_sigma_m000}, we show the surface mass density
maps centred on 4 random haloes within mass bins with bin edges
specified by
$\log_{10} m_\mathrm{200c} /\msun \in [13.5, 14.0, 14.5, 15.0, 15.5]$
for reference simulation M000 at $z=0.43$. Clearly, the downsampling
of the particle catalogue results in emptier and noisier mass maps.
Every particle in the simulation has a $p=0.01$ chance of being
included in the downsampled particle catalogue. As a result, particle
catalogues of downsampled haloes will include a shot-noise
contribution of $pN$, resulting in a fractional uncertainty on the
final 3D halo mass of $\delta m / m = \sqrt{pN}^{-1}$, which is
$\approx [20, 10, 6, 3, 2] \, \percent$ for haloes located at the mass
bin edges. Since the spherical overdensity halo masses were saved on
the fly, the downsampling does not affect the halo mass catalogues.
The aperture masses, however, \emph{are} affected by the particle
downsampling. We show the distribution of the fractional aperture mass
uncertainty due to the finite number of particles for different 3D
halo mass bins in Fig.~\ref{fig:sigma_delta_m_m000}. We show the
fractional uncertainty,
$\sigma_{\log \Delta M} = \sigma_{\Delta M} / \Delta M$, for
$\Rin=1.0 \, \cmpc$, since this aperture size is similar to the virial
radius for haloes with $10^{13.5} < m_\mathrm{200c} /\msun < 10^{14}$.
We calculate the uncertainty by adding the shot noise contributions to
$M(<\Rin)$ and $M_\mathrm{bg}(<\Rin)$ in quadrature. Even though the
individual contributions to the aperture mass in
Eq.~\eqref{eq:delta_m} can be determined at high accuracy due to the
extra particles included along the line-of-sight, their difference has
a large fractional uncertainty. Hence, we will limit our halo sample
to haloes with $m_\mathrm{200c} > 10^{14} \, \msun$ whose aperture
masses can be determined with a median fractional uncertainty of
$\lesssim 15 \, \percent$ from the available particle data. We note
that even though the median uncertainty of the mass bin
$10^{14.0} < m_\mathrm{200c} / \msun < 10^{14.25}$ is
$\lesssim 15 \, \percent$, there are also significant outliers.

\begin{figure*}
  \centering
  \includegraphics[width=0.97\textwidth]{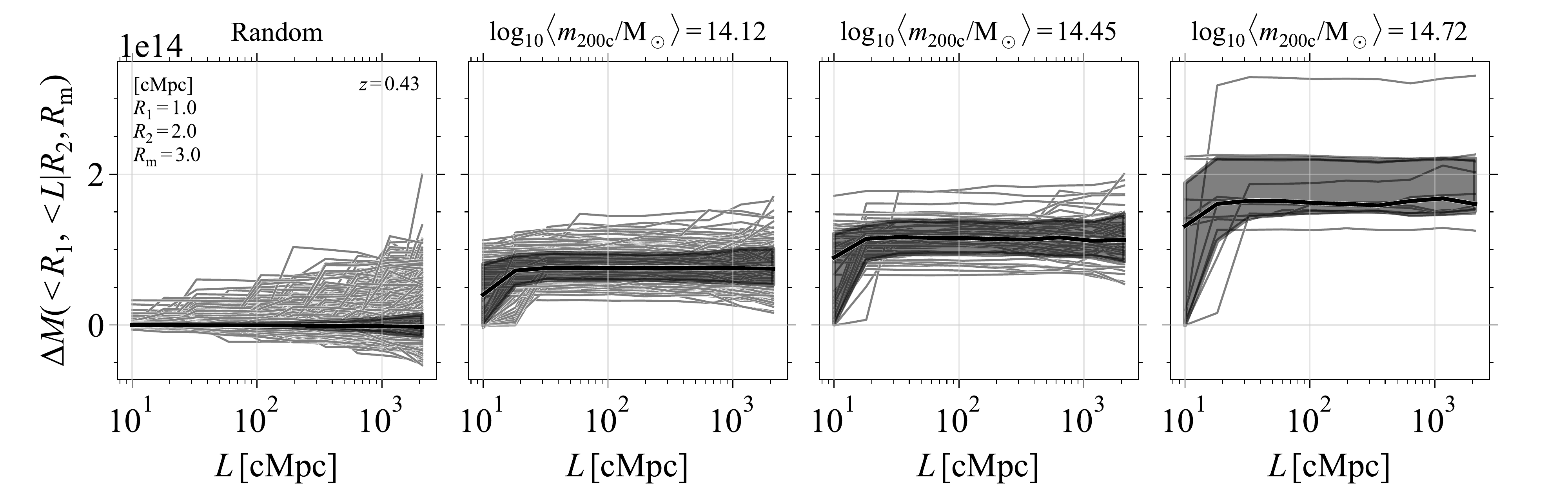}
  \caption{Convergence of the aperture mass, $\Delta M$, with the
    line-of-sight depth, $L$, centred on random positions (\emph{first
      column}), and haloes within increasing $m_\mathrm{200c}$ bins
    (\emph{second to fourth columns}). Light grey lines indicate the
    individual positions/haloes, which were chosen to have
    $x$-coordinates within $\pm 5 \, \cmpc$ of the centre of the
    $x$-axis along which we project. The median and 16th to 84th
    percentile scatter are indicated with thick black lines and the
    shaded region, respectively. The median aperture mass along random
    lines-of-sight is zero, as expected, with a slight increase in the
    scatter for larger line-of-sight integration lengths.
    Lines-of-sight centred on haloes generally converge within
    $\approx 30 \, \cmpc$ along the line-of-sight, with a large
    scatter that increases slightly with increasing $L$. The aperture
    mass for individual haloes can increase or decrease significantly
    when encountering a massive structure along the line-of-sight
    within $\Rin$ or $\Rout < R < \Rmax$, respectively.}
  \label{fig:delta_m_los}
\end{figure*}
It is important to verify that the background subtraction in the
aperture mass definition, Eq.~\eqref{eq:delta_m}, actually makes the
aperture mass independent of the line-of-sight integration depth. In
Fig.~\ref{fig:delta_m_los}, we show the calculated aperture masses as
a function of the line-of-sight integration length, $L$, centred on
$10000$ random positions (first column) or on all haloes within
different halo mass bins that have $x$-coordinates that are within
$\pm 5 \, \cmpc$ of the midpoint of the $x$-axis of the simulation box
(second to fourth columns) for simulation M000 at $z=0.43$. When
centring on random positions, the aperture masses are consistent with
zero since the average surface mass densities within $\Rin$ and the
control annulus are equal. The scatter in the aperture masses for
randomly-positioned apertures, which is equivalent to measuring the
cosmic shear on the scale of the aperture, increases with the
line-of-sight integration depth, since larger modes contribute to the
dispersion $\langle \Delta M^2(<\Rin, <L|\Rout, \Rmax) \rangle$
\citep[see e.g.][]{schneider1998}. This effect is also present when
centring on haloes, but since the cosmic shear introduces a fixed
scatter, the effect is relatively smaller for more massive haloes
\citep{hoekstra2001}. For haloes, the average aperture mass generally
converges to its final value within $\approx 30 \, \cmpc$. However,
the individual halo trajectories along the line-of-sight can increase
or decrease significantly when encountering massive structures within
$\Rin$ or $\Rout < R < \Rmax$, respectively. Hence, we confirm that
the aperture mass measurements are converged with respect to the
line-of-sight integration length of $L=2100 \, \cmpc$.

The aperture mass measurements in the simulations automatically
include the intrinsic scatter due to halo triaxiality and
substructure, and due to both correlated and uncorrelated large-scale
structures. We do not include observational uncertainties since these
will depend on the survey of interest. One source of observational
systematic uncertainty is the shear map generation, which relies on
the accuracy of the shape measurements of the background source
galaxies and the determination of their redshift distribution
\citep[e.g.][]{vonderlinden2014, hoekstra2015}. Another source of
uncertainty is the centring of the aperture on the halo. In the
simulations, we centre the surface mass density maps exactly on the
potential minimum of the spherical overdensity, but observationally
this centre cannot be identified so unambiguously. However,
\citet{hoekstra2012} showed that deprojected mass estimates derived
from aperture mass measurements within large apertures corresponding
to overdensity radii with $\Delta < 1000$, are only affected by
$\lesssim 5 \, \percent$ for miscentring radii up to
$0.5 \, h_{70}^{-1} \cmpc$. For reference, the distribution of the
offset, $\Delta R$, between the SZ signal peak and the location of the
brightest cluster galaxy position shows that the bulk of clusters
($\approx 95 \, \percent$) are well centred with
$\sigma_{\Delta R} \lesssim 0.2 R_\mathrm{500c}$, which is smaller
than $0.5 \, h_{70}^{-1} \cmpc$ for all clusters with
$m_\mathrm{500c} \lesssim 5\times 10^{15} \msun$, while the remaining
clusters show a larger dispersion
$\sigma_{\Delta R} \approx 0.7 R_\mathrm{500c}$ \citep[see
e.g.][]{saro2015, bleem2020}. In the same vein as the results of
\citet{hoekstra2012}, aperture masses measured within apertures
considerably larger than the miscentring radius of the cluster should
not be significantly affected by miscentring. Hence, ignoring
miscentring does not change the conclusions of our work. Next, we will
show the dependence of halo aperture masses on the 3D halo mass and
the aperture.

\subsection{Aperture mass behaviour}\label{sec:aperture_behaviour}
\begin{figure}
  \centering
  \includegraphics[width=\columnwidth]{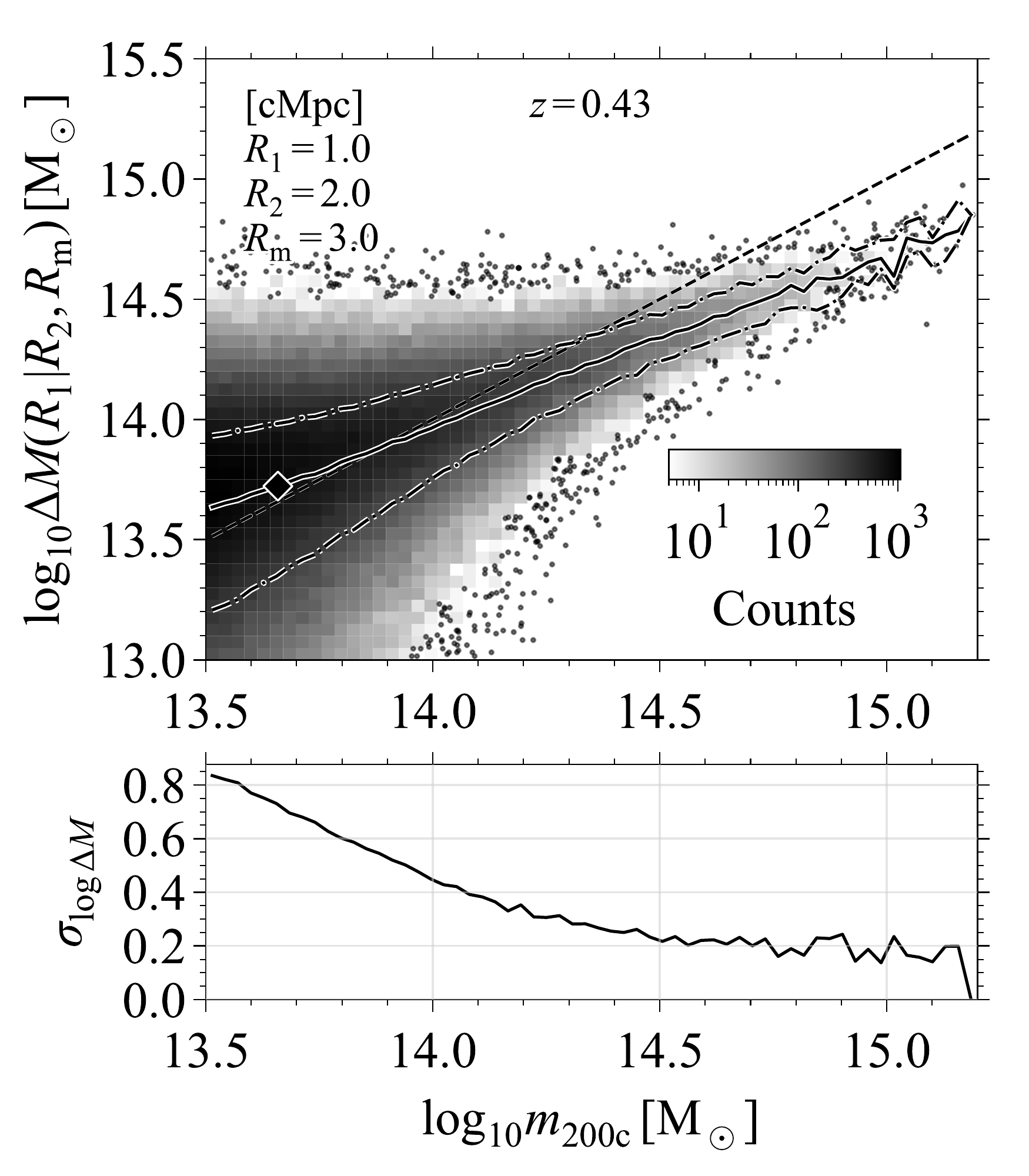}
  \caption{\emph{Top panel}: The distribution of aperture masses,
    $\Delta M(<\Rin=1.0 \, \cmpc|\Rout=2.0 \, \cmpc, \Rmax=3.0 \,
    \cmpc)$, as a function of the 3D spherical overdensity mass
    $m_\mathrm{200c}$ for simulation M011 at $z=0.43$. The dashed line
    indicates the one-to-one relation, the solid line indicates the
    median relation, and the dash-dotted lines the 16th and 84th
    percentile scatter. The diamond indicates the 3D halo mass for
    which $r_\mathrm{200c}=\Rin$. The large scatter in $\Delta M$ at
    fixed $m_\mathrm{200c}$ is caused by the large variation in the
    matter distribution along the line-of-sight. \emph{Bottom panel}:
    The logarithmic scatter in the aperture mass distribution at fixed
    $m_\mathrm{200c}$, calculated as half the difference between the
    84th and the 16th percentiles. The scatter decreases from
    $\sigma_{\log \Delta M} \approx 0.45$ at
    $m_\mathrm{200c} = 10^{14} \, \msun$ to $\lesssim 0.2$ for
    $m_\mathrm{200c} > 10^{14.5} \, \msun$.}
  \label{fig:M_vs_m_dist}
\end{figure}
Since halo properties are mostly studied as a function of their 3D
mass, we show the distribution of aperture masses for
$\Rin=1.0 \, \cmpc$, $\Rout=2 \, \cmpc$, and $\Rmax=3 \, \cmpc$ as a
function of the 3D halo mass, $m_\mathrm{200c}$, at $z=0.43$ in the
M011 simulation in the top panel of Fig.~\ref{fig:M_vs_m_dist}. The
median $\Delta M$--$m_\mathrm{200c}$ relation, indicated with the
solid line, is slightly shallower than one-to-one: the aperture mass
for haloes with $r_\mathrm{200c} \gtrsim (\lesssim) \Rin$ is smaller
(larger) than $m_\mathrm{200c}$ since the halo mass represents a
larger (smaller) fraction of the total aperture mass. For simulation
M011, the halo radius $r_\mathrm{200c} = \Rin = 1.0 \, \cmpc$ for
$m_\mathrm{200c} \approx 10^{13.65} \, \msun$. Haloes at fixed
$m_\mathrm{200c}$ can have greatly differing aperture masses due to
differences in the matter distribution along the line-of-sight of
haloes at fixed 3D mass (see also Fig.~\ref{fig:delta_m_los}). For
low-mass haloes the scatter around the median relation increases
significantly since mass outside the halo contributes relatively more
to the mass within the aperture.

In the bottom panel of Fig.~\ref{fig:M_vs_m_dist}, we show the
logarithmic scatter around the median $\Delta M$--$m_\mathrm{200c}$
relation. We calculate the scatter as half the difference between the
84th and the 16th percentile of $\log \Delta M$. The scatter increases
strongly for low-mass haloes, partially due to the particle
downsampling of the halo catalogues shown in
Fig.~\ref{fig:sigma_delta_m_m000}, but also since matter outside the
halo contributes more to the aperture mass. The intrinsic scatter in
the aperture mass at fixed halo mass decreases from
$\sigma_{\log \Delta M} \approx 0.45$ for
$m_\mathrm{200c} = 10^{14} \, \msun$ to $\lesssim 0.2$ for
$m_\mathrm{200c} > 10^{14.5} \, \msun$, which is similar to the
scatter in the weak lensing-inferred 3D halo mass at fixed halo mass
due to triaxiality and substructure (see
Fig.~\ref{fig:sigma_delta_m_obs} and
Section~\ref{sec:aperture_uncertainty} for a comparison with the mock
weak lensing analysis from \citealt{bahe2012b}). The scatter at high
halo masses is dominated by differences in the projected structure
along the line-of-sight to the halo, both correlated and uncorrelated,
since the downsampling has a negligible effect on high-mass haloes.

\begin{figure*}
  \centering
  \includegraphics[width=\textwidth]{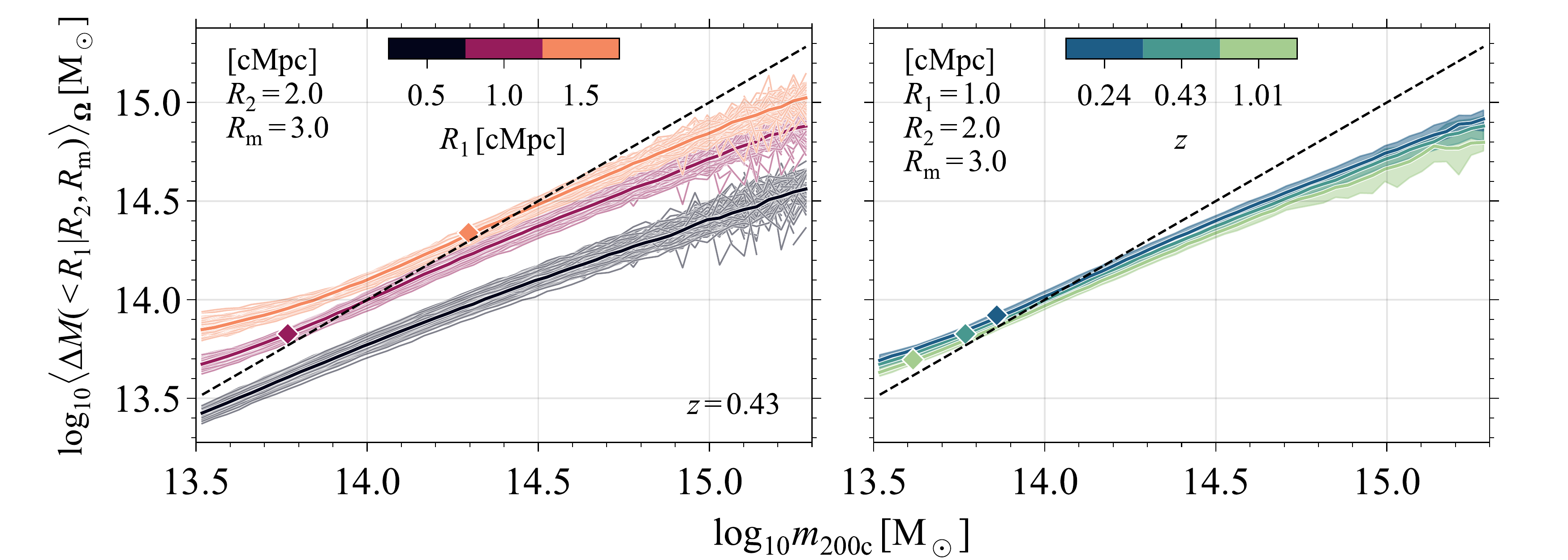}
  \caption{\emph{Left panel:} The relation between the median
    projected aperture masses of all cosmologies in the hypercube,
    $\langle \Delta M(<\Rin|\Rout, \Rmax) \rangle_\mathbf{\Omega}$,
    within different apertures $\Rin$ (thick coloured lines) and the
    3D spherical overdensity mass $m_\mathrm{200c}$. The thin,
    transparent lines show the results for individual simulations. The
    black, dashed line indicates the one-to-one relation. The coloured
    diamonds show the halo mass for which $r_\mathrm{200c} = \Rin$.
    The reference annulus for all aperture mass measurements spans the
    region between $\Rout=2\, \cmpc$ and $\Rmax=3\,\cmpc$. Masses
    measured within larger apertures more closely match the 3D masses
    of more massive haloes. \emph{Right panel:} The redshift evolution
    of the relation between the median aperture mass of all
    cosmologies and $m_\mathrm{200c}$. At fixed $m_\mathrm{200c}$, the
    ratio of the virial radius, $r_\mathrm{200c}$ and the comoving
    aperture radius decreases with time due to increasing contribution
    of dark energy to the critical density. The extra contribution of
    the halo outskirts within the fixed comoving aperture increases
    the measured aperture masses with time.}
  \label{fig:M_vs_m_avg}
\end{figure*}
Since different apertures are naturally tuned to detect haloes of
different mass and size, we show the median relation between the
aperture mass, $\Delta M$, measured in different apertures and the 3D
halo mass, $m_\mathrm{200c}$, for all cosmologies in the hypercube in
the left panel of Fig.~\ref{fig:M_vs_m_avg}. Smaller apertures more
closely capture the 3D mass of lower-mass haloes, however, as is clear
from Fig.~\ref{fig:M_vs_m_dist}, there is a large scatter around the
median relation due to the differing matter distributions along the
line-of-sight to different haloes. For higher-mass haloes, measuring
the mass in different apertures allows the characterization of the
halo density profile, since the matter belonging to the halo dominates
the total aperture mass out to larger apertures.

In the right-hand panel of Fig.~\ref{fig:M_vs_m_avg}, we show the
redshift evolution of the aperture mass within a fixed aperture of
$\Rin=1\, \cmpc$. Since we measure within fixed comoving apertures,
the uncorrelated large-scale structure contribution to both $M(<\Rin)$
and $M_\mathrm{bg}(<\Rin)$ should be the same on average. Hence, the
redshift evolution is dominated by the local overdensity changes
around the halo. At fixed $m_\mathrm{200c}$, the virial radius
$r_\mathrm{200c}$ will increase less rapidly with increasing time than
the aperture radius does as the critical density---and also
$r_\mathrm{200c}(z)$---approaches a constant in the dark
energy-dominated era. As a result, the aperture mass increases with
time, since more matter outside of the halo is included within the
same comoving aperture at fixed halo mass. For angular apertures,
there would be an additional change due to the changing angular
diameter distance. For a halo mass defined with respect to the mean
matter density, such as $m_\mathrm{200m}$, the virial radius and the
comoving aperture radius do not evolve with redshift at fixed halo
mass and, hence, the redshift evolution would be set by the change in
the halo density profile.

\begin{figure*}
  \centering
  \includegraphics[width=\textwidth]{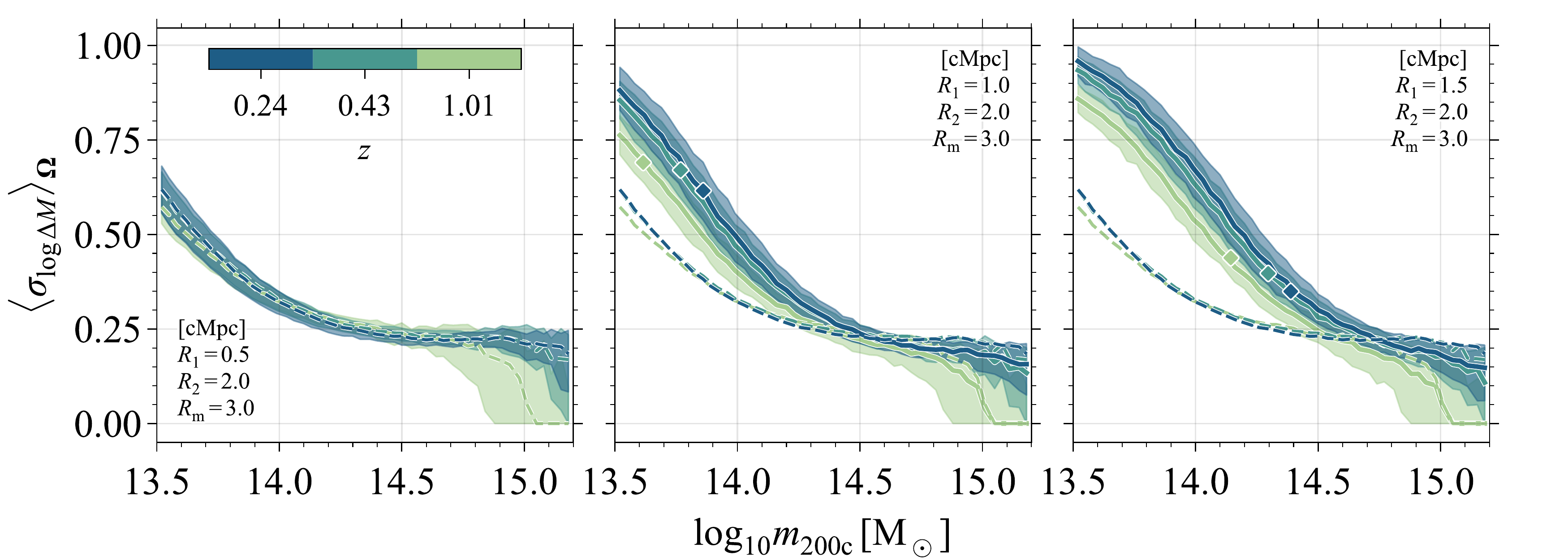}
  \caption{The median redshift evolution of the scatter in the
    $\Delta M$--$m_\mathrm{200c}$ distribution measured in apertures
    $\Rin \in[0.5, 1.0, 1.5] \, \cmpc$ for all simulations (\emph{left
      to right columns}). Coloured lines and shaded regions indicate
    the median and the 16th to 84th percentile scatter for all
    cosmologies at different redshifts. The $\Rin=0.5\, \cmpc$
    distribution is indicated with dashed lines and repeated in the
    other panels. The coloured diamonds show the median halo mass for
    which $r_\mathrm{200c} = \Rin$ (these masses are smaller than
    $10^{14} \, \msun$ for $\Rin < 1.5 \, \cmpc$). At fixed
    $m_\mathrm{200c}$ and $\Rin$, the scatter increases significantly
    with time for haloes whose virial radius, $r_\mathrm{200c}$, is
    not significantly larger than the aperture (low-mass haloes) or
    whose number density increases (high-mass haloes). Increasing
    $\Rin$ at fixed $m_\mathrm{200c}$ increases the scatter when the
    virial radius becomes comparable to the aperture due to the
    increased sensitivity to matter outside the halo.}
  \label{fig:sigma_delta_m_avg}
\end{figure*}
To study how the scatter in $\Delta M$ at fixed $m_\mathrm{200c}$
changes with cosmology and redshift, we show the redshift evolution of
the median scatter, $\sigma_{\log \Delta M}$, of all cosmologies in
the Mira--Titan suite for the different apertures in the panels of
Fig.~\ref{fig:sigma_delta_m_avg}. The shaded regions show the 16th to
84th percentile scatter. We indicate the median halo mass for which
$r_\mathrm{200c} = \Rin$ with a coloured diamond. The overall trends
are the same as in the bottom panel of Fig.~\ref{fig:M_vs_m_dist},
i.e. less scatter for higher-mass haloes. Within the smallest
aperture, $\Rin = 0.5 \, \cmpc$, there is very little redshift
evolution: the aperture is significantly smaller than
$r_\mathrm{200c}$ for all halo masses shown, and the halo matter
dominates the aperture mass. For all apertures, the increase in the
scatter with time for the most massive haloes results mainly from
their increasing number density with time. For the most massive
haloes, the scatter only changes by $\approx \pm 5 \, \percent$ for
different aperture sizes, as can be seen by comparing the dashed lines
(which are for $\Rin = 0.5 \, \cmpc$ in every panel) with the results
for larger apertures in the middle and rightmost panels of
Fig.~\ref{fig:sigma_delta_m_avg}. For lower-mass haloes, however, the
scatter is more sensitive to the aperture and increases when the halo
radius becomes comparable to the aperture.

So far, we have shown that aperture masses can be measured easily in
simulations and that they correlate strongly with the true, 3D halo
mass, albeit with a large intrinsic scatter due to their sensitivity
to the matter along the line-of-sight to the halo. Paradoxically, this
could give the aperture mass an advantage in the context of cluster
cosmology since it means that the line-of-sight structure contributes
to the aperture mass signal, not its noise. We will investigate the
possible strengths and difficulties of aperture mass calibrations for
cluster cosmology next.

\subsection{Uncertainties}\label{sec:aperture_uncertainty}
For cluster cosmology, it is crucial that cluster masses inferred from
observations can be calibrated accurately, that is without bias and,
ideally, also with small uncertainties. Due to the exponential
sensitivity of the halo abundance to the halo mass, biases and
uncertainties that are not accounted for in the cluster mass
measurement can introduce catastrophic biases in the inferred
cosmological parameters. Consequently, minimizing the uncertainty in
the mass--observable relation can dramatically increase the
constraining power of cluster surveys. Previously, we have shown that
the intrinsic scatter between the aperture mass and the 3D halo mass
can be large, particularly for low-mass haloes. We will now consider
the strengths and the difficulties of aperture masses for cluster
cosmology.

Taking Eq.~\eqref{eq:N_obs} as our guide, we see that the uncertainty
in the mass--observable relation is due to the uncertainty in the
relation between the measured observable and the measured aperture
mass, $P(\mathcal{O}|\mathcal{M}_\mathrm{obs})$, and the observational
uncertainty between the measured aperture mass and the true halo
aperture mass, $P(\mathcal{M}_\mathrm{obs}|\mathcal{M})$, sometimes
referred to in the literature as the intrinsic uncertainty
\citep[e.g.][]{becker2011}. First, we will look into the intrinsic
measurement uncertainty of the halo aperture mass, comparing it to
that of 3D halo masses.

The stringent requirements on the accuracy of the shear measurements
for future surveys mean that the finite number of background galaxies
used to sample the shear field and the source redshift distribution
set the baseline, minimum uncertainty for any weak lensing mass
measurement \citep[e.g.][]{kohlinger2015}. The source redshift
distribution determines the critical surface mass density that enables
the conversion from measured weak lensing shear to surface mass
density. This uncertainty will affect any weak lensing mass
measurement similarly, so we do not include it here. The uncertainty
of aperture mass measurements is then fully determined by the galaxy
shape noise, as shown by \citet{schneider1996}. In comparison, 3D halo
masses inferred from deprojected weak lensing observations are
intrinsically highly sensitive to the large variation in the
line-of-sight matter distribution at fixed, true 3D halo mass.

To quantify the intrinsic measurement uncertainties for 3D halo masses
of individual clusters, we look at the literature. \citet{bahe2012b}
have estimated the uncertainty of the
$P(\mathcal{M}_\mathrm{obs}|\mathcal{M})$ scaling relation by
generating mock weak lensing observations of clusters with
$m_\mathrm{200c} > 10^{14} \, \msun$ at $z \approx 0.2$, a shape noise
of $\sigma_\mathrm{gal}=0.2$, and with a mean lensed background galaxy
number density $\mean{n}_\mathrm{gal} = 30 \, \arcmin^{-2}$ for
sources at $z=1$. This set-up assumes perfect knowledge of the source
redshift distribution and the critical surface mass density. They find
a large uncertainty of $\sigma_{\log m_\mathrm{obs}} = 0.45 \, (0.25)$
for haloes with $m_\mathrm{200c}= 10^{14} \, (10^{15}) \, \msun$ when
inferring $m_\mathrm{obs}$ from fitting NFW density profiles to the
observed lensing shear. Importantly, \citet{bahe2012b} only include
the local, correlated large-scale structure within $10 \, \cmpc$ of
the halo when generating the lensing signal. However, uncorrelated
large-scale structures add to the scatter of the true lensing signal
\citep[e.g.][]{hoekstra2001, hoekstra2003}. Hence, their results
should be considered a lower limit on the true scatter in the inferred
3D halo masses. \citet{becker2011} similarly find an uncertainty of
$\sigma_{\log m_\mathrm{obs}} \approx 0.3$ for a mock sample with
$m_\mathrm{200c} > 10^{14.5} \, \mh$ that does include the cosmic
noise due to uncorrelated large-scale structure.

On the other hand, for the same set-up as \citet{bahe2012b}, weak
lensing aperture masses are only affected by the shape noise due to
the finite number of galaxies used to sample the shear field. More
specifically, the uncertainty is given by Eq.~\eqref{eq:sigma_delta_m}
in Appendix~\ref{app:weak_lensing}. We derive a fixed uncertainty
$\sigma_{\Delta M_\mathrm{obs}}=1.16 \times 10^{13} \, \msun$ for
$\Rin=0.5 \, \cmpc$, $\Rout = 2 \, \cmpc$, and $\Rmax = 3 \, \cmpc$.
For reference, from Fig.~\ref{fig:M_vs_m_avg} we see that
$\Delta M(m_\mathrm{200c} = 10^{14} \, \msun, \Rin=0.5 \, \cmpc)
\approx 10^{13.75} \msun$, implying a fractional uncertainty
$\sigma_{\log \Delta M_\mathrm{obs}} \approx 0.2$, i.e. more than $2$
times smaller than the fractional uncertainty in the 3D mass and
without any dependence on an assumed density profile. Importantly, the
fractional uncertainty scales inversely with the halo aperture mass,
giving fractional uncertainties of $\approx 0.1$ and $0.05$ for
$\Delta M/\msun = 10^{14}$ and $10^{14.5}$, respectively.

\begin{figure}
  \centering
  \includegraphics[width=\columnwidth]{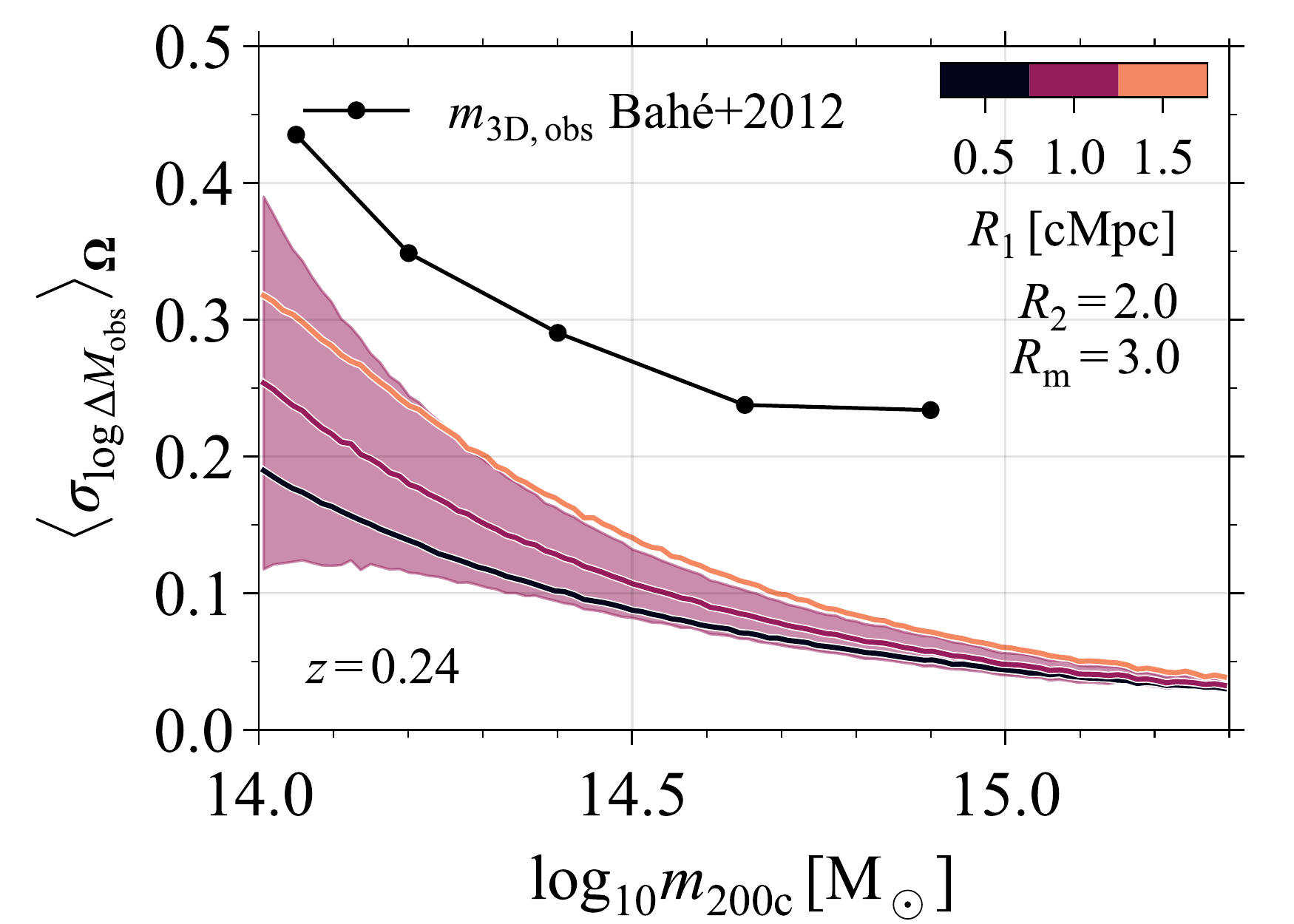}
  \caption{The median observational fractional uncertainty in the
    aperture mass at fixed halo mass within different apertures for a
    lensing cluster at $z=0.24$ and source galaxies at $z=1$ with a
    mean background density of
    $\mean{n}_\mathrm{gal} = 30 \, \arcmin^{-2}$ and shape noise
    $\sigma_\mathrm{gal} = 0.2$. The thick, coloured lines indicate
    the median uncertainty over all cosmologies for $m_\mathrm{200c}$
    and within different apertures. The shaded region shows the
    variation of the observational uncertainty for $\Rin = 1 \, \cmpc$
    due to the median scatter in $\Delta M$ at fixed $m_\mathrm{200c}$
    for all cosmologies, shown in Fig.~\ref{fig:sigma_delta_m_avg}.
    The black points show the scatter in the 3D masses inferred from
    mock weak lensing observations by \citet{bahe2012b}. Smaller
    apertures have a lower observational uncertainty due to the larger
    number of background galaxies as the masses are measured within
    $\Rin < R \leq \Rmax$. Aperture masses can be determined more
    precisely than 3D masses over the full halo mass range.}
  \label{fig:sigma_delta_m_obs}
\end{figure}
In Fig.~\ref{fig:sigma_delta_m_obs}, we show the aperture radius
dependence of the median fractional observational uncertainty,
$\sigma_{\log \Delta M_\mathrm{obs}}$, at fixed halo mass,
$m_\mathrm{200c}$, calculated from Eq.~\eqref{eq:sigma_delta_m}, for a
lensing cluster at $z=0.24$ and source galaxies at $z=1$ with
background density $n_\mathrm{gal}=30 \, \arcmin^{-2}$ and shape noise
$\sigma_\mathrm{gal}=0.2$, similar to \citet{bahe2012b}. The aperture
mass uncertainty in Eq.~\eqref{eq:sigma_delta_m} additionally depends
on the chosen filter, that is the aperture radii $\Rin$, $\Rout$, and
$\Rmax$. To obtain the fractional uncertainty, we divide
$\sigma_{\Delta M_\mathrm{obs}}$ from Eq.~\eqref{eq:sigma_delta_m} by
the aperture mass, $\Delta M$. For $\Rin = 1 \, \cmpc$, we indicate
the median uncertainty in the aperture mass at fixed $m_\mathrm{200c}$
over all cosmologies (the solid line in the middle panel of
Fig.~\ref{fig:sigma_delta_m_avg}) as the shaded region. For
comparison, we show the observational uncertainty in 3D halo masses
inferred from the mock weak lensing observations of \citet{bahe2012b}.
Over the entire halo mass range, the aperture mass can be determined
at least $2$ times more precisely than the 3D halo mass for apertures
similar to the halo radius. Increasing the inner aperture radius,
$\Rin$, increases the observational uncertainty since the weak lensing
signal is inferred from the smaller number of galaxies within $\Rin$
and $\Rmax$. Hence, aperture masses can be measured more cleanly from
observations than 3D halo masses since the line-of-sight structure
contributes to the signal as opposed to the noise.

The uncertainty in aperture mass calibrations for cluster surveys with
baryonic observables, such as the galaxy overdensity, the SZ signal or
the X-ray luminosity, will also depend on the relation between the
observable, $\mathcal{O}$, and the measured aperture mass,
$\Delta M_\mathrm{obs}$. As mentioned before, this relation depends
solely on observational properties of the clusters and the uncertainty
will be highly sensitive to the observable $\mathcal{O}$ under
consideration.

A particularly ill-suited scenario for aperture masses would be an
observable that is not sensitive to projection effects, such as the
X-ray luminosity or the thermal energy of the hot gas, $Y_X$. These
observables depend strongly on the gas density and predominantly trace
the cluster core. Due to the tight correlation with small scatter
between the X-ray luminosity and the 3D halo mass, $m$, the
uncertainty in $P(\mathcal{O}| \Delta M_\mathrm{obs})$ can be
approximated by $P(m|\Delta M_\mathrm{obs})$. As can be seen from the
spread in $m_\mathrm{200c}$ at fixed $\Delta M$ in the top panel of
Fig.~\ref{fig:M_vs_m_dist}, this uncertainty is considerable. Such an
observable is ideal for 3D halo mass calibrations. However, the
uncertainty between the observable, $\mathcal{O}$, and the true halo
mass, $m$, will still be limited by the uncertainty floor in
$P(m_\mathrm{obs}|m)$, set by the deprojection of the lensing profile.

In the best-case scenario for aperture masses, the observable closely
traces the total projected mass with small uncertainty.
\citet{andreon2014} show that the richness is such an observable when
measured within the same aperture as the weak lensing aperture mass.
Other studies also find that the stellar mass fraction, when measured
sufficiently far away from the brightest cluster galaxy, is
approximately constant in groups and clusters
\citep[e.g.][]{bahcall2014, budzynski2014, zu2015, wang2018d}. For
observables related to the stellar mass of clusters, aperture masses
provide mass calibrations with low uncertainty and without any model
dependence which is ideal for cluster cosmology.

We would also expect the SZ signal to be sensitive to projection
effects since it is independent of redshift, and since its pressure
dependence allows it to probe larger scales. However, the steep
scaling of the SZ signal with the 3D halo mass due to its scaling with
the gas temperature and density, means that low-mass haloes will
constitute an approximately constant background that can be corrected
for \citep[e.g.][]{angulo2012, lebrun2015}. Hence, the SZ signal is
likely less sensitive to projection effects than the cluster stellar
mass, but more sensitive than cluster X-ray properties.

A full comparison between the performance of aperture and 3D mass
calibrations for different survey observables would require generating
mock surveys and mimicking the aperture mass measurement and the 3D
mass inference from mock weak lensing observations, which is beyond
the scope of this work.

All in all, halo aperture masses provide clear advantages for cluster
cosmology. The direct connection between aperture masses measured from
simulations and observations make them practically independent from
assumptions about the density profile of clusters. Moreover, the
relation between the cluster observable of interest and the true
cluster aperture mass cleanly separates in a purely observational
scaling relation and an intrinsic measurement uncertainty between the
observed and the true aperture mass, which can be calibrated using
simulations. Next, we turn our attention to the final ingredient for
cluster cosmology in Eq.~\eqref{eq:N_obs}: the aperture mass function.

\section{Halo aperture mass function}
\label{sec:aperture_mass_function}
Having introduced the aperture mass and compared it to the 3D halo
mass, we now study the aperture mass function. We show how the
aperture mass function depends on the aperture mass in
Section~\ref{sec:n_M_behaviour}. Then, we briefly explain how we fit a
Gaussian process emulator to capture the cosmology dependence of the
aperture mass function in Section~\ref{sec:n_M_emulator}, leaving the
details of the implementation to Appendix~\ref{app:emulation} and the
verification to Appendix~\ref{app:emulator_performance}. Finally, we
discuss the cosmology sensitivity of the aperture mass function in
Section~\ref{sec:n_M_cosmo}.

\subsection{Aperture mass function behaviour}\label{sec:n_M_behaviour}
\begin{figure*}
  \centering
  \includegraphics[height=0.5\textwidth]{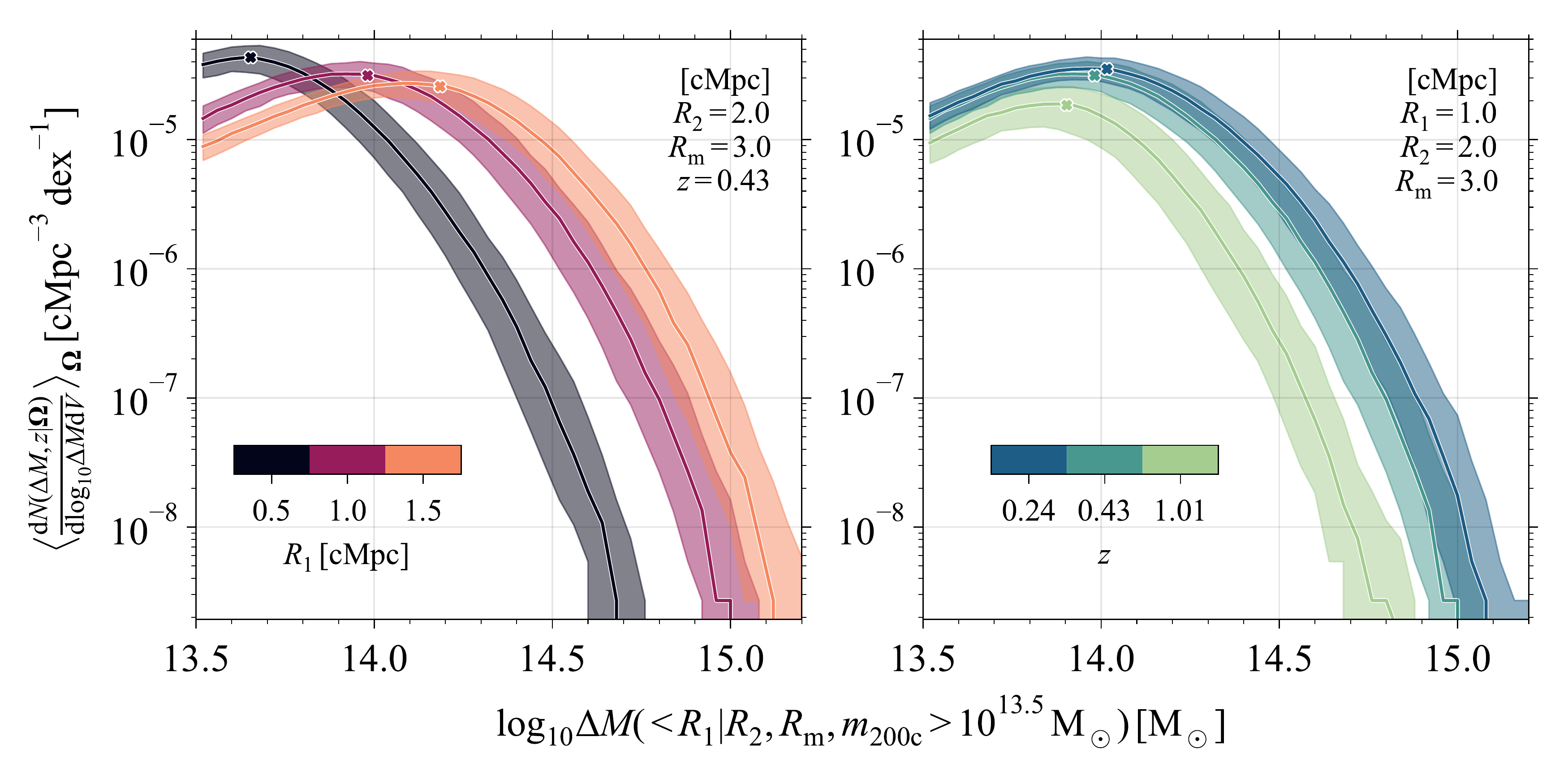}
  \caption{\emph{Left panel:} The median aperture mass function, for a fixed comoving volume, $n_V$, of all
    cosmologies in the hypercube for aperture masses measured within
    different apertures $\Rin$ (thick coloured lines) at $z=0.43$ for
    all haloes with $m_\mathrm{200c} > 10^{13.5} \, \msun$. The shaded
    regions show the 16th to 84th percentile scatter. The reference
    annulus for all aperture mass measurements spans the region
    between $\Rout=2\, \cmpc$ and $\Rmax=3\,\cmpc$. Larger apertures
    result in higher aperture masses and shift the aperture mass
    function to the right. The number density decreases for low
    aperture masses if a significant fraction of the haloes has 3D
    masses near the selection limit, indicated by the crosses that
    show the 84th percentile aperture mass for haloes with
    $m_\mathrm{200c,lim} = 10^{13.5} \, \msun$. \emph{Right panel:}
    The redshift evolution of the median aperture mass function with
    $\Rin=1 \, \cmpc$. The number density increases with time as
    haloes grow more massive. The peak of the aperture mass function
    shifts to larger values with time due to the increased scatter at
    the fixed 3D halo mass limit.}
  \label{fig:n_M_sims}
\end{figure*}
We compute the aperture mass function by dividing the number of haloes
in mass bins of $\log_{10} \Delta M$ by the simulated volume and the
bin width. The number density, $n$, dependent on the cosmological
parameters, $\mathbf{\Omega}_i$, can be defined either as a function
of the comoving volume, $V$,
\begin{align}
  \label{eq:n_comoving}
  n_V(\Delta M, z, \mathbf{\Omega}_i)  = \frac{\diff N(\Delta M, z, \mathbf{\Omega}_i)}{\diff V(z, \mathbf{\Omega}_i) \diff \log_{10} \Delta M} \,
\end{align}
or as a function of the probed survey volume
\begin{align}
  \label{eq:n_angular}
  n_\Omega(\Delta M, z, \mathbf{\Omega}_i) & = \frac{\diff N(\Delta M, z, \mathbf{\Omega}_i)}{\diff \Omega(z, \mathbf{\Omega}_i) \diff z \diff \log_{10} \Delta M} \, .
\end{align}
We have introduced the cosmology-dependent differential solid angle,
$\diff \Omega$, and the redshift range, $\diff z$. For cosmological
simulations, $n_V$ naturally matches the data since we can divide the
mass-binned number counts directly by the comoving simulation volume.
The growth of structure from the initial density field fixes the
cosmology dependence of the volumetric number density, $n_V$. The
cosmology dependence of the \emph{observed} halo number density,
however, receives an additional geometric contribution since we
observe our past lightcone. We obtain the observed number density from
the volumetric number density as
\begin{align}
  \label{eq:nV2nOmega}
  n_\mathrm{\Omega}(\Delta M, z, \mathbf{\Omega}_i) & = n_V(\Delta M, z, \mathbf{\Omega}_i) \frac{\diff V(z, \mathbf{\Omega}_i)}{\diff \Omega \diff z} \, ,
\end{align}
where the geometric conversion depends on the comoving distance and
the transverse comoving distance at redshift $z$ for the assumed
cosmology. The conversion scales the amplitude of the volumetric
aperture mass in a cosmology and redshift-dependent way. The same
geometric factor also applies to the simulated 3D halo mass function.

Since the weak lensing aperture mass receives contributions from
structure along the past lightcone weighted by the lensing kernel,
technically, the scatter in the aperture mass at fixed halo mass adds
a geometry sensitivity to the volumetric aperture mass function.
However, as we have shown in Fig.~\ref{fig:delta_m_los}, for
higher-mass haloes this scatter becomes less important compared to the
intrinsic scatter due to the differing matter distribution close
($L \lesssim 30 \, \cmpc$) to the cluster. Hence, neglecting the past
lightcone should not significantly change our conclusions.

In what follows, we will initially show results for $n_V$ as is
generally done for the 3D halo mass function in the literature to aid
in the interpretation of our results. However, only $n_\Omega$
includes the full cosmology dependence of both the aperture mass
function and the 3D halo mass function. We will use $n_\Omega$ to
investigate the cosmology sensitivity of the aperture mass function in
Section~\ref{sec:n_M_cosmo}.

In the left-hand panel of Fig.~\ref{fig:n_M_sims}, we show the median
aperture mass function, $n_V$, and its 16th to 84th percentile scatter
for all cosmologies in the parameter hypercube and aperture masses
measured within different apertures. All aperture masses have been
computed with the same control annulus between $\Rout = 2 \, \cmpc$
and $\Rmax = 3 \, \cmpc$, and only haloes with
$m_\mathrm{200c} > m_\mathrm{200c,lim}=10^{13.5} \, \msun$ are
included within the sample. Since larger apertures will result in
higher aperture masses for the same halo, increasing the aperture size
shifts the aperture mass function to higher aperture masses. The
aperture mass function decreases towards both high and low aperture
masses. The former is caused by the rarity of high-mass haloes and the
latter by the halo mass selection of the sample and the large scatter
in aperture mass at fixed halo mass. When a significant fraction of
the haloes at fixed aperture mass has 3D masses near the selection
limit, the number density starts decreasing. We show this by
highlighting the 84th percentile aperture mass for haloes with 3D
masses at the selection limit with a cross. These crosses coincide
almost perfectly with the peak in the aperture mass function. The
right-hand panel of Fig.~\ref{fig:n_M_sims} shows that the aperture
mass function increases with redshift as more massive haloes form,
just like the traditional halo mass function does. The peak of the
aperture mass function shifts towards higher aperture masses with time
due to the increased scatter at the fixed 3D halo mass limit (see
Fig.~\ref{fig:sigma_delta_m_avg}).

\begin{figure}
  \centering
  \includegraphics[width=\columnwidth]{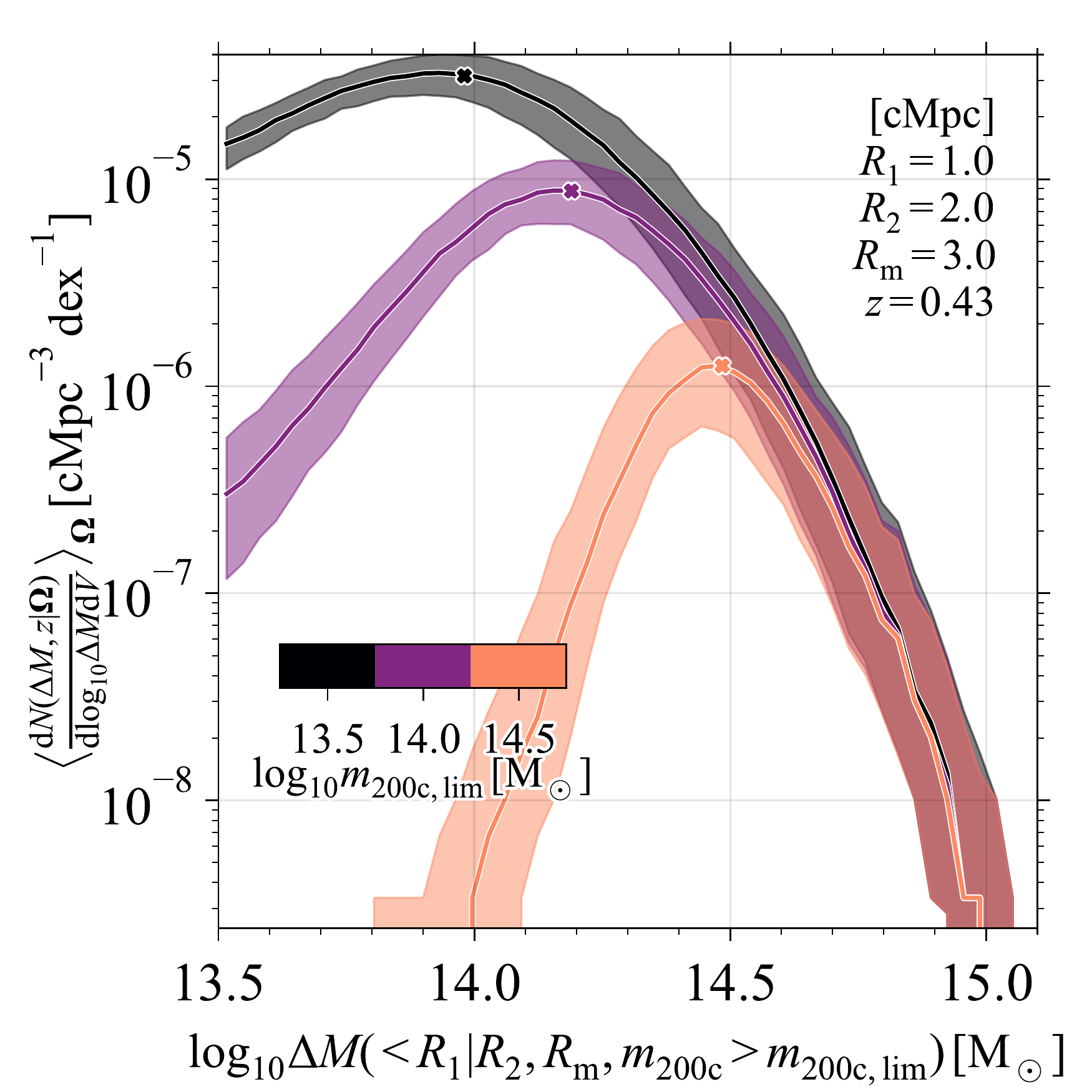}
  \caption{The change in the median aperture mass function for a fixed comoving volume, $n_V$, for all
    cosmologies at fixed aperture size when increasing the mass
    selection limit, $m_\mathrm{200c,lim}$. The thick, coloured lines
    show the different mass limits, $m_\mathrm{200c,lim}$. The crosses
    indicate the 84th percentile aperture mass for haloes with
    $m_\mathrm{200c} = m_\mathrm{200c,lim}$. The scatter in the
    aperture mass for haloes at the mass limit sets the peak of the
    aperture mass function.}
  \label{fig:n_M_m200c_lim}
\end{figure}
In Fig.~\ref{fig:n_M_m200c_lim}, we show how the aperture mass
function changes when increasing the 3D mass limit,
$m_\mathrm{200c,lim} / \msun$, from $10^{13.5}$ to $10^{14.5}$. The
number density for the largest aperture mass haloes is not strongly
affected since the scatter in the aperture mass at the mass limit
decreases with increasing mass limit. For all mass limits, the cross
indicates the 84th percentile aperture mass for haloes with 3D masses
at the selection limit. Since the $\Delta M$--$m_\mathrm{200c}$
relation is sublinear, the median aperture mass at
$m_\mathrm{200c,lim}$, and, therefore, also the peak mass increase
less strongly than the 3D halo mass when increasing
$m_\mathrm{200c,lim}$. The number density for aperture masses beyond
the peak is still affected by the mass limit, albeit less so. Hence,
for aperture mass cosmological analyses, it will be important to
select clusters using observables that either have small scatter with
respect to the aperture mass, or whose scatter is well-understood.

We stress that the haloes in Fig.~\ref{fig:n_M_m200c_lim} are selected
solely based on their 3D halo mass. However, low-mass haloes that
scatter to much higher aperture masses than the median relation for
their halo mass, are either part of the correlated structure or chance
alignments with a massive cluster. In realistic observational
scenarios, such haloes would not be part of the cluster sample, as
they would blend in with the larger cluster. However, this also
requires such haloes to be excluded from the theoretical aperture mass
function calculation. The same problem applies to the 3D mass
function; end-to-end pipelines are needed to model such effects.

\begin{figure}
  \centering
  \includegraphics[width=\columnwidth]{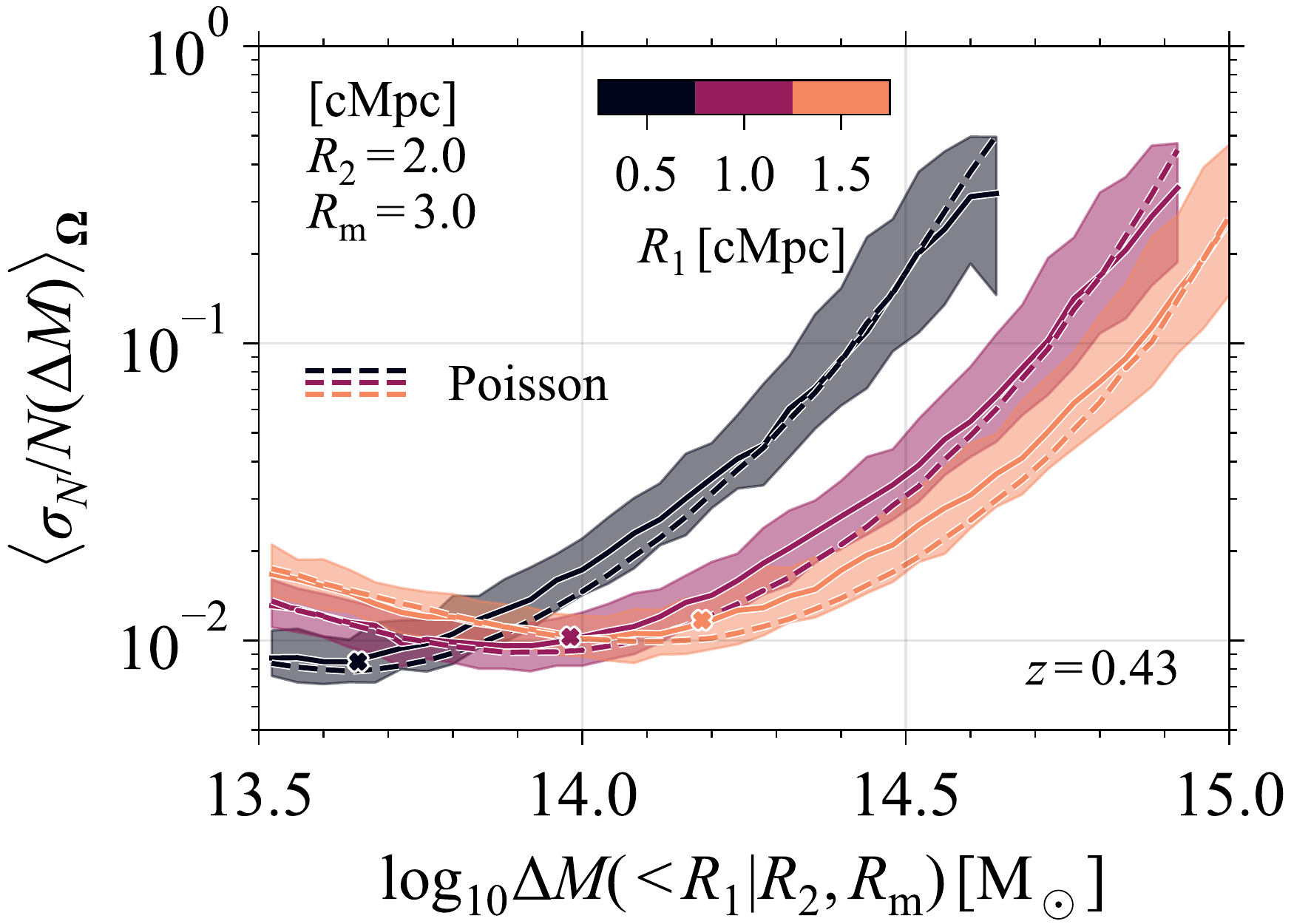}
  \caption{The median fractional variance in the aperture mass
    function for different apertures for all cosmologies at $z=0.43$
    (solid, coloured lines). All haloes are selected to have
    $m_\mathrm{200c} > 10^{13.5} \, \msun$. The reference annulus for
    all aperture mass measurements spans the region between
    $\Rout=2\, \cmpc$ and $\Rmax=3\,\cmpc$. The shaded regions show
    the 16th to 84th percentile scatter and the dashed lines show the
    median shot-noise expectation. The crosses indicate the peak of
    the aperture mass function. The aperture mass function variance
    generally exceeds the shot-noise.}
  \label{fig:n_M_variance}
\end{figure}
Finally, we investigate the sample variance of the aperture mass
function, which we will need to accurately calibrate the emulator.
Since large-scale modes can locally and coherently boost or suppress
the number counts, the variance of the aperture mass function needs to
be estimated by resampling the data over sufficiently large volumes
that include the inherent correlation structure. \citet{crocce2010}
and \citet{smith2011b} have shown that the 3D halo mass function
variance is dominated by Poisson noise at high halo masses, and that a
jackknife-type resampling can recover the true variance accurately.
For this reason, we use bootstrap resampling to divide the projected
mass maps into $(n, n)$ subregions. We then compute the aperture mass
function variance for $10,000$ halo samples generated by including
$n^2=25$ randomly chosen, possibly repeating, subareas. This way, we
can estimate the sample variance of the aperture mass function for
cluster samples obtained from an equal simulation volume.

We show the bootstrapped fractional aperture mass variance in
Fig.~\ref{fig:n_M_variance}. We also include the Poisson expectation
based on the number of haloes at fixed aperture mass. We find that the
sample variance of the aperture mass function exceeds the Poisson
expectation by up to a factor of $\approx 1.5$, except for the lowest-
and highest-aperture mass haloes. We will use the bootstrapped
variance estimates for the individual simulations when fitting the
aperture mass function emulator in the following Section.

\subsection{Emulating the aperture mass function}\label{sec:n_M_emulator}
We construct an emulator to infer the general cosmology dependence of
the aperture mass function from the available grid of cosmological
parameters. Usually, emulators fit some compressed form of the true
underlying data, such as the cosmology dependence of either the
parameters of a theoretical fitting function
\citep[e.g.][]{mcclintock2019a} or the weights of the principal
components of either the data or some functional approximation
\citep[e.g.][]{bocquet2020}. However, all these methods assume that
those compressed models accurately capture the underlying halo mass
function behaviour for all masses. While this assumption can be
checked as long as haloes are abundant, it might not hold in the
exponentially declining tail which contains important cosmological
information, potentially resulting in confident but inaccurate
predictions.

We therefore fit a Gaussian process \emph{directly} to the simulated
data at each redshift independently, only assuming Gaussian
correlations in the latent function and a discrete likelihood for the
\emph{observed} number counts. Previously, fitting a Gaussian process
directly to large datasets with non-Gaussian likelihoods was not
feasible: there was no well-understood and unified way to both account
for general, non-Gaussian likelihoods, and deal with the
computationally intensive inversion of the covariance matrix in the
model optimization. However, since the work of \citet{titsias2009} and
\citet{hensman2014}, this is no longer an issue. We gain a subtle but
important advantage by modelling the number counts directly with a
Gaussian process: the high-mass tail of cosmological models with no
observed clusters can be fit consistently with the correct likelihood
and without assuming any functional form for the aperture mass
function.

We provide a detailed description of our emulator implementation and
the performance in Appendices~\ref{app:emulation}
and~\ref{app:emulator_performance}, respectively, but detail the main
insights here. Briefly, we will fit the normalized aperture mass
function
\begin{equation}
  \label{eq:f_latent}
  f(\vect{x}_i=(\Delta M, \mathbf{\Omega}_i)^T) = \log n(\Delta M, \mathbf{\Omega}_i) - \log \amean{n(\Delta M, \mathbf{\Omega}_i)}_\mathbf{\Omega} \, ,
\end{equation}
to reduce the dynamic range and the impact of the peak in the aperture
mass function on the emulator calibration. We have checked that
training the emulator on $n_V$ and $n_\Omega$, defined in
Eqs.~\eqref{eq:n_comoving} and~\eqref{eq:n_angular}, respectively,
gives consistent performance. Then, we assume a Gaussian process prior
for the mean and the variance of $f$
\begin{align}
  \label{eq:f_mean}
  \mathds{E}[f(\vect{x}_i)] &= \mu \\
  \label{eq:f_var}
  \mathrm{Var}[f(\vect{x}_i), f(\vect{x}_j)] & = k(\vect{x}_i, \vect{x}_j) \, ,
\end{align}
where $k(\vect{x}_i, \vect{x}_j)$ is the covariance function between
inputs $\vect{x}_i$ and $\vect{x}_j$. We will be using the radial
basis function (or squared exponential) kernel for $k$:
\begin{equation}
  \label{eq:rbf}
  k(\vect{x}, \vect{x}^\prime) = \sigma^2 \prod_{i=0}^{d} \exp\left( -\frac{((\vect{x})_i - (\vect{x}^\prime)_i)^2}{2\ell_{i}^2} \right) \, ,
\end{equation}
where $i$ runs over the $d=9$ dimensions of $\vect{x}$ and each
dimension has its own covariance lengthscale $\ell_i$, resulting in
hyperparameters $\theta = (\mu, \sigma^2, \bm{\ell})$. The
hyperparameters, $\theta$, can be optimized to accurately capture the
cosmology dependence of the aperture mass function, assuming the
likelihood of the simulated number counts, $(\vect{x}_i, N_i)$, given
the model, $f(\vect{x}_i)$.

We leave the details of optimizing this Gaussian process to
Appendix~\ref{app:emulation}, but the scalable, variational inference
method developed by \citet{titsias2009} and \citet{hensman2014} allows
us to fit directly to the large, simulated dataset, assuming a
discrete likelihood that naturally matches the simulated number
counts, meaning that we do not need to assume any functional form for
the aperture mass function.

We find that the Gaussian process emulator is able to predict most of
the simulated aperture mass functions to within $\pm 2 \, \percent$ in
the high-abundance regime and to within the shot-noise for
high-aperture masses (see Fig.~\ref{fig:n_M_perf}). The emulator also
generalizes well in a leave-one-out-test as it is generally able to
predict most simulations within $\pm 5 \, \percent$ when not including
them in the emulator calibration (see Fig.~\ref{fig:n_M_loo}).

At this point, we are satisfied with the emulator performance in
capturing the underlying cosmology dependence of the aperture mass
function. However, we want to reiterate that our goal has not been to
calibrate the emulator to the level of accuracy required for future
surveys. Such an emulator needs to be calibrated specifically to the
survey specifications such as the chosen angular aperture size, the
probed redshift range, the selection function of the observable, and
needs to compute the aperture masses from the full past lightcone. We
require the emulator only to be able to investigate how varying
individual cosmological parameters affects the aperture mass function.

\subsection{Cosmology dependence of the aperture mass function}\label{sec:n_M_cosmo}
\begin{figure*}
  \centering
  \includegraphics[width=\textwidth]{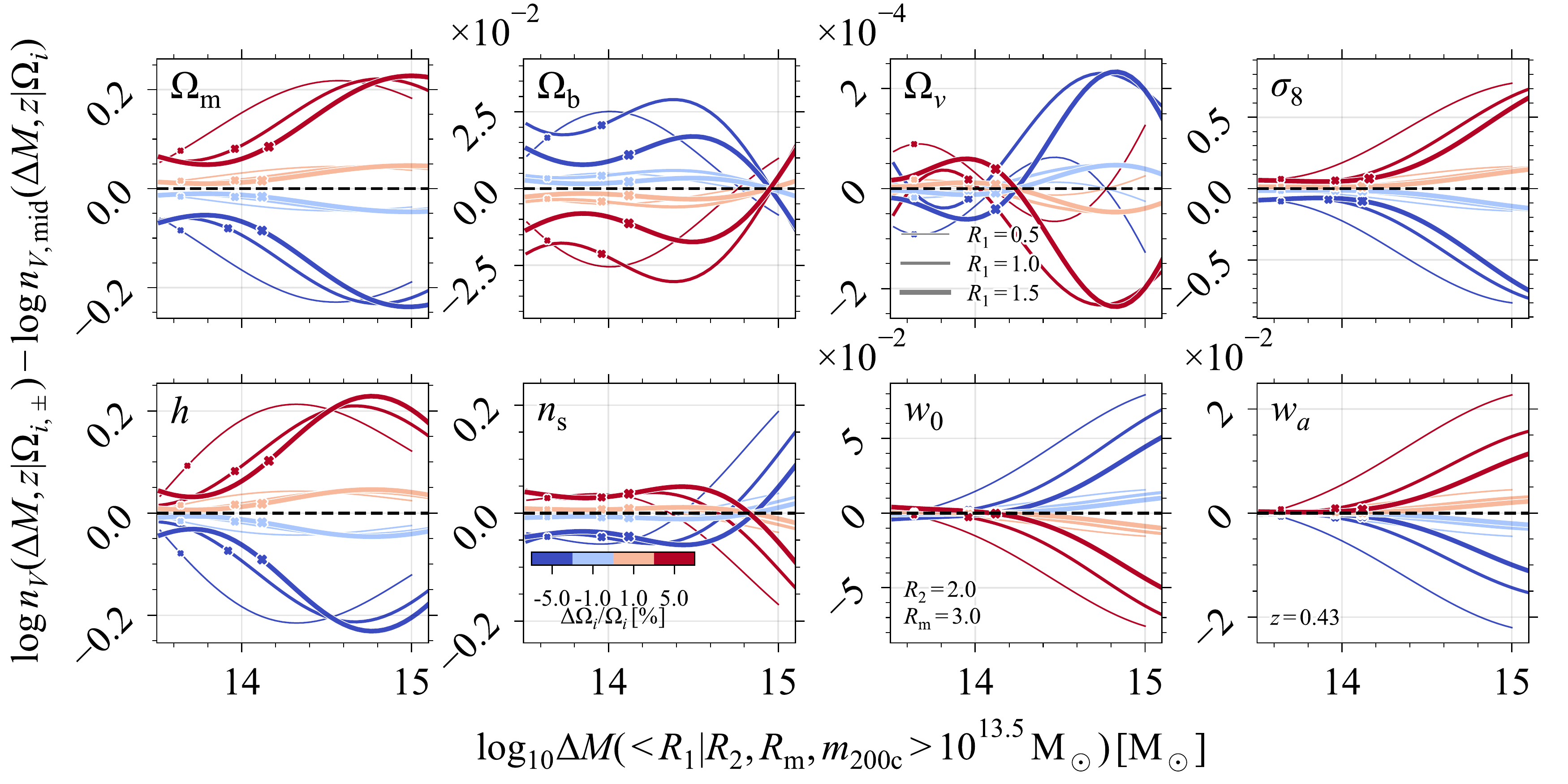}
  \caption{The sensitivity of the aperture mass function for a fixed comoving volume, $n_V$, to changing
    individual cosmological parameter values by $\pm 1, 5 \, \percent$
    at $z=0.43$. The cosmological parameter being varied is indicated
    in the top-left corner of each panel (note the different $y$-axis
    scaling for the different parameters). We assume
    $w_a=\pm 0.01, \pm 0.05$ since the fiducial value is 0. Coloured
    lines indicate the fractional change in the individual
    cosmological parameters with respect to the fiducial
    \citet{planckcollaboration2020} cosmology, with the line thickness
    varying from thin to thick for $\Rin=0.5, 1.0, 1.5 \, \cmpc$. All
    aperture masses were measured with
    $(\Rout, \Rmax) = (2, 3) \, \cmpc$ and all haloes have
    $m_\mathrm{200c} > 10^{13.5} \, \msun$. The peak of the aperture
    mass function is indicated with a cross. Increasing the aperture
    size mainly shifts the aperture mass function to higher masses.
    The amplitude also changes noticeably for different aperture sizes
    when varying $\Ob$ and $\On$. The aperture mass function is most
    sensitive to changes in $\sigma_8$, $\Om$, and $h$, with
    additional sensitivities to the scalar spectral index of the
    initial power spectrum, $n_\mathrm{s}$, and the dark energy
    equation-of-state parameters $w_0$ and $w_a$.}
  \label{fig:n_M_cosmo_R1}
\end{figure*}
We can use the calibrated emulator to investigate the cosmological
sensitivity of the aperture mass function. Previously,
\citet{marian2009, marian2010} showed that the aperture mass function
for a filter that optimizes the cluster SNR, closely follows the
cosmology dependence of the 3D mass function, suggesting a similar
cosmology sensitivity. However, their chosen filter required assuming
a typical density profile for clusters, which we have been careful to
avoid.

Fig.~\ref{fig:n_M_cosmo_R1} shows the sensitivity of the volumetric
aperture mass function to changes in individual cosmological
parameters (different panels) and the aperture (different line
thickness) at fixed redshift. We reiterate that the full cosmology
dependence of the \emph{observed} aperture mass function also depends
on the geometry through the volume of the past lightcone, as
Eq.~\eqref{eq:nV2nOmega} shows. We adopt a fiducial
\citet{planckcollaboration2020} cosmology with
$\mathbf{\Omega} \equiv \{\Om, \Ob, \On, \sigma_8, h, n_\mathrm{s},
w_0, w_a\} = \{0.315, 0.049, 0.0014, 0.811, 0.674, 0.965, -1, 0\}$,
with $\On$ corresponding to $M_\nu=0.06 \, \mathrm{eV}$, and
separately vary each of the cosmological parameters by $\pm 1$ and
$5 \, \percent$ (different colours). For $w_a$, we assume fixed values
$\pm 0.01$ and $\pm 0.05$, since the fiducial value is 0. In agreement
with the 3D halo mass function, to which we explicitly compare in
Fig.~\ref{fig:n_m_ap_vs_n_m200c_cosmo}, the shape of the aperture mass
function at fixed redshift is most sensitive to changes in $\sigma_8$
and $\Om$, with a $\pm 1 \, \percent$ change in $\sigma_8$ ($\Om$)
resulting in $> 10 \, \percent$ (up to $5 \, \percent$) changes in the
aperture mass function. Besides $\Om$ and $\sigma_8$, the aperture
mass function is also sensitive to both the dimensionless Hubble
parameter, $h$, and the scalar spectral index of the linear power
spectrum, $n_\mathrm{s}$. The equation-of-state parameters, $w_0$ and
$w_a$, mainly affect the abundance of high-aperture mass haloes.
Increasing the aperture size shifts the aperture mass function to
larger aperture masses. However, apart from this approximate shift for
different aperture sizes, the amplitude of the aperture mass function
also changes noticeably for $\Ob$ and $\On$.

\begin{figure*}
  \centering
  \includegraphics[width=\textwidth]{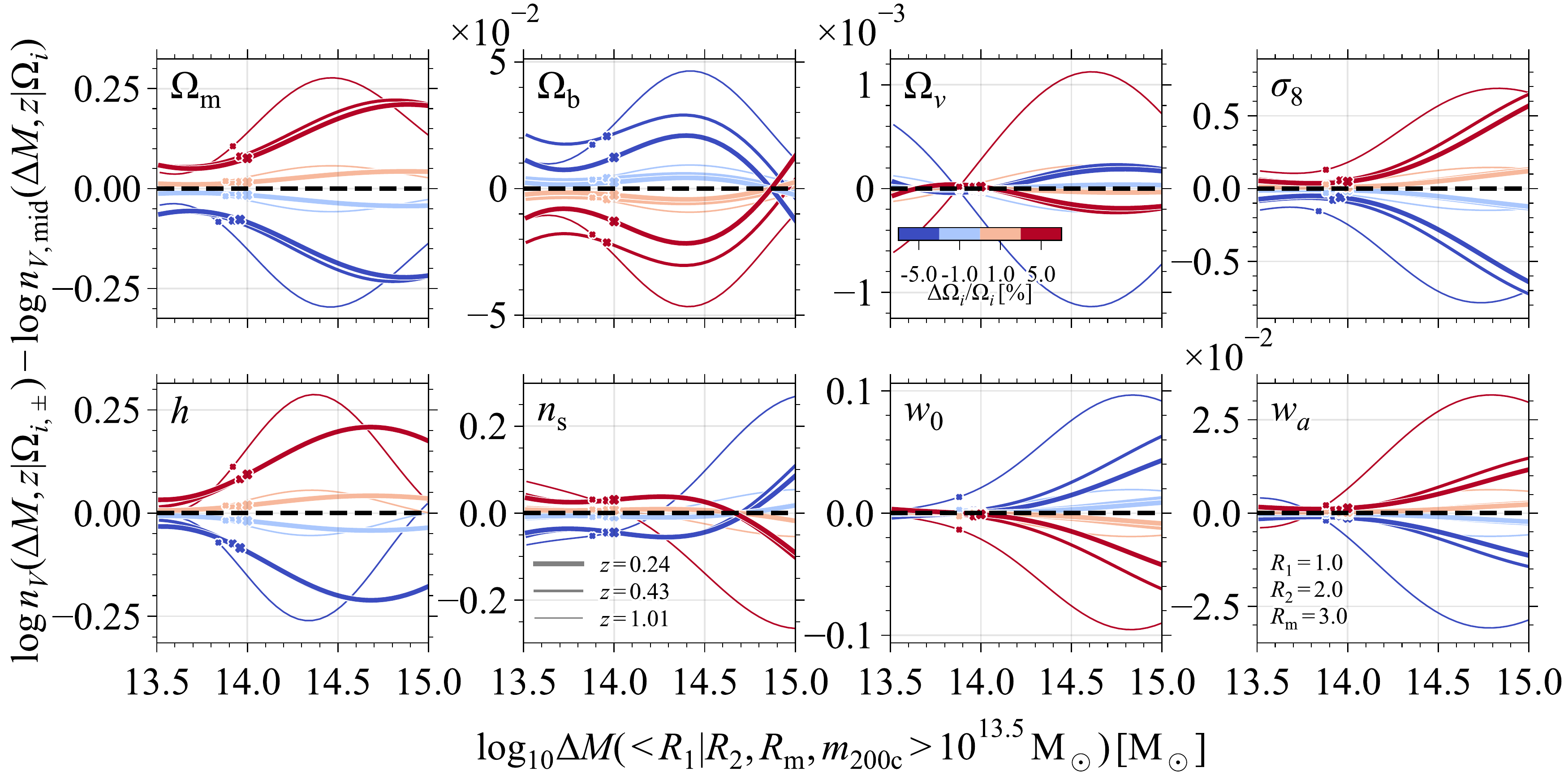}
  \caption{The sensitivity of the aperture mass function for a fixed comoving volume, $n_V$, to changing
    individual cosmological parameter values by $\pm 1, 5 \, \percent$
    for $\Rin=1 \, \cmpc$ and different redshifts. We assume
    $w_a=\pm 0.01, \pm 0.05$ since the fiducial value is 0. The
    cosmological parameter being varied is indicated in the top-left
    corner of each panel. Coloured lines indicate the fractional
    change in the individual cosmological parameters with respect to
    the fiducial \citet{planckcollaboration2020} cosmology, with the
    line thickness varying from thick to thin for the redshifts
    $z = 0.24, 0.43, 1.01$. All aperture masses were measured with
    $(\Rin, \Rout, \Rmax) = (1, 2, 3) \, \cmpc$, and all haloes have
    $m_\mathrm{200c} > 10^{13.5} \, \msun$. The peak of the aperture
    mass functions is indicated with a cross. The peak of the mass
    function shifts to higher masses for lower redshifts. The aperture
    mass function is most sensitive to changes in $\sigma_8$, $\Om$,
    and $h$. The relative impact of changing the cosmological
    parameters on the abundance increases with redshift.}
  \label{fig:n_M_cosmo_z}
\end{figure*}
In Fig.~\ref{fig:n_M_cosmo_z}, we show the cosmology sensitivity of
the aperture mass function for masses measured within
$\Rin = 1 \, \cmpc$ at different redshifts. At all redshifts, the
aperture mass function is most sensitive to changes in $\sigma_8$,
$\Om$, and $h$. For most cosmological parameter changes, the abundance
changes more strongly at higher redshifts. Noticeably, the dark energy
equation-of-state parameters affect the halo abundance more
significantly at higher redshifts. The peak of the aperture mass
function, which is indicated with a cross, shifts to higher aperture
masses with decreasing redshift.

\begin{figure*}
  \centering
  \includegraphics[width=\textwidth]{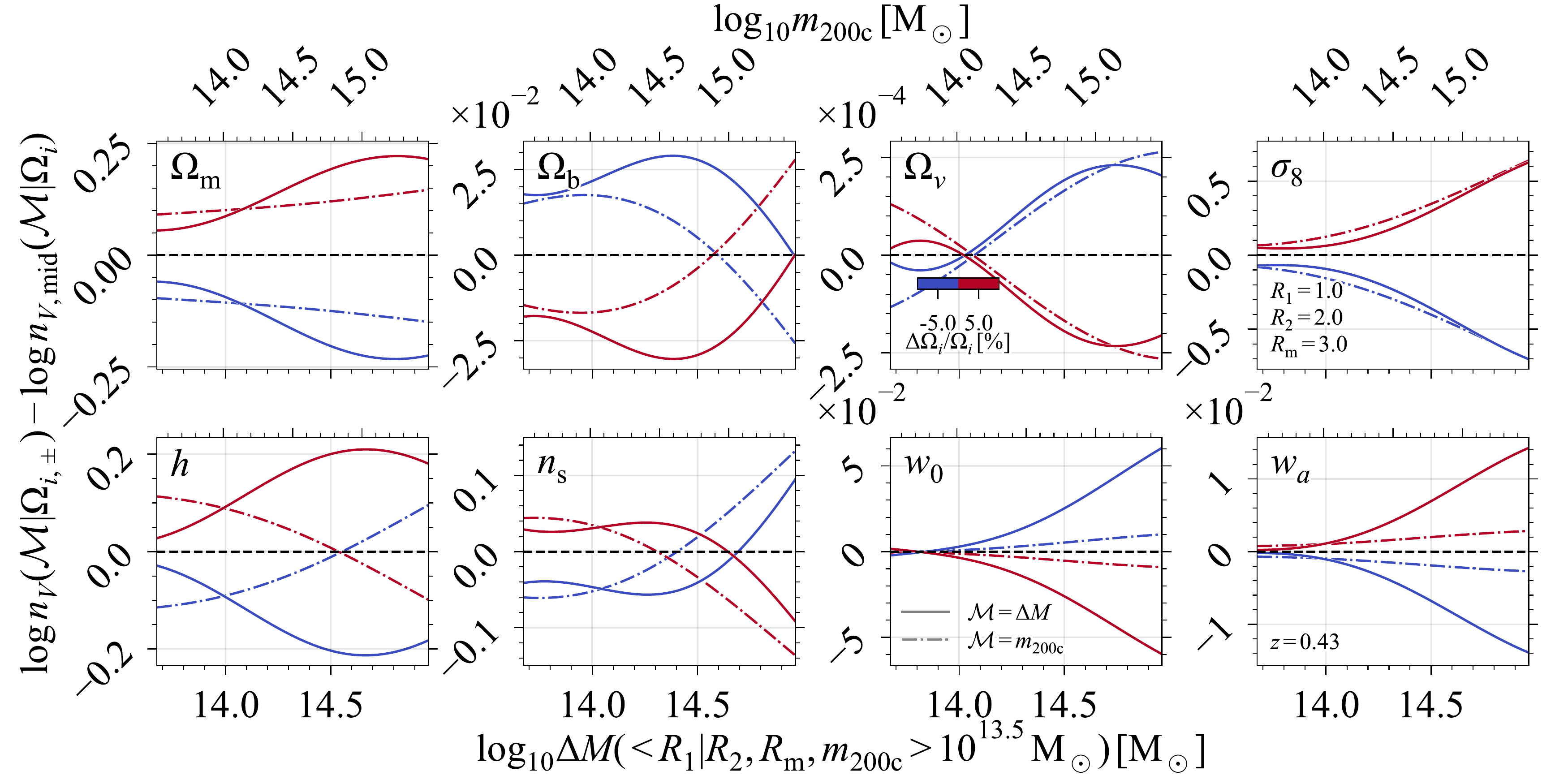}
  \caption{Comparison between the cosmology sensitivity of the 3D halo
    mass function (dash-dotted lines) and the aperture mass function
    (solid lines), for a fixed comoving volume, to changing individual cosmological parameter values
    by $\pm 5 \, \percent$ at $z=0.43$. We assume $w_a=\pm 0.05$ since
    the fiducial value is 0. The aperture mass function is plotted for
    the median aperture mass at $m_\mathrm{200c}$ for all the
    simulations in the hypercube, and for apertures
    $(\Rin, \Rout, \Rmax) = (1, 2, 3) \, \cmpc$. All haloes have
    $m_\mathrm{200c} > 10^{13.5} \, \msun$. The cosmological
    parameters are indicated in the top-left corner of each panel.
    Coloured lines indicate the fractional change in the individual
    cosmological parameters with respect to the fiducial
    \citet{planckcollaboration2020} cosmology. Both the 3D mass and
    the aperture mass function show similar sensitivity to changes in
    the $\sigma_8$, $n_\mathrm{s}$ and $\On$. The aperture mass
    function is more sensitive to changes in $\Om$, $h$, the dark
    energy equation-of-state parameters, $w_0$ and $w_a$, and $\Ob$.}
  \label{fig:n_m_ap_vs_n_m200c_cosmo}
\end{figure*}
The dominant cosmology dependence of the aperture mass function can be
understood from the 3D halo mass function, since
\begin{equation}
  \label{eq:n_M_from_n_m}
  n(\Delta M, z|\mathbf{\Omega}) = \int_0^\infty \diff m_\mathrm{200c} \, n(m_\mathrm{200c}, z|\mathbf{\Omega}) P(\Delta M, z|m_\mathrm{200c}, \mathbf{\Omega}) \, .
\end{equation}
The large scatter in aperture mass at fixed 3D halo mass does cause
differences in the detailed mass dependence. In
Fig.~\ref{fig:n_m_ap_vs_n_m200c_cosmo}, we compare the cosmology
sensitivity of the 3D halo mass function (dash-dotted lines) and
aperture mass function (solid lines) for the median aperture mass at
$m_\mathrm{200c}$ for all cosmologies in the hypercube,
$\amean{\Delta M | m_\mathrm{200c}}_\mathbf{\Omega}$. The individual
cosmological parameters vary by $\pm 5 \, \percent$ around the
\citet{planckcollaboration2020} best-fit parameters (coloured lines in
the different panels). For the 3D halo mass function, the peak height
of haloes determines their abundance, with more significant peaks
being less abundant. Increasing $\sigma_8$ while fixing the remaining
cosmological parameters boosts the average variance on all scales
equally, which decreases the peak height at all halo masses and
results in an increased abundance, as can be seen in the top-right
panel of Fig.~\ref{fig:n_m_ap_vs_n_m200c_cosmo}. In the exponentially
declining tail, the constant decrease in the peak height increases the
abundance more dramatically. The aperture mass function follows these
trends.

When changing the other cosmological parameters, it is important to
remember that we fix $\sigma_8$, implying that the initial
normalization of the matter power spectrum, $A_\mathrm{s}$, does
change. Fixing $\sigma_8$ instead of $A_\mathrm{s}$ reduces the impact
of changing the other cosmological parameters on the mass function.
Increasing $\Om$ in a flat universe will result in deeper dark matter
potential wells, a faster growth of structure, and a delayed onset of
dark energy domination. The peak height decreases for all haloes,
resulting in higher abundances. The top-left panel of
Fig.~\ref{fig:n_m_ap_vs_n_m200c_cosmo} shows that the abundance of
low-aperture mass haloes changes less than the 3D halo mass function
for low halo masses due to the increasing incompleteness at fixed, low
aperture mass (see Fig.~\ref{fig:n_M_m200c_lim}). At high aperture
masses the large scatter in aperture mass at fixed $m_\mathrm{200c}$
results in a larger sensitivity of the aperture mass function compared
to the 3D halo mass function due to the contribution of abundant
low-mass haloes.

Increasing $h$ at fixed $\Om$ increases the density which results in
faster structure formation and makes haloes at fixed $m_\mathrm{200c}$
more compact, decreasing their peak height and increasing their
abundance. The aperture mass function is significantly more sensitive
to changes in $h$ than the 3D halo mass function. Increasing the
scalar spectral index, $n_\mathrm{s}$, at fixed $\sigma_8$ shifts the
power from large to small scales, resulting in more low-mass and fewer
high-mass haloes for both the 3D and the aperture mass function.
Finally, increasing the magnitude of the equation-of-state parameter
of dark energy, $w_0$, dampens the growth of the most massive haloes,
reducing their abundance. Again, the aperture mass function is more
sensitive to these changes than the 3D halo mass function.

\begin{figure*}
  \centering
  \includegraphics[width=\textwidth]{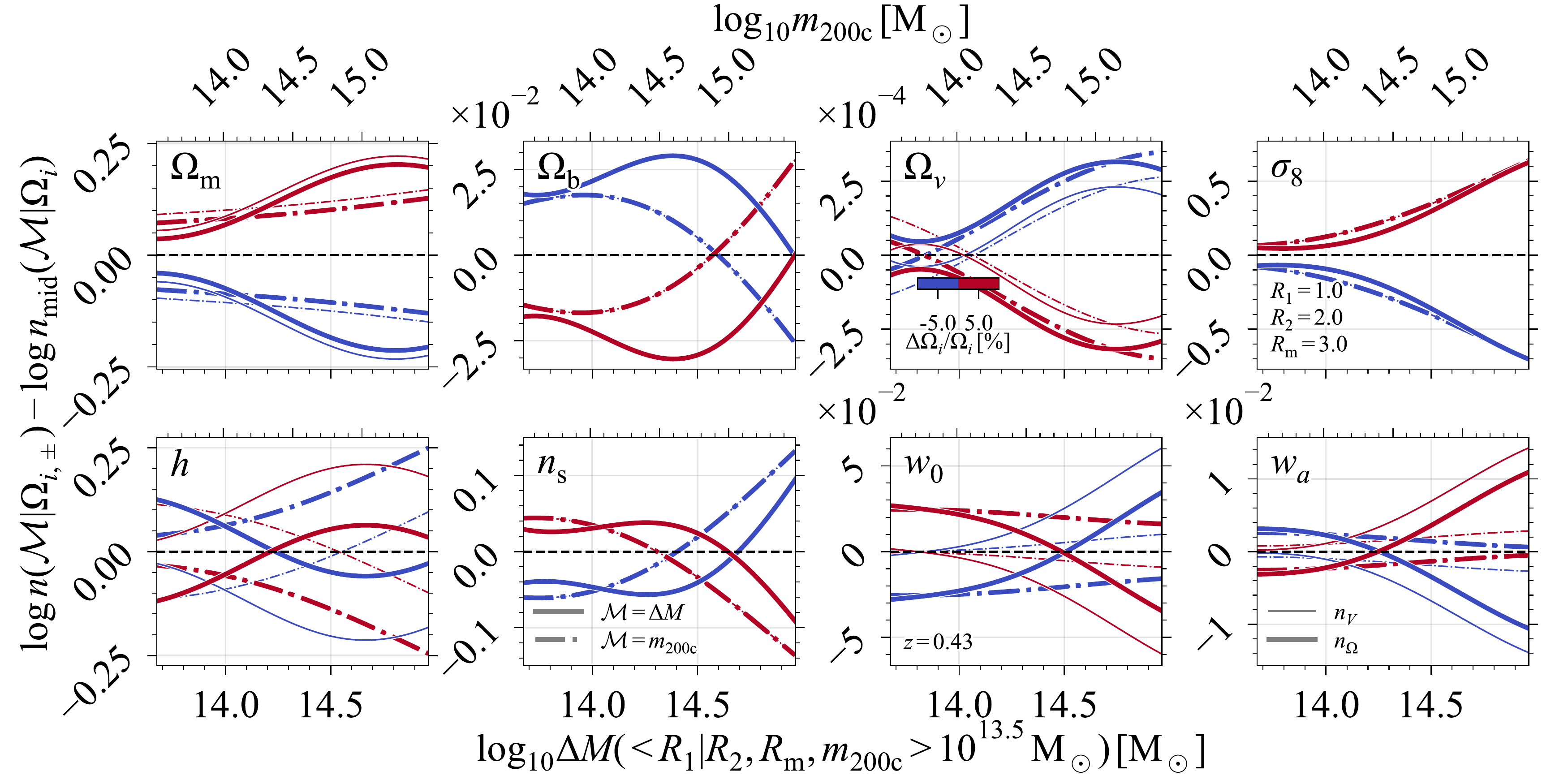}
  \caption{Comparison between the cosmology sensitivity of the 3D halo
    mass function (dash-dotted lines) and the aperture mass function
    (solid lines) with ($n_\Omega$, thick lines) and without ($n_V$,
    thin lines) including the cosmology dependence of the survey solid angle and redshift interval at $z=0.43$. The cosmological parameter being varied
    by $\pm \, 5 \, \percent$ (coloured lines) with respect to the
    fiducial \citet{planckcollaboration2020} cosmology is indicated in
    the top-left corner of each panel. We assume $w_a=\pm 0.05$ since
    the fiducial value is 0. The aperture mass function is plotted for
    the median aperture mass at $m_\mathrm{200c}$ for all the
    simulations in the hypercube, and for apertures
    $(\Rin, \Rout, \Rmax) = (1, 2, 3) \, \cmpc$. All haloes have
    $m_\mathrm{200c} > 10^{13.5} \, \msun$. The probed comoving volume
    for a fixed observed area is mainly sensitive to $h$, $\Om$, $w_0$
    and $w_a$. Compared to the 3D halo mass function, the aperture mass function is more sensitive to changes in $\Om$ and $w_a$, similarly sensitive to $\sigma_8$ and $w_0$, and less sensitive to $h$.}
  \label{fig:n_m_ap_vs_n_m200c_comoving_vs_angular}
\end{figure*}
Finally, in Fig.~\ref{fig:n_m_ap_vs_n_m200c_comoving_vs_angular}, we
compare the volumetric mass functions, defined in
Eq.~\eqref{eq:n_comoving} (thin lines), to the observed mass functions
including the cosmology-dependent volume of the past lightcone,
defined in Eq.~\eqref{eq:n_angular} (thick lines), for both the 3D
halo mass function (dash-dotted lines) and the aperture mass function
(solid lines). Changing the background evolution of the Universe
modifies the number of observed haloes per fixed solid angle,
$\diff \Omega$, and redshift interval, $\diff z$, due to the change in
the probed comoving volume. The background evolution does not depend
on $\sigma_8$, $n_\mathrm{s}$, and $\Ob$ (since $\Om$ is fixed).

The background evolution is most sensitive to changes in the Hubble
parameter. Increasing (decreasing) $h$ reduces (increases) the
distance to redshift $z$. As a result, a fixed survey area at redshift
$z$ will probe a smaller (larger) comoving volume. Hence, we would
observe fewer (more) haloes for a fixed volumetric number density. The
bottom-left panel of
Fig.~\ref{fig:n_m_ap_vs_n_m200c_comoving_vs_angular} shows that the
decrease in the probed volume is larger than the increase in the
volumetric number density due to the increased matter density.
Changing $h$ results in the largest difference between the observed
and the volumetric mass functions, making the 3D halo mass function
more sensitive to changes in $h$, and the aperture mass function less
sensitive.

Increasing the matter density, $\Om$, similarly reduces the probed
volume for a fixed survey area at fixed redshift. This suppresses the
observed number density, $n_\Omega$, compared to the volumetric number
density, $n_V$, for both the 3D halo mass function and the aperture
mass function. The aperture mass function is still more sensitive to
changes in $\Om$ than the 3D halo mass function. The comoving volume
for a fixed area on the sky increases (decreases) significantly when
increasing (decreasing) the magnitude of $w_0$, resulting in more
(fewer) observed haloes. This geometric effect is stronger than the
decrease (increase) in the volumetric number density due to the less
(more) efficient structure formation. Increasing (decreasing) $w_a$
decreases (increases) the magnitude of $w(z)$ for $z > 0$, which in
turn lowers (raises) the observed number density compared to the
volumetric number density. The aperture mass function becomes less
sensitive to changes in $w_0$ and $w_a$. However, compared to the 3D
halo mass function, the total sensitivity to changes in $w_a$ remains
higher and the sensitivity to $w_0$ becomes similar.

Providing a detailed comparison between the performance of aperture
masses and 3D halo masses in a cluster cosmology analysis is more
complicated than investigating the percentage differences in the mass
functions given a difference in the cosmological parameters.
Eq.~\eqref{eq:N_obs} shows that the number counts depend on the
integral over the mass function taking into account the uncertainty in
the mass--observable relation. Even though we have shown that the
intrinsic measurement uncertainty in aperture mass measurements is
much lower than that in 3D halo mass inference, the total uncertainty
in the mass--observable relation still depends on the scatter between
the survey selection observable and the measured aperture or 3D mass.
For surveys that do not select clusters based on their weak lensing
shear signal, the scatter in the observable at fixed aperture mass can
still result in a significant total uncertainty in the aperture
mass--observable relation. Hence, a comparison between 3D and aperture
mass calibrations in a full cosmological analysis also needs to take
into account the survey observable.

In conclusion, the sensitivity of the aperture mass function to small
changes in the cosmological parameters opens the possibility of
calibrating cluster masses with weak lensing aperture masses,
bypassing the modelling uncertainty introduced when deprojecting the
observations.

\section{Discussion}\label{sec:discussion}
We have provided arguments for calibrating cluster masses with weak
lensing aperture masses. As long as we do not have predictions for the
halo abundance directly as a function of the survey observable, such
as the galaxy overdensity, the X-ray luminosity, or the SZ signal,
cluster cosmology needs to follow a two step process. Assuming that
the selection function has been accounted for, the mass--observable
relation needs to be calibrated, and the cosmology dependence of the
mass function needs to be understood. Eq.~\eqref{eq:N_obs} shows that
the mass calibration requires both the calibration between the
observable and the mass inferred from observations, and the
calibration between the inferred mass and the theoretical mass used in
the mass function.

To more closely match weak lensing observations, it makes sense to
calibrate cluster masses with the projected aperture mass, which can
also be measured in simulations. The mass calibration then separates
cleanly into a purely observational relation between the measured
aperture mass and the observable, and a calibration between the
theoretical and the measured aperture mass. This clean separation does
not hold for 3D cluster masses, which can only be inferred by
deprojecting the observations under the assumption of a density
profile. Any mismatch between the assumed and the true cluster density
profile biases the inferred 3D masses. The large variation in the
matter distribution along the lines-of-sight to different clusters
adds further uncertainty.

We showed that aperture masses correlate strongly with the 3D mass,
albeit with large scatter due to the matter along the line-of-sight.
We found that the aperture masses can be measured much more precisely
than 3D masses, since the precision is only limited by the shape noise
of the background galaxies. Next, we calibrated an emulator to
reproduce the cosmology dependence of the aperture mass function,
finding that it is also highly sensitive to variations in the
cosmological parameters. Now we will discuss some of the difficulties
that arise in cluster cosmology, and how they affect the aperture mass
specifically. We will also position our contribution within the wider
literature.

\subsection{Impact of the selection function}
\label{sec:selection_effects}
One vital ingredient of a cluster cosmology analysis that we did not
discuss in this paper is the selection function of the cluster sample.
The completeness, i.e. the fraction of all clusters that is detected,
and the purity, i.e. the fraction of detections that are real
clusters, of the cluster sample should be as high as possible
\citep[e.g.][]{allen2011, aguena2018}. We have studied the aperture
mass function in the idealized setting of perfect purity since we have
centred directly on the known clusters in the simulations. Our halo
sample becomes increasingly incomplete for aperture mass bins that
contain a significant fraction of haloes with 3D masses near our
selection limit, as can be seen from Fig.~\ref{fig:n_M_m200c_lim}.
Future aperture mass function emulators should thus ensure that they
can reliably measure aperture masses for haloes with masses
significantly below the expected detection limit of the survey, which
we were unable to do due to the downsampling inherent to the
Mira--Titan particle catalogues (although this is not a problem in
principle for simulations).

Since haloes with masses below the mean expected mass at the
observable selection limit can scatter above the signal threshold, the
completeness of the cluster sample near the selection limit depends on
the scatter of the mass--observable relation
\citep[e.g.][]{mantz2019}. The main benefit of aperture masses is the
ease with which they can be measured both in simulations and in
observations, which significantly decreases the intrinsic measurement
uncertainty in the mass calibration,
$P(\mathcal{M}_\mathrm{obs}| \mathcal{M})$, compared to 3D masses, as
we showed in Section~\ref{sec:aperture_uncertainty}. However, this
gain can be lost if the observable used to select clusters has a
significantly larger scatter at fixed aperture mass compared to its
scatter at fixed 3D mass. Hence, aperture masses could greatly
increase the performance of cluster surveys based on observables that
correlate with the aperture mass with small uncertainty. This will be
the case for observables that are more sensitive to projection
effects, such as the SZ signal \citep[e.g.][]{hallman2007}, galaxy
overdensities \citep[e.g.][]{vanhaarlem1997, erickson2011}, and,
naturally, the shear signal.

The purity of the halo sample will depend sensitively on the cluster
detection method, with shear-selected samples only reaching a maximum
purity of $\approx 85 \, \percent$ since chance line-of-sight
alignments can generate a significant signal due to the broadness of
the lensing kernel \citep[e.g.][]{hennawi2005}. The purity of other
detection methods that are also susceptible to such projection
effects, such as the SZ signal or the galaxy overdensity, will need to
be modelled in simulations. Baryonic observables that predominantly
trace the inner cluster density profile, such as the X-ray luminosity,
on the other hand, should reach higher purity
\citep[e.g.][]{voit2001}. However, samples selected from these
observables are necessarily more sensitive to the halo density
profile, introducing possible detection biases near the selection
limit \citep[e.g.][]{chon2017}.

We highlight one final important point about the synergy between
observed and simulated aperture mass measurements. Since the detection
bias for observables such as the SZ signal and the galaxy overdensity
is in large part due to projection effects
\citep[e.g.][]{shirasaki2016, zhang2022}, this bias is naturally
included in aperture masses measured in simulations. Hence, emulators
calibrated on a cluster sample generated by mimicking the survey
selection in the simulations will naturally include the survey
detection bias while providing aperture mass measurements that are
directly comparable to those measured observationally.

\subsection{Impact of systematic uncertainties}\label{sec:obs_systematics}
In a realistic cosmological analysis, different observational
systematic effects need to be taken into account. Any weak lensing
mass measurement will be sensitive to the systematic errors in the
shape measurements, the redshift distribution of the sources,
contamination of the lensing signal due to uncertainty in the
photometric redshift determination of cluster galaxies, and
miscentring of the cluster \citep[e.g.][]{vonderlinden2014,
  hoekstra2015}.

The main advantage of aperture masses over 3D masses is that no
density profile needs to be assumed in the analysis, eliminating the
impact of this modelling uncertainty. The aperture mass within $\Rin$
is actually measured from the lensing signal of galaxies
\emph{outside} $\Rin$, significantly reducing the impact of sources of
systematic error near the cluster centre, such as miscentring and
contamination \citep[e.g.][]{mandelbaum2010a}. The optimal choice of
$\Rin$ balances the reduced contamination of the lensing signal by
cluster galaxies when increasing $\Rin$ against the increase in the
statistical uncertainty due to the reduced number of background
galaxies. Since the bulk of the haloes have miscentring radii
$< 0.2 R_\mathrm{500c}$ \citep[e.g.][]{saro2015, bleem2020}, apertures
can be chosen large enough such that the mass within the aperture
should only be slightly affected, while limiting the increased
statistical uncertainty.

Another advantage stems from the fact that aperture masses can always
be computed unambiguously, even for triaxial and merging systems. As
long as the choice of aperture in the mass function and the
observations is consistent, the mass measurement should yield similar
results. Moreover, since emulators can be calibrated for different
aperture sizes, the consistency of the inferred cosmology for an
analysis using different apertures can pinpoint possible biases in the
cosmological analysis.

A limitation of our preliminary study is the fact that we did not
construct convergence maps from the full past lightcone. The lensing
efficiency of matter structures at redshift $z_\mathrm{l}$ for source
galaxies at a fixed redshift $z_\mathrm{s}$,
$\epsilon(z_\mathrm{l}, z_\mathrm{s}) = D_\mathrm{l} D_\mathrm{ls} /
D_\mathrm{s}$, is very broad. This means that matter over a
significant range of redshifts can contribute to the lensing signal of
a given background galaxy. A full line-of-sight in simulation M000
with $L=2100 \, \cmpc$ at $z=0.5$ corresponds to a redshift range
$z \approx [0.2, 0.85]$. Hence, projecting the mass along the
simulation volume at fixed $z$ does not take into account the time
evolution of the included structures or the change in the angular
diameter distance across the length of the box. As such, aperture mass
functions should really be calibrated on simulation lightcone outputs,
not on single snapshots. This makes the analysis more complex since
the resulting lensing maps need to be reconstructed for different
source redshifts, $z_\mathrm{s}$.

Finally, since we have used gravity-only simulations, we have not
included the impact of baryonic physics on the aperture mass function.
For 3D halo mass functions, it is well established that the mass of
haloes with $m_\mathrm{200m,dmo} \lesssim 10^{14.5} \, \msun$
decreases significantly due to galaxy formation physics processes
\citep[e.g.][]{velliscig2014}. We expect baryonic physics to also
impact the cluster aperture mass measurements, albeit less
significantly due to the projected nature of the measurement
\citep[e.g.][]{debackere2021}. We study the impact of baryonic physics
on the aperture mass measurements in a companion paper
\citet{debackere2022}.

\subsection{Comparison to previous work}\label{sec:comparison_literature}
The abundance of clusters is a powerful probe of the cosmological
evolution of the Universe, so an active effort is underway to minimize
the impact of mass calibration uncertainties. For example,
\citet{grandis2021d} directly calibrate the mass--observable relation,
$P(\mathcal{O}|\mathcal{M})$, using simulations. They generate lensing
profiles from hydrodynamical simulations which they fit with NFW
density profiles with a fixed concentration and assuming a miscentring
distribution. They then calibrate the resulting relation between the
best-fit NFW mass and the true mass of the matched cluster in DMO
simulations. This method then converts a weak lensing-inferred 3D halo
mass into the 3D halo mass of the matching DMO halo, allowing the use
of 3D halo mass function emulators calibrated on DMO simulations. This
method is still explicitly limited by the scatter between the inferred
and the true 3D halo mass due to the assumed density profile.

\citet{cromer2021b} improve the accuracy of weak lensing-inferred 3D
halo masses by fitting the lensing shear with an emulated cluster
density profile that includes a phenomenological contribution due to
baryons. Their model results in more accurate cluster mass estimates,
but, again, relies on the ultimately inaccurate assumption that the
complex cluster density profile can be modelled accurately with
simplified, spherically symmetric profiles.

\citet{marian2009, marian2010} carry out analyses that are the most
similar to ours. They generate lensing maps for different slabs in DMO
simulations to which they apply a hierarchical peak finder that
extracts the aperture mass within a filter designed to optimally
detect the cluster signal. They show that the resulting peak abundance
function has a similar cosmological sensitivity as the 3D mass
function. Similarly to us, they find that the peak aperture masses
show a large scatter at fixed halo mass. However, at the time of their
work, large suites of cosmological simulations and emulators were not
yet available. Hence, they resorted to constructing an analytic
framework to extract cosmological information from weak lensing peak
counts.

Another option is to neglect the cluster selection entirely and use
the distribution of shear peaks as a function of their signal-to-noise
ratio to constrain the cosmology \citep[see e.g.][]{wang2009a,
  dietrich2010, kratochvil2010}. However, since the evolution of
clusters with time contains a wealth of cosmological information,
stronger cosmological constraints can be obtained by including
redshift information for the observed peaks, as suggested by
\citet{hennawi2005}. The main difficulty with these shear-selected
analyses is that a significant fraction of the high significance peaks
arises from chance line-of-sight alignments due the broadness of the
lensing kernel \citep[e.g.][]{hennawi2005, yang2011}. In recent
studies, \citet{hamana2015}, \citet{shan2018} and \citet{martinet2018}
have used peaks identified from weak lensing observations to constrain
the matter density and clustering of the Universe.

We locate our work in between peak abundance studies and cluster
analyses based on 3D cluster masses: our method corresponds to an
idealized survey that selects clusters based on a secondary observable
that perfectly correlates with the 3D halo mass, while the cluster
masses are determined through aperture masses which would in practice
be derived from weak lensing observations. Hence, our work is very
similar to a standard cluster cosmology analysis, as worked out in
detail by \citet{mantz2010a,mantz2010}, but now using the aperture
mass to calibrate the cluster masses. In such an analysis, one assumes
a functional form for the mass--observable relation, which gets
calibrated simultaneously with the cosmology-dependent aperture mass
function by forward modelling the observed cluster abundance as a
function of the observable signal, taking into account the selection
function of the observable for a given survey. Importantly, any
cosmology dependence in the mass--observable relation needs to be
taken into account implying that the cosmology dependence of both
$P(\Delta M_\mathrm{obs}| \Delta M, \mathbf{\Omega}, z)$ and
$P(\mathcal{O}|\Delta M_\mathrm{obs}, \mathbf{\Omega}, z)$ need to be
calibrated from mock observations in realistic cosmological and,
preferably, hydrodynamical simulations \citep[e.g.][]{dietrich2019}.

\section{Conclusions}\label{sec:conclusions}
We have argued that cluster cosmology analyses can decrease their
sensitivity to modelling assumptions by using weak lensing-like excess
aperture mass measurements to calibrate cluster masses. As long as
predictions for the cosmology-dependent abundance of clusters as a
function of their observed signal are not available, cluster cosmology
necessarily relies on an accurately determined and well-understood
mass--observable relation and a theoretical prediction for the
cosmology dependence of the mass function. Only suites of large-volume
simulations with varying cosmological parameters can predict the mass
function at the accuracy required for future surveys. If we are using
simulations, however, we might as well predict the aperture mass
function instead of (or along with) the 3D halo mass function.

Aperture masses are a natural choice for cluster mass calibrations
since they can be measured accurately both in observations and in
simulations, with an uncertainty determined solely by the background
galaxy shape noise in the weak lensing observations. In contrast, 3D
halo masses can only be inferred by deprojecting observations assuming
a density profile. The mismatch between the assumed density profile
and the true, triaxial halo density profile, including substructure
and correlated matter, and the neglected matter along the
line-of-sight, introduce a model-dependent bias and scatter in the
inferred mass.

We used the Mira--Titan suite of large-volume, DMO simulations to
measure the excess projected mass of clusters within fixed aperture
sizes of $\Rin = 0.5, 1.0, 1.5 \, \cmpc$ with a background subtraction
calculated in an outer annulus between $2 < R / \cmpc< 3$. We studied
the behaviour of these aperture masses and the corresponding aperture
mass function. We showed that the aperture mass correlates strongly
with the 3D halo mass, with aperture masses being larger (smaller)
than the halo virial mass when measured within apertures larger
(smaller) than the virial radius (Fig.~\ref{fig:M_vs_m_avg}). The
aperture mass exhibits large scatter at fixed halo mass when the halo
virial radius is not significantly larger than the aperture due to the
contribution of matter outside the halo
(Fig.~\ref{fig:sigma_delta_m_avg}). Advantageously, the uncertainty in
the \emph{measurement} of the aperture mass is between 2 to 3 times
smaller than that of the inferred 3D mass
(Fig.~\ref{fig:sigma_delta_m_obs}). This is because the measurement
uncertainty depends only on the background galaxy shape noise in the
weak lensing observations, and since line-of-sight structures
contribute to the aperture mass signal whereas they introduce noise in
the deprojection for 3D masses.

We did not investigate the scatter between the survey observable and
the aperture mass since the Mira--Titan suite does not include
hydrodynamics to model the complex baryonic processes related to
galaxy formation. However, we argued that observables such as the
galaxy overdensity, and the shear should correlate strongly with
aperture masses with small uncertainty, since they are also sensitive
to line-of-sight matter structures beyond the halo. X-ray
luminosities, on the other hand, due to their steep scaling with the
3D halo mass, may show large scatter at fixed aperture mass, similarly
to the 3D halo mass. The uncertainty between the observable and the
measured aperture mass will ultimately determine the scatter in the
mass--observable relation, given the small intrinsic scatter of the
measured aperture mass with respect to the true aperture mass.
Investigating the uncertainty between the observable and the measured
aperture mass in hydrodynamical simulations is a fruitful direction
for future research.

We used the Mira--Titan hypercube of DMO simulations to calibrate a
Gaussian process emulator to \emph{directly} emulate the cosmology
dependence of the aperture mass function given the simulated number
counts and their likelihood, i.e. without assuming an underlying,
dimensionality-reducing model for the simulation data. This is
possible thanks to advances in Gaussian process modelling, allowing
for the efficient optimization of large datasets and non-Gaussian
likelihoods. We argued that this gives an advantage over usual
emulators since the high-mass tail of the emulator will only depend on
the simulation data and the assumed likelihood, \emph{not} on the
assumed mass dependence for the assumed data model. We showed that the
emulator can accurately reproduce most of the simulations to within
$2 \, \percent$ or within the bootstrapped variance at high-aperture
masses (Fig.~\ref{fig:n_M_perf}).

Isolating the influence of structure formation on the halo abundance,
we found that, compared to the 3D halo mass function, the aperture
mass function is similarly sensitive to changes in $\sigma_8$ and
$n_\mathrm{s}$, and more sensitive to changes in $\Om$, $h$, $w_0$ and
$w_a$ (Fig.~\ref{fig:n_m_ap_vs_n_m200c_cosmo}). Even
$\pm 1 \, \percent$ changes in $\Om$, $\sigma_8$, and $h$ result in
$> 10 \, \percent$ changes in the expected halo number density at
fixed redshift (Fig.~\ref{fig:n_M_cosmo_R1}). Including the cosmology
dependence of the volume probed by the past lightcone, we found that,
compared to the 3D halo mass function, the aperture mass function is
more sensitive to changes in $\Om$ and $w_a$, similarly sensitive to
changes in $\sigma_8$ and $w_0$ and slightly less sensitive to changes
in $h$ (Fig~\ref{fig:n_m_ap_vs_n_m200c_comoving_vs_angular}). We
stress that a detailed comparison between the performance of the
aperture mass function compared to the 3D halo mass function also
needs to take into account the survey observable. Importantly, since
emulators can easily be calibrated for multiple apertures, the
consistency of the inferred cosmology for an analysis using different
apertures can provide useful insights into possible biases in the
cosmological analysis.

In the future, it will be possible to emulate cluster surveys using
lightcones output from hydrodynamical simulations, mimicking the
observable measurement and selection directly while skipping the mass
calibration step (given that one can trust the simulation predictions
at the accuracy required for future surveys, or marginalize over the
simulation uncertainty). To validate the fidelity of such simulations,
aperture masses provide the best choice to test the simulated
mass--observable relations. Since no such simulations are currently
available, however, we argue that our approach provides a valuable
intermediate step. Emulators of the aperture mass function, which is
closer to the data than the 3D halo mass function, can already be
trained, minimizing the impact of uncertain modelling assumptions on
cluster cosmology analyses.

\section*{Acknowledgements}
We would like to thank the referee for a clear report that helped
clarify the aim of our work. This work is part of the research
programme Athena with project number 184.034.002 and Vici grants
639.043.409 and 639.043.512, which are financed by the Dutch Research
Council (NWO). Argonne National Laboratory's work was supported under
the U.S. Department of Energy contract DE-AC02-06CH11357. This
research was supported in part by DOE HEP's Computational HEP program.
This research used resources of the Argonne Leadership Computing
Facility at the Argonne National Laboratory, which is supported by the
Office of Science of the U.S. Department of Energy under Contract No.
DE-AC02-06CH11357. This work also used resources of the Oak Ridge
Leadership Computing Facility, which is a DOE Office of Science User
Facility supported under Contract DE-AC05-00OR22725.

\section*{Data availability}
The data used in this paper is available upon request to the first author.

%%%%%%%%%%%%%%%%%%%%%%%%%%%%%%%%%%%%%%%%%%%%%%%%%% 

%%%%%%%%%%%%%%%%%%%% REFERENCES %%%%%%%%%%%%%%%%%%

% The best way to enter references is to use BibTeX:
\bibliographystyle{mnras}
\bibliography{aperture_mass_function.bbl}

\begin{thebibliography}{}
\makeatletter
\relax
\def\mn@urlcharsother{\let\do\@makeother \do\$\do\&\do\#\do\^\do\_\do\%\do\~}
\def\mn@doi{\begingroup\mn@urlcharsother \@ifnextchar [ {\mn@doi@}
  {\mn@doi@[]}}
\def\mn@doi@[#1]#2{\def\@tempa{#1}\ifx\@tempa\@empty \href
  {http://dx.doi.org/#2} {doi:#2}\else \href {http://dx.doi.org/#2} {#1}\fi
  \endgroup}
\def\mn@eprint#1#2{\mn@eprint@#1:#2::\@nil}
\def\mn@eprint@arXiv#1{\href {http://arxiv.org/abs/#1} {{\tt arXiv:#1}}}
\def\mn@eprint@dblp#1{\href {http://dblp.uni-trier.de/rec/bibtex/#1.xml}
  {dblp:#1}}
\def\mn@eprint@#1:#2:#3:#4\@nil{\def\@tempa {#1}\def\@tempb {#2}\def\@tempc
  {#3}\ifx \@tempc \@empty \let \@tempc \@tempb \let \@tempb \@tempa \fi \ifx
  \@tempb \@empty \def\@tempb {arXiv}\fi \@ifundefined
  {mn@eprint@\@tempb}{\@tempb:\@tempc}{\expandafter \expandafter \csname
  mn@eprint@\@tempb\endcsname \expandafter{\@tempc}}}

\bibitem[\protect\citeauthoryear{Aguena \& Lima}{Aguena \&
  Lima}{2018}]{aguena2018}
Aguena M.,  Lima M.,  2018, \mn@doi [Phys. Rev. D]
  {10.1103/PhysRevD.98.123529}, 98, 123529

\bibitem[\protect\citeauthoryear{Allen, Evrard  \& Mantz}{Allen
  et~al.}{2011}]{allen2011}
Allen S.~W.,  Evrard A.~E.,   Mantz A.~B.,  2011, \mn@doi [Annu. Rev. Astron.
  Astrophys.] {10.1146/annurev-astro-081710-102514}, 49, 409

\bibitem[\protect\citeauthoryear{Andreon \& Congdon}{Andreon \&
  Congdon}{2014}]{andreon2014}
Andreon S.,  Congdon P.,  2014, \mn@doi [A\&A] {10.1051/0004-6361/201423616},
  568, A23

\bibitem[\protect\citeauthoryear{Angulo, Springel, White, Jenkins, Baugh  \&
  Frenk}{Angulo et~al.}{2012}]{angulo2012}
Angulo R.~E.,  Springel V.,  White S. D.~M.,  Jenkins A.,  Baugh C.~M.,   Frenk
  C.~S.,  2012, \mn@doi [Mon. Not. R. Astron. Soc.]
  {10.1111/j.1365-2966.2012.21830.x}, 426, 2046

\bibitem[\protect\citeauthoryear{Applegate et~al.,}{Applegate
  et~al.}{2014}]{applegate2014}
Applegate D.~E.,  et~al., 2014, \mn@doi [Mon. Not. R. Astron. Soc.]
  {10.1093/mnras/stt2129}, 439, 48

\bibitem[\protect\citeauthoryear{Bahcall \& Kulier}{Bahcall \&
  Kulier}{2014}]{bahcall2014}
Bahcall N.~A.,  Kulier A.,  2014, \mn@doi [Monthly Notices of the Royal
  Astronomical Society] {10.1093/mnras/stu107}, 439, 2505

\bibitem[\protect\citeauthoryear{Bah{\'e}, Mccarthy  \& King}{Bah{\'e}
  et~al.}{2012}]{bahe2012b}
Bah{\'e} Y.~M.,  Mccarthy I.~G.,   King L.~J.,  2012, \mn@doi [Mon. Not. R.
  Astron. Soc.] {10.1111/j.1365-2966.2011.20364.x}, 421, 1073

\bibitem[\protect\citeauthoryear{Bartelmann \& Schneider}{Bartelmann \&
  Schneider}{2001}]{bartelmann2001a}
Bartelmann M.,  Schneider P.,  2001, \mn@doi [Physics Reports]
  {10.1016/S0370-1573(00)00082-X}, 340, 291

\bibitem[\protect\citeauthoryear{Becker \& Kravtsov}{Becker \&
  Kravtsov}{2011}]{becker2011}
Becker M.~R.,  Kravtsov A.~V.,  2011, \mn@doi [Astrophys. J.]
  {10.1088/0004-637X/740/1/25}, 740

\bibitem[\protect\citeauthoryear{Bhattacharya, Heitmann, White, Luki{\'c},
  Wagner  \& Habib}{Bhattacharya et~al.}{2011}]{bhattacharya2011}
Bhattacharya S.,  Heitmann K.,  White M.,  Luki{\'c} Z.,  Wagner C.,   Habib
  S.,  2011, \mn@doi [Astrophys. J.] {10.1088/0004-637X/732/2/122}, 732, 122

\bibitem[\protect\citeauthoryear{Bleem et~al.,}{Bleem et~al.}{2020}]{bleem2020}
Bleem L.~E.,  et~al., 2020, \mn@doi [Astrophys. J. Suppl. Ser.]
  {10.3847/1538-4365/ab6993}, 247, 25

\bibitem[\protect\citeauthoryear{Bocquet, Heitmann, Habib, Lawrence, Uram,
  Frontiere, Pope  \& Finkel}{Bocquet et~al.}{2020}]{bocquet2020}
Bocquet S.,  Heitmann K.,  Habib S.,  Lawrence E.,  Uram T.,  Frontiere N.,
  Pope A.,   Finkel H.,  2020, \mn@doi [Astrophys. J.]
  {10.3847/1538-4357/abac5c}, 901, 5

\bibitem[\protect\citeauthoryear{Bond, Cole, Efstathiou  \& Kaiser}{Bond
  et~al.}{1991}]{bond1991}
Bond J.~R.,  Cole S.,  Efstathiou G.,   Kaiser N.,  1991, \mn@doi [Astrophys.
  J.] {10.1086/170520}, 379, 440

\bibitem[\protect\citeauthoryear{Budzynski, Koposov, McCarthy  \&
  Belokurov}{Budzynski et~al.}{2014}]{budzynski2014}
Budzynski J.~M.,  Koposov S.~E.,  McCarthy I.~G.,   Belokurov V.,  2014,
  \mn@doi [Mon. Not. R. Astron. Soc.] {10.1093/mnras/stt1965}, 437, 1362

\bibitem[\protect\citeauthoryear{Chon \& B{\"o}hringer}{Chon \&
  B{\"o}hringer}{2017}]{chon2017}
Chon G.,  B{\"o}hringer H.,  2017, \mn@doi [Astron. Astrophys.]
  {10.1051/0004-6361/201731854}, 606, L4

\bibitem[\protect\citeauthoryear{Clowe, Luppino, Kaiser, Henry  \& Gioia}{Clowe
  et~al.}{1998}]{clowe1998}
Clowe D.,  Luppino G.~A.,  Kaiser N.,  Henry J.~P.,   Gioia I.~M.,  1998,
  \mn@doi [Astrophys. J.] {10.1086/311285}, 497, L61

\bibitem[\protect\citeauthoryear{Crocce, Fosalba, Castander  \&
  Gazta{\~n}aga}{Crocce et~al.}{2010}]{crocce2010}
Crocce M.,  Fosalba P.,  Castander F.~J.,   Gazta{\~n}aga E.,  2010, \mn@doi
  [Monthly Notices of the Royal Astronomical Society]
  {10.1111/j.1365-2966.2009.16194.x}, 403, 1353

\bibitem[\protect\citeauthoryear{Cromer, Battaglia, Miyatake  \& Simet}{Cromer
  et~al.}{2021}]{cromer2021b}
Cromer D.,  Battaglia N.,  Miyatake H.,   Simet M.,  2021, arXiv:2104.06925
  [astro-ph]

\bibitem[\protect\citeauthoryear{{DES Collaboration} et~al.,}{{DES
  Collaboration} et~al.}{2020}]{descollaboration2020}
{DES Collaboration} et~al., 2020, \mn@doi [Phys. Rev. D]
  {10.1103/PhysRevD.102.023509}, 102, 023509

\bibitem[\protect\citeauthoryear{Debackere, Schaye  \& Hoekstra}{Debackere
  et~al.}{2021}]{debackere2021}
Debackere S. N.~B.,  Schaye J.,   Hoekstra H.,  2021, \mn@doi [Monthly Notices
  of the Royal Astronomical Society] {10.1093/mnras/stab1326}, 505, 593

\bibitem[\protect\citeauthoryear{Debackere, Hoekstra  \& Schaye}{Debackere
  et~al.}{2022}]{debackere2022}
Debackere S. N.~B.,  Hoekstra H.,   Schaye J.,  2022, Galaxy Cluster Aperture
  Masses Are More Robust to Baryonic Effects than {{3D}} Halo Masses
  (\mn@eprint {arXiv} {2205.08424}), \mn@doi{10.48550/arXiv.2205.08424}

\bibitem[\protect\citeauthoryear{Despali, Giocoli, Angulo, Tormen, Sheth, Baso
  \& Moscardini}{Despali et~al.}{2016}]{despali2016}
Despali G.,  Giocoli C.,  Angulo R.~E.,  Tormen G.,  Sheth R.~K.,  Baso G.,
  Moscardini L.,  2016, \mn@doi [Mon. Not. R. Astron. Soc.]
  {10.1093/mnras/stv2842}, 456, 2486

\bibitem[\protect\citeauthoryear{Diemer}{Diemer}{2020}]{diemer2020a}
Diemer B.,  2020, \mn@doi [ApJ] {10.3847/1538-4357/abbf52}, 903, 87

\bibitem[\protect\citeauthoryear{Dietrich \& Hartlap}{Dietrich \&
  Hartlap}{2010}]{dietrich2010}
Dietrich J.~P.,  Hartlap J.,  2010, \mn@doi [Mon. Not. R. Astron. Soc.]
  {10.1111/j.1365-2966.2009.15948.x}, 402, 1049

\bibitem[\protect\citeauthoryear{Dietrich et~al.,}{Dietrich
  et~al.}{2019}]{dietrich2019}
Dietrich J.~P.,  et~al., 2019, \mn@doi [Mon. Not. R. Astron. Soc.]
  {10.1093/mnras/sty3088}, 483, 2871

\bibitem[\protect\citeauthoryear{Erickson, Cunha  \& Evrard}{Erickson
  et~al.}{2011}]{erickson2011}
Erickson B. M.~S.,  Cunha C.~E.,   Evrard A.~E.,  2011, \mn@doi [Phys. Rev. D]
  {10.1103/PhysRevD.84.103506}, 84, 103506

\bibitem[\protect\citeauthoryear{Gardner, Pleiss, Bindel, Weinberger  \&
  Wilson}{Gardner et~al.}{2021}]{gardner2021}
Gardner J.~R.,  Pleiss G.,  Bindel D.,  Weinberger K.~Q.,   Wilson A.~G.,
  2021, arXiv:1809.11165 [cs, stat]

\bibitem[\protect\citeauthoryear{Grandis, Bocquet, Mohr, Klein  \&
  Dolag}{Grandis et~al.}{2021}]{grandis2021d}
Grandis S.,  Bocquet S.,  Mohr J.~J.,  Klein M.,   Dolag K.,  2021, \mn@doi
  [Monthly Notices of the Royal Astronomical Society] {10.1093/mnras/stab2414},
  507, 5671

\bibitem[\protect\citeauthoryear{Habib et~al.,}{Habib et~al.}{2016}]{habib2016}
Habib S.,  et~al., 2016, \mn@doi [New Astronomy]
  {10.1016/j.newast.2015.06.003}, 42, 49

\bibitem[\protect\citeauthoryear{Haiman, Mohr  \& Holder}{Haiman
  et~al.}{2001}]{Haiman2001}
Haiman Z.,  Mohr J.~J.,   Holder G.~P.,  2001, \mn@doi [Astrophys. J.]
  {10.1086/320939}, 553, 545

\bibitem[\protect\citeauthoryear{Hallman, O'Shea, Burns, Norman, Harkness  \&
  Wagner}{Hallman et~al.}{2007}]{hallman2007}
Hallman E.~J.,  O'Shea B.~W.,  Burns J.~O.,  Norman M.~L.,  Harkness R.,
  Wagner R.,  2007, \mn@doi [ApJ] {10.1086/522912}, 671, 27

\bibitem[\protect\citeauthoryear{Hamana, Sakurai, Koike  \& Miller}{Hamana
  et~al.}{2015}]{hamana2015}
Hamana T.,  Sakurai J.,  Koike M.,   Miller L.,  2015, \mn@doi [Publications of
  the Astronomical Society of Japan] {10.1093/pasj/psv034}, 67, 34

\bibitem[\protect\citeauthoryear{Heitmann et~al.,}{Heitmann
  et~al.}{2016}]{heitmann2016}
Heitmann K.,  et~al., 2016, \mn@doi [Astrophys. J.]
  {10.3847/0004-637X/820/2/108}, 820, 108

\bibitem[\protect\citeauthoryear{Heitmann et~al.,}{Heitmann
  et~al.}{2019}]{heitmann2019}
Heitmann K.,  et~al., 2019, \mn@doi [ApJS] {10.3847/1538-4365/ab3724}, 244, 17

\bibitem[\protect\citeauthoryear{Hennawi \& Spergel}{Hennawi \&
  Spergel}{2005}]{hennawi2005}
Hennawi J.~F.,  Spergel D.~N.,  2005, \mn@doi [Astrophys. J.] {10.1086/428749},
  624, 59

\bibitem[\protect\citeauthoryear{Hensman, Matthews  \& Ghahramani}{Hensman
  et~al.}{2014}]{hensman2014}
Hensman J.,  Matthews A.,   Ghahramani Z.,  2014, arXiv:1411.2005 [stat]

\bibitem[\protect\citeauthoryear{Henson, Barnes, Kay, McCarthy  \&
  Schaye}{Henson et~al.}{2017}]{henson2017}
Henson M.~A.,  Barnes D.~J.,  Kay S.~T.,  McCarthy I.~G.,   Schaye J.,  2017,
  \mn@doi [Mon. Not. R. Astron. Soc.] {10.1093/mnras/stw2899}, 465, 3361

\bibitem[\protect\citeauthoryear{Hoekstra}{Hoekstra}{2001}]{hoekstra2001}
Hoekstra H.,  2001, \mn@doi [Astron. Astrophys.] {10.1051/0004-6361:20010293},
  370, 743

\bibitem[\protect\citeauthoryear{Hoekstra}{Hoekstra}{2003}]{hoekstra2003}
Hoekstra H.,  2003, \mn@doi [Mon. Not. R. Astron. Soc.]
  {10.1046/j.1365-8711.2003.06264.x}, 339, 1155

\bibitem[\protect\citeauthoryear{Hoekstra, Mahdavi, Babul  \&
  Bildfell}{Hoekstra et~al.}{2012}]{hoekstra2012}
Hoekstra H.,  Mahdavi A.,  Babul A.,   Bildfell C.,  2012, \mn@doi [Monthly
  Notices of the Royal Astronomical Society]
  {10.1111/j.1365-2966.2012.22072.x}, 427, 1298

\bibitem[\protect\citeauthoryear{Hoekstra, Bartelmann, Dahle, Israel, Limousin
  \& Meneghetti}{Hoekstra et~al.}{2013}]{hoekstra2013}
Hoekstra H.,  Bartelmann M.,  Dahle H.,  Israel H.,  Limousin M.,   Meneghetti
  M.,  2013, \mn@doi [Space Sci. Rev.] {10.1007/s11214-013-9978-5}, 177, 75

\bibitem[\protect\citeauthoryear{Hoekstra, Herbonnet, Muzzin, Babul, Mahdavi,
  Viola  \& Cacciato}{Hoekstra et~al.}{2015}]{hoekstra2015}
Hoekstra H.,  Herbonnet R.,  Muzzin A.,  Babul A.,  Mahdavi A.,  Viola M.,
  Cacciato M.,  2015, \mn@doi [Monthly Notices of the Royal Astronomical
  Society] {10.1093/mnras/stv275}, 449, 685

\bibitem[\protect\citeauthoryear{K{\"o}hlinger, Hoekstra  \&
  Eriksen}{K{\"o}hlinger et~al.}{2015}]{kohlinger2015}
K{\"o}hlinger F.,  Hoekstra H.,   Eriksen M.,  2015, \mn@doi [Mon. Not. R.
  Astron. Soc.] {10.1093/mnras/stv1852}, 453, 3107

\bibitem[\protect\citeauthoryear{Kratochvil, Haiman  \& May}{Kratochvil
  et~al.}{2010}]{kratochvil2010}
Kratochvil J.~M.,  Haiman Z.,   May M.,  2010, \mn@doi [Physical Review D]
  {10.1103/PhysRevD.81.043519}, 81, 043519

\bibitem[\protect\citeauthoryear{Lawrence et~al.,}{Lawrence
  et~al.}{2017}]{lawrence2017}
Lawrence E.,  et~al., 2017, \mn@doi [Astrophys. J.] {10.3847/1538-4357/aa86a9},
  847, 50

\bibitem[\protect\citeauthoryear{Le~Brun, McCarthy  \& Melin}{Le~Brun
  et~al.}{2015}]{lebrun2015}
Le~Brun A. M.~C.,  McCarthy I.~G.,   Melin J.~B.,  2015, \mn@doi [Mon. Not. R.
  Astron. Soc.] {10.1093/mnras/stv1172}, 451, 3868

\bibitem[\protect\citeauthoryear{Mandelbaum, Seljak, Baldauf  \&
  Smith}{Mandelbaum et~al.}{2010}]{mandelbaum2010a}
Mandelbaum R.,  Seljak U.,  Baldauf T.,   Smith R.~E.,  2010, \mn@doi [Monthly
  Notices of the Royal Astronomical Society]
  {10.1111/j.1365-2966.2010.16619.x}, 405, 2078

\bibitem[\protect\citeauthoryear{Mantz}{Mantz}{2019}]{mantz2019}
Mantz A.~B.,  2019, \mn@doi [Mon. Not. R. Astron. Soc.] {10.1093/mnras/stz320},
  485, 4863

\bibitem[\protect\citeauthoryear{Mantz, Allen, Ebeling, Rapetti  \&
  {Drlica-Wagner}}{Mantz et~al.}{2010a}]{mantz2010a}
Mantz A.,  Allen S.~W.,  Ebeling H.,  Rapetti D.,   {Drlica-Wagner} A.,  2010a,
  \mn@doi [Monthly Notices of the Royal Astronomical Society]
  {10.1111/j.1365-2966.2010.16993.x}, 406, 1773

\bibitem[\protect\citeauthoryear{Mantz, Allen, Rapetti  \& Ebeling}{Mantz
  et~al.}{2010b}]{mantz2010}
Mantz A.,  Allen S.~W.,  Rapetti D.,   Ebeling H.,  2010b, \mn@doi [Mon. Not.
  R. Astron. Soc.] {10.1111/j.1365-2966.2010.16992.x}, 406, 1759

\bibitem[\protect\citeauthoryear{Marian, Smith  \& Bernstein}{Marian
  et~al.}{2009}]{marian2009}
Marian L.,  Smith R.~E.,   Bernstein G.~M.,  2009, \mn@doi [Astrophys. J.]
  {10.1088/0004-637X/698/1/L33}, 698, 33

\bibitem[\protect\citeauthoryear{Marian, Smith  \& Bernstein}{Marian
  et~al.}{2010}]{marian2010}
Marian L.,  Smith R.~E.,   Bernstein G.~M.,  2010, \mn@doi [Astrophys. J.]
  {10.1088/0004-637X/709/1/286}, 709, 286

\bibitem[\protect\citeauthoryear{Martinet et~al.,}{Martinet
  et~al.}{2018}]{martinet2018}
Martinet N.,  et~al., 2018, \mn@doi [Monthly Notices of the Royal Astronomical
  Society] {10.1093/mnras/stx2793}, 474, 712

\bibitem[\protect\citeauthoryear{Matthews}{Matthews}{2017}]{matthews2017}
Matthews A. G. d.~G.,  2017, Thesis, University of Cambridge,
  \mn@doi{10.17863/CAM.25348}

\bibitem[\protect\citeauthoryear{Matthews, Hensman, Turner  \&
  Ghahramani}{Matthews et~al.}{2016}]{matthews2016}
Matthews A. G. d.~G.,  Hensman J.,  Turner R.,   Ghahramani Z.,  2016, in
  Proceedings of the 19th {{International Conference}} on {{Artificial
  Intelligence}} and {{Statistics}}. {PMLR}, pp 231--239

\bibitem[\protect\citeauthoryear{McClintock et~al.,}{McClintock
  et~al.}{2019}]{mcclintock2019a}
McClintock T.,  et~al., 2019, \mn@doi [Astrophys. J.]
  {10.3847/1538-4357/aaf568}, 872, 53

\bibitem[\protect\citeauthoryear{Nishimichi et~al.,}{Nishimichi
  et~al.}{2019}]{nishimichi2019}
Nishimichi T.,  et~al., 2019, \mn@doi [Astrophys. J.]
  {10.3847/1538-4357/ab3719}, 884, 29

\bibitem[\protect\citeauthoryear{Oguri \& Hamana}{Oguri \&
  Hamana}{2011}]{oguri2011a}
Oguri M.,  Hamana T.,  2011, \mn@doi [Mon. Not. R. Astron. Soc.]
  {10.1111/j.1365-2966.2011.18481.x}, 414, 1851

\bibitem[\protect\citeauthoryear{{Planck Collaboration} et~al.,}{{Planck
  Collaboration} et~al.}{2020}]{planckcollaboration2020}
{Planck Collaboration} et~al., 2020, \mn@doi [Astron. Astrophys.]
  {10.1051/0004-6361/201833910}, 641, A6

\bibitem[\protect\citeauthoryear{Pratt, Arnaud, Biviano, Eckert, Ettori, Nagai,
  Okabe  \& Reiprich}{Pratt et~al.}{2019}]{pratt2019}
Pratt G.~W.,  Arnaud M.,  Biviano A.,  Eckert D.,  Ettori S.,  Nagai D.,  Okabe
  N.,   Reiprich T.~H.,  2019, \mn@doi [Space Science Reviews]
  {10.1007/s11214-019-0591-0}, 215, 25

\bibitem[\protect\citeauthoryear{Press \& Schechter}{Press \&
  Schechter}{1974}]{press1974}
Press W.~H.,  Schechter P.,  1974, \mn@doi [Astrophys. J.] {10.1086/152650},
  187, 425

\bibitem[\protect\citeauthoryear{Rasmussen \& Williams}{Rasmussen \&
  Williams}{2006}]{rasmussen2006}
Rasmussen C.~E.,  Williams C. K.~I.,  2006, Gaussian Processes for Machine
  Learning.
Adaptive Computation and Machine Learning, {MIT Press}, {Cambridge, Mass}

\bibitem[\protect\citeauthoryear{Reblinsky \& Bartelmann}{Reblinsky \&
  Bartelmann}{1999}]{reblinsky1999a}
Reblinsky K.,  Bartelmann M.,  1999, Astron. Astrophys., 345, 1

\bibitem[\protect\citeauthoryear{Saro et~al.,}{Saro et~al.}{2015}]{saro2015}
Saro A.,  et~al., 2015, \mn@doi [Monthly Notices of the Royal Astronomical
  Society] {10.1093/mnras/stv2141}, 454, 2305

\bibitem[\protect\citeauthoryear{Sartoris et~al.,}{Sartoris
  et~al.}{2016}]{sartoris2016}
Sartoris B.,  et~al., 2016, \mn@doi [Mon. Not. R. Astron. Soc.]
  {10.1093/mnras/stw630}, 459, 1764

\bibitem[\protect\citeauthoryear{Schneider}{Schneider}{1996}]{schneider1996}
Schneider P.,  1996, \mn@doi [Mon. Not. R. Astron. Soc.]
  {10.1093/mnras/283.3.837}, 283, 837

\bibitem[\protect\citeauthoryear{Schneider, Van~Waerbeke, Jain  \&
  Kruse}{Schneider et~al.}{1998}]{schneider1998}
Schneider P.,  Van~Waerbeke L.,  Jain B.,   Kruse G.,  1998, \mn@doi [Mon. Not.
  R. Astron. Soc.] {10.1046/j.1365-8711.1998.01422.x}, 296, 873

\bibitem[\protect\citeauthoryear{Shan et~al.,}{Shan et~al.}{2018}]{shan2018}
Shan H.,  et~al., 2018, \mn@doi [Monthly Notices of the Royal Astronomical
  Society] {10.1093/mnras/stx2837}, 474, 1116

\bibitem[\protect\citeauthoryear{Shirasaki, Nagai  \& Lau}{Shirasaki
  et~al.}{2016}]{shirasaki2016}
Shirasaki M.,  Nagai D.,   Lau E.~T.,  2016, \mn@doi [Monthly Notices of the
  Royal Astronomical Society] {10.1093/mnras/stw1263}, 460, 3913

\bibitem[\protect\citeauthoryear{Smith \& Marian}{Smith \&
  Marian}{2011}]{smith2011b}
Smith R.~E.,  Marian L.,  2011, \mn@doi [Monthly Notices of the Royal
  Astronomical Society] {10.1111/j.1365-2966.2011.19525.x}, 418, 729

\bibitem[\protect\citeauthoryear{Tinker, Kravtsov, Klypin, Abazajian, Warren,
  Yepes, Gottl{\"o}ber  \& Holz}{Tinker et~al.}{2008}]{tinker2008}
Tinker J.,  Kravtsov A.~V.,  Klypin A.,  Abazajian K.,  Warren M.,  Yepes G.,
  Gottl{\"o}ber S.,   Holz D.~E.,  2008, \mn@doi [Astrophys. J.]
  {10.1086/591439}, 688, 709

\bibitem[\protect\citeauthoryear{Titsias}{Titsias}{2009}]{titsias2009}
Titsias M.,  2009, in Proceedings of the {{Twelth International Conference}} on
  {{Artificial Intelligence}} and {{Statistics}}. {PMLR}, pp 567--574

\bibitem[\protect\citeauthoryear{Upadhye, Biswas, Pope, Heitmann, Habib, Finkel
   \& Frontiere}{Upadhye et~al.}{2014}]{upadhye2014}
Upadhye A.,  Biswas R.,  Pope A.,  Heitmann K.,  Habib S.,  Finkel H.,
  Frontiere N.,  2014, \mn@doi [Phys. Rev. D] {10.1103/PhysRevD.89.103515}, 89,
  103515

\bibitem[\protect\citeauthoryear{Velliscig, {van Daalen}, Schaye, McCarthy,
  Cacciato, Le~Brun  \& Vecchia}{Velliscig et~al.}{2014}]{velliscig2014}
Velliscig M.,  {van Daalen} M.~P.,  Schaye J.,  McCarthy I.~G.,  Cacciato M.,
  Le~Brun A.~M.,   Vecchia C.~D.,  2014, \mn@doi [Mon. Not. R. Astron. Soc.]
  {10.1093/mnras/stu1044}, 442, 2641

\bibitem[\protect\citeauthoryear{Voit, Evrard  \& Bryan}{Voit
  et~al.}{2001}]{voit2001}
Voit G.~M.,  Evrard A.~E.,   Bryan G.~L.,  2001, \mn@doi [ApJ]
  {10.1086/319102}, 548, L123

\bibitem[\protect\citeauthoryear{{Von der Linden} et~al.,}{{Von der Linden}
  et~al.}{2014}]{vonderlinden2014}
{Von der Linden} A.,  et~al., 2014, \mn@doi [Mon. Not. R. Astron. Soc.]
  {10.1093/mnras/stt1945}, 439, 2

\bibitem[\protect\citeauthoryear{Wang, Haiman  \& May}{Wang
  et~al.}{2009}]{wang2009a}
Wang S.,  Haiman Z.,   May M.,  2009, \mn@doi [The Astrophysical Journal]
  {10.1088/0004-637X/691/1/547}, 691, 547

\bibitem[\protect\citeauthoryear{Wang et~al.,}{Wang et~al.}{2018}]{wang2018d}
Wang C.,  et~al., 2018, \mn@doi [Monthly Notices of the Royal Astronomical
  Society] {10.1093/mnras/sty073}, 475, 4020

\bibitem[\protect\citeauthoryear{Yang, Kratochvil, Wang, Lim, Haiman  \&
  May}{Yang et~al.}{2011}]{yang2011}
Yang X.,  Kratochvil J.~M.,  Wang S.,  Lim E.~A.,  Haiman Z.,   May M.,  2011,
  \mn@doi [Phys. Rev. D] {10.1103/PhysRevD.84.043529}, 84, 043529

\bibitem[\protect\citeauthoryear{Zhang \& Annis}{Zhang \&
  Annis}{2022}]{zhang2022}
Zhang Y.,  Annis J.,  2022, \mn@doi [Monthly Notices of the Royal Astronomical
  Society: Letters] {10.1093/mnrasl/slac002}, 511, L30

\bibitem[\protect\citeauthoryear{Zu \& Mandelbaum}{Zu \&
  Mandelbaum}{2015}]{zu2015}
Zu Y.,  Mandelbaum R.,  2015, \mn@doi [Mon. Not. R. Astron. Soc.]
  {10.1093/mnras/stv2062}, 454, 1161

\bibitem[\protect\citeauthoryear{{van Haarlem}, Frenk  \& White}{{van Haarlem}
  et~al.}{1997}]{vanhaarlem1997}
{van Haarlem} M.~P.,  Frenk C.~S.,   White S. D.~M.,  1997, \mn@doi [Monthly
  Notices of the Royal Astronomical Society] {10.1093/mnras/287.4.817}, 287,
  817

\makeatother
\end{thebibliography}

%%%%%%%%%%%%%%%%%%%%%%%%%%%%%%%%%%%%%%%%%%%%%%%%%%

%%%%%%%%%%%%%%%%%%%% APPENDICES %%%%%%%%%%%%%%%%%%

\appendix
\section{Weak lensing measurements of the aperture
  mass}\label{app:weak_lensing}
In this appendix we show how aperture mass measurements from weak
lensing observations relate directly to aperture masses measured from
simulations. Overdensities in the mass distribution modify the light
propagation from background galaxies depending on the projected
distance from the overdensity, distorting the galaxy shapes. By
measuring the average shape distortion of a large sample of background
galaxies within some annular region, we can derive the total mass
contained within that annulus without making any assumptions about the
mass distribution.

In general, weak lensing-derived aperture masses are filtered
measurements of the surface mass density centred on a position
$\bm{\theta}_0$, with a filter function
$U(\bm{\theta} - \bm{\theta}_0)$. We follow the notation of
\citet{bartelmann2001a} and write
\begin{equation}
  \label{eq:m_ap_kappa} M_\mathrm{ap}(\bm{\theta}_0) = \int \diff^2 \bm{\theta} \, 
  U(\bm{\theta} - \bm{\theta}_0) \kappa(\bm{\theta})
  \, .
\end{equation}
We have introduced the convergence
\begin{equation}
  \label{eq:lensing_k} \kappa(\theta) = \frac{\Sigma(\theta)}{\Sc} \, ,
\end{equation}
where the critical surface mass density $\Sc$, which sets the
magnitude of the lensing, is a physical constant given by
\begin{equation}
  \label{eq:sigma_crit} \Sc = \frac{c^2}{4 \pi G} \frac{1}{\beta \Dl}
\, ,
\end{equation}
which depends on the angular diameter distance to the lens, $\Dl$, and
the lensing efficiency, $\beta=\mathrm{max}(0, \Dls / \Ds)$, for a
source at angular diameter distance $\Ds$ from the observer and $\Dls$
from the lens. There is no lensing signal ($\beta=0$) when the source
is in front of the lens, i.e. $\Dls<0$.

For a radial, compensated filter obeying the relation
\begin{equation}
  \label{eq:u_compensated} \int \diff \theta \, \theta U(\theta) = 0
  \, ,
\end{equation}
Eq.~\eqref{eq:m_ap_kappa} can be rewritten in terms of the tangential
shear as
\begin{equation}
  \label{eq:m_ap_gammat} M_\mathrm{ap}(\bm{\theta}_0) = \int \diff^2 \bm{\theta} \, 
  Q(|\bm{\theta} - \bm{\theta}_0|) \gammat(\bm{\theta}|\bm{\theta}_0)
  \, ,
\end{equation}
where the tangential shear is defined as
\begin{equation}
  \label{eq:lensing_gt}
  \gammat(\theta) = \frac{\mean{\Sigma}(\leq \theta) - \Sigma(\theta)}{\Sc} \, ,
\end{equation}
and the new filter function $Q(\theta)$ is related to the surface mass
density filter $U(\theta)$ as
\begin{equation}
  \label{eq:q_from_u} Q(\theta) = \frac{2}{\theta^2}
  \int_0^\theta \diff \theta^\prime \, \theta^\prime U(\theta^\prime) - U(\theta) \, .
\end{equation}
Choosing filters $U(\theta)$ that are constant within some small inner
aperture $\tin$ will result in $Q(\theta) = 0$ for $\theta < \tin$.
Similarly, compensated filters with $U(\theta) = 0$ outside $\tmax$
give $Q(\theta) = 0$ for $\theta > \tmax$. Hence, aperture masses can
be measured from the tangential shear within some finite region
$\tin < \theta < \tmax$ for carefully chosen filters $U$. The region
can be chosen with $\tin$ large enough to avoid the contamination from
the densely populated cluster core and, importantly, to ensure
measurements within the weak lensing regime. Generally, gravitational
lensing does not measure the tangential shear directly, but is instead
sensitive to the reduced shear
\begin{equation}
  \label{eq:shear_red} \gt(\theta) = \frac{\gammat(\theta)}{1 -
\kappa(\theta)} \, .
\end{equation}
However, if $\tin$ is chosen large enough, then $\kappa(\theta) \ll 1$
and the weak lensing assumption $\gt \approx \gammat$ holds.

Since galaxy ellipticities are an unbiased estimator of the local
shear field in the weak lensing regime, the aperture mass can be
estimated directly by summing over the observed galaxy ellipticities
\citep{schneider1996}. Assuming the mean number density of lensed
background galaxies, $\mean{n}_\mathrm{gal}$, we get
\begin{equation}
  \label{eq:zeta_c_i}
  M_\mathrm{ap}(\bm{\theta}_0) = \frac{1}{\mean{n}_\mathrm{gal}} \sum_i Q(|\bm{\theta}_i - \bm{\theta}_0|) \gammat(\bm{\theta}_i) \, .
\end{equation}
The uncertainty in this aperture mass measurement depends only on the
shape noise due to the finite number of galaxies sampling the shear
field. For an average uncertainty $\sigma_\mathrm{gal}$ in the shear
measurement $\gammat$ of an individual galaxy, and a background galaxy
number density $\mean{n}_\mathrm{gal}$, the uncertainty in
$M_\mathrm{ap}$ is
\begin{equation}
  \label{eq:sigma_delta_m}
  \sigma^2_{M_\mathrm{ap}(\bm{\theta}_0)} = \frac{\sigma^2_\mathrm{gal}}{\mean{n}_\mathrm{gal}} \sum_i Q^2(|\bm{\theta}_i - \bm{\theta}_0|) \, .
\end{equation}

The aperture masses that we have used in this paper are directly
related to the $\zc$-statistic, introduced by \citet{clowe1998}, which
can be measured from the tangential shear as
\begin{equation}
  \label{eq:zeta_c_gamma}
  \zc(\tin) = 2\int_{\tin}^{\tout} \diff \ln \theta
    \amean{\gammat} + \frac{2}{1 - \tout^2/\tmax^2}
    \int_{\tout}^{\tmax} \diff \ln \theta \amean{\gammat} \, .
\end{equation}
We have introduced the tangentially averaged tangential shear,
$\amean{\gammat}$, defined as
\begin{align}
  \amean{\gammat}(\theta) &= \frac{1}{2\pi} \oint \diff \phi \, \gammat(\theta, \phi) \, .
\end{align}
Eq.~\eqref{eq:zeta_c_gamma} implies a filter function
\begin{equation}
  \label{eq:zeta_c_q}
  Q_{\zc}(\theta) =
  \begin{cases}
    \frac{1}{\pi \theta^2} & \mathrm{for} \, \tin < \theta \leq \tout \\
    \frac{1}{\pi \theta^2} \frac{\tmax^2}{\tmax^2 - \tout^2} & \mathrm{for} \, \tout < \theta \leq \tmax \\
    0 & \mathrm{elsewhere} \, .
  \end{cases}
\end{equation}
We can readily obtain $\Delta M$ from $\zc$ as
\begin{equation}
  \label{eq:delta_m_from_zeta_c}
  \Delta M(<\tin|\tout, \tmax) = \Sc \zc(<\tin)\pi \tin^2 \, .
\end{equation}

\section{Scalable Gaussian processes for non-Gaussian likelihoods}\label{app:emulation}
We start by introducing our notation. For each of the 100 cosmologies,
$\mathbf{\Omega}_i$, simulated in Mira--Titan, we have calculated the
aperture mass function $n(\Delta M, \mathbf{\Omega}_i)$ on a
log-spaced grid of 50 points with
$\log_{10} \Delta M/\msun \in [13.5, 15.5]$ for the redshifts
$z\in \{0.1, 0.24, 0.43, 0.66, 1.0, 1.6, 2.0\}$. For a set of input
locations and observations $\{(\vect{x}_i, N_i)|i=1, \ldots, n\}$,
with $n=100 \times 50$ ($100$ cosmologies with $50$ mass bins each),
we group the $1 \times d$-dimensional input vectors $\vect{x}_i^T$
containing the cosmological parameters and the mass bin, into the rows
of the $n \times d$ matrix $\X$, i.e. $\X_i=\vect{x}_i^T$, and the
measured number counts for each redshift $z_j$ into the
$n$-dimensional vector $\vect{N}_j$. We will drop the subscript $j$ in
what follows, since the procedure will be the same for each redshift
with only the input measurements differing.

Given the large dynamic range and the peaked nature of the aperture
mass function, we do not model the number counts directly. Instead, we
predict the number density normalized to the mean value over all
cosmologies in the grid
\begin{equation}
  \label{eq:gp_latent}
  f(\vect{x}_i) = \log n(\vect{x}_i) - \log \langle n(\Delta M_l) \rangle_\mathbf{\Omega} \, ,
\end{equation}
with $\vect{x}_i^T=(\mathbf{\Omega}_k^T, \Delta M_l)$, a vector
containing the aperture mass for different cosmologies. We stress that
a single cosmology, $\mathbf{\Omega}_k$, has 50 mass bins,
$\Delta M_l$, and we normalize the aperture mass function with the
mean over all cosmologies for each mass bin. This normalization
reduces the dynamic range of the latent function $f(\vect{x}_i)$ to
values approximately between -1 and 1. We can easily recover the
predicted number counts from $f(\vect{x}_i)$ by converting it to
$n(\vect{x}_i)$ using Eq.~\eqref{eq:gp_latent}, and multiplying by the
volume element and the bin-spacing in $\Delta M$. As long as the mean
number density $\langle n(M_l) \rangle_\mathbf{\Omega} > 0$ in
Eq.~\eqref{eq:gp_latent}, the high-mass tail of cosmological models
with no observed clusters can be fit consistently with the correct
likelihood and without assuming any functional form for the aperture
mass function.

To fit this model to the simulated mass functions, we need to assume
the likelihood of the simulated data. Since the number counts are
discrete observations with exponential cosmology sensitivity in the
low-number count, high-mass tail, we cannot assume a Gaussian
likelihood that does not accurately describe low number counts. We
cannot assume a Poisson likelihood either, since, as shown in
Fig.~\ref{fig:n_M_variance}, the dispersion of the aperture mass
function exceeds the Poisson value. Hence, we assume a negative
binomial likelihood for the data $N_i$ given the model
$f(\vect{x}_i)$. The probability density function of the negative
binomial distribution can be written in terms of the mean, $\mu$, and
the variance, $\alpha \mu$, where $\alpha>1$ captures the
overdispersion compared to the Poisson distribution. In our case, we
write the likelihood of the simulated number counts, $N_i$, given the
model, $f(\vect{x}_i)$, as
\begin{equation}
  \label{eq:data_likelihood}
  p(N_i|f(\vect{x}_i)) = \mathcal{NB}(N_i|N(\vect{x}_i), \alpha_i) \, ,
\end{equation}
where $N(\vect{x}_i)$ is the number of haloes inferred from
$f(\vect{x}_i)$, and $\alpha_i$ is calculated as the ratio between the
bootstrapped variance and the observed number of haloes in the mass
bin. Standard Gaussian processes cannot be solved analytically for
data with non-Gaussian likelihoods, so we will use the approximate,
variational inference Gaussian process method from \citet{hensman2014}
and implemented in
\texttt{GPyTorch}\footnote{\url{https://github.com/cornellius-gp/gpytorch}}
\citep{gardner2021}.

The Gaussian process assumption models the latent function in
Eq.~\eqref{eq:gp_latent} as (following the notation of
\citealp{rasmussen2006})
\begin{align}
  \label{eq:gp_prior}
  f(\vect{x}) & \sim \mathcal{GP}(\mu, k(\vect{x}, \vect{x}^\prime|\theta)) \, ,
\end{align}
which is shorthand for
\begin{align}
  \label{eq:gp_mean}
  & \mathds{E}[f(\vect{x})] = \mu \, ,\\
  \label{eq:gp_var}
  & \mathrm{Var}[f(\vect{x}), f(\vect{x}^\prime)] = k(\vect{x}, \vect{x}^\prime|\theta) \, ,
\end{align}
and means that the values of $f$ are fully determined by the mean,
$\mu$, and the covariance function
$k(\vect{x}, \vect{x}^\prime|\theta)$ between different inputs
$\vect{x}$ and $\vect{x}^\prime$. We will be using the radial basis
function (or squared exponential) kernel for $k$:
\begin{equation}
  \label{eq:gp_rbf}
  k(\vect{x}, \vect{x}^\prime|\theta) = \sigma^2 \prod_{i=0}^{d} \exp\left( -\frac{((\vect{x})_i - (\vect{x}^\prime)_i)^2}{2\ell_{i}^2} \right) \, ,
\end{equation}
where $i$ runs over the $d=9$ dimensions of $\vect{x}$ and each
dimension has its own covariance lengthscale $\ell_i$, resulting in
hyperparameter $\theta = (\mu, \sigma^2, \bm{\ell})$.

The power of Gaussian process regression stems from the conditioning
property of Gaussian distributions. In what follows, we assume
$\mu = 0$ for simplicity. Given the assumed joint Gaussian
distribution between function values at $\X$ and $\Xs$,
$p(\vect{f}, \vect{f}^*)$, which we can write as
\begin{equation}
  \label{eq:mvn_joint}
  p(\vect{f}, \vect{f}^*) = p\left(
    \begin{bmatrix}
      \vect{f} \\ \vect{f}^*
    \end{bmatrix}
  \right) = \mathcal{N}\left(
    \begin{bmatrix}
      \vect{0} \\ \vect{0}
    \end{bmatrix} ,
    \begin{bmatrix}
      \KXX  & \KXXs \\
      \KXsX & \KXsXs \\
    \end{bmatrix}
  \right) \, ,
\end{equation}
the conditional distribution $p(\vect{f}^*|\vect{f})$ is a new
Gaussian distribution given by
\begin{equation}
  \label{eq:mvn_conditioning}
  p(\vect{f}^* | \vect{f}, \theta) = \mathcal{N}\left( \KXsX \KXX^{-1} \vect{f}, \KXsXs - \KXsX \KXX^{-1} \KXXs \right) \, .
\end{equation}
Here we have introduced the $n \times n$ covariance matrix $\KXX$,
with $(\KXX)_{ij}=k(\vect{x}_i, \vect{x}_j)$ and $k$ given by
Eq.~\eqref{eq:gp_rbf}, containing the covariance between different
input points in $\X$. Importantly, the probability distribution of
$\vect{f}^*$ for an arbitrary input location $\Xs$ depends solely on
the finite set of \emph{measured} inputs $\X$. Clearly, the accuracy
of the prediction $f(\vect{x}^*)$ depends on the distance to the
nearest measured input $\vect{x}$ in $\X$ and the lengthscale
hyperparameter $\bm{\ell}$, with the function values $f^*$ regressing
to the mean $0$ and prior uncertainty $k(\vect{x}^*, \vect{x}^*)$ for
$\K_{\vect{x}^*\X} \to \vect{0}^T$. We can use
$p(\vect{f}^*|\vect{f}, \theta)$ to predict $\vect{N}(\Xs)$.

The optimal hyperparameters, $\theta$, for the simulated data,
$\vect{N}$, are found by maximizing
\begin{equation}
  \label{eq:gp_hyper_params}
  p(\theta|\vect{N}) = \frac{p(\theta) p(\vect{N}|\theta)}{p(\vect{N})} \, .
\end{equation}
where we introduced the marginal likelihood
\begin{equation}
  \label{eq:gp_mll}
  p(\vect{N}|\theta) = \int p(\vect{N}|\vect{f}) p(\vect{f}|\theta) \diff \vect{f} \, ,
\end{equation}
which cannot be solved analytically in the case of a negative binomial
likelihood.

This standard Gaussian process encounters two major difficulties.
First, the $\KXX^{-1}$-term in Eq.~\eqref{eq:mvn_conditioning} becomes
computationally expensive for datasets with large $n$. Second,
non-Gaussian likelihoods require approximations to optimize
Eq.~\eqref{eq:gp_hyper_params}, since no closed-form analytical
solution exists. Both of these problems have been solved by the sparse
method using inducing variables and the variational free energy as
introduced by \citet{titsias2009} and applied to non-Gaussian
likelihoods by \citet{hensman2014} and formalized by
\citet{matthews2016}. We will briefly introduce the necessary
ingredients for this method.

The idea behind the method of \citet{titsias2009} is to introduce both
an extra set of $m \ll n$ inducing (or pseudo) inputs $\Z$ of the
Gaussian process such that $\vect{f}(\Z) \equiv \vect{u}$ and an
approximate distribution over these function values,
$q_\psi(\vect{u})$. The inducing point locations $\Z$ and the
parameters $\psi$ of the approximate distribution family will be
chosen in such a way that they optimally capture the true posterior
probability of the Gaussian process, i.e.
$q_\psi(\vect{f}) \simeq p(\vect{f}|\vect{N})$. Assuming a Gaussian
distribution for $q(\vect{u})$ with
\begin{equation}
  \label{eq:gp_q_dist}
  q_\psi(\vect{u}) = \mathcal{N}(\vect{m}, \mathsf{S}) \, ,
\end{equation}
we get $\psi=(\vect{m}, \mathsf{S})$ and we calculate the full
approximate distribution as
\begin{equation}
  \label{eq:gp_qf}
  q_\psi(\vect{f}, \vect{u}) = p(\vect{f}|\vect{u}) q_\psi(\vect{u}) \, .
\end{equation}
The optimization of $(\Z, \vect{m}, \mathsf{S})$ now needs to ensure
that
\begin{align}
  \nonumber
  p(\vect{f}|\vect{N}) & \simeq \int p(\vect{f}|\vect{u}) q(\vect{u}) \diff \vect{u} \\
  \nonumber
                       & \Updownarrow p(\vect{f}|\vect{u}) = \mathcal{N}(\K_{\X\Z}\K_{\Z\Z}^{-1}
                         \vect{u}, \mathsf{D}_{\X\X}) \\
  \label{eq:gp_inducing_pred}
                         & = \mathcal{N}(\K_{\X\Z}\K_{\Z\Z}^{-1}
                           \vect{m}, \mathsf{D}_{\X\X} + \K_{\X\Z}\K_{\Z\Z}^{-1}\mathsf{S} \K_{\Z\Z}^{-1}\K_{\Z\X}) \, ,
\end{align}
with $\mathsf{D}_{\X\X} = \KXX - \K_{\X\Z} \K_{\Z\Z}^{-1} \K_{\Z\X}$,
due to the conditioning property of Eq.~\eqref{eq:mvn_conditioning}
(see Chapter 4.3 of \citealp{matthews2017} for detailed explanations).
Evaluating this expression only requires the inverted $m \times m$
matrix $K_{\Z\Z}^{-1}$, significantly reducing the computational cost
of making model predictions.

To determine the optimal values $(\Z, \vect{m}, \mathsf{S})$, we
minimize the difference between the approximate distribution
$q_\psi(\vect{f}, \vect{u})$ and the model posterior
$p(\vect{f}, \vect{u}|\vect{N})$ through the Kullback-Leibler (KL)
divergence
\begin{align}
  \label{eq:gp_inducing_kl}
  \mathcal{KL}[q(\vect{f}, \vect{u})||p(\vect{f}, \vect{u}|\vect{N})] = -\int q(\vect{f}, \vect{u}) \log\left( \frac{p(\vect{f}, \vect{u}|\vect{N})}{q(\vect{f}, \vect{u})} \right) \diff \vect{f} \diff \vect{u} \, .
\end{align}
Defining this equation as $\mathcal{K}$, we use Bayes' theorem to
rewrite
$p(\vect{f}, \vect{u}|\vect{N}) =
p(\vect{N}|\vect{f})p(\vect{f}|\vect{u}) p(\vect{u}) /
p(\vect{N|\theta})$, making use of the fact that the observations are
only conditionally dependent on their corresponding function values
$\vect{f}$. Also filling in Eq.~\eqref{eq:gp_qf}, we then find
\begin{align}
  \nonumber
  \mathcal{K} = & -\int p(\vect{f}|\vect{u})q(\vect{u}) \log\left( \frac{p(\vect{N}|\vect{f}) p(\vect{u})}{p(\vect{N|\theta})q(\vect{u})} \right) \diff \vect{f} \diff \vect{u} \\
  \nonumber
  = & -\int q(\vect{f}) \log p(\vect{N}|\vect{f}) \diff \vect{f} 
      + \int p(\vect{f}|\vect{u}) q(\vect{u}) \log p(\vect{N}|\theta) \diff \vect{f} \diff \vect{u} \\
  \nonumber
                & + \int p(\vect{f}|\vect{u}) q(\vect{u}) \log\left( \frac{q(\vect{u})}{p(\vect{u})} \right) \diff \vect{f} \diff \vect{u} \\
  \label{eq:gp_elbo_full}
  \mathcal{K} = & \log p(\vect{N}|\theta) - \mathds{E}_{q(\vect{f})}\left[ \log p(\vect{N}|\vect{f}) \right] + \mathcal{KL}[q(\vect{u})||p(\vect{u})] \, .
\end{align}
We can rearrange terms in this expression and use the fact that the KL
divergence is strictly positive to arrive at the variational evidence
lower bound (ELBO), which provides a lower bound on the marginal
likelihood---also called the model evidence---as the name suggests
\begin{equation}
  \label{eq:gp_elbo_bound}
  \log p(\vect{N}|\theta) \geq \mathcal{L}_\mathrm{ELBO} = \mathds{E}_{q(\vect{f})}\left[ \log p(\vect{N}|\vect{f}) \right] -  \mathcal{KL}[q(\vect{u})||p(\vect{u})] \, .
\end{equation}
Equality for this equation holds exactly when
Eq.~\eqref{eq:gp_inducing_kl} equals zero, which is the case when
$q(\vect{f}, \vect{u}) = p(\vect{f}, \vect{u}|\vect{N})$. Assuming no
covariance between $f_i$ and $N_{j\neq i}$, the likelihood factors and
we have
\begin{equation}
  \label{eq:gp_elbo}
  \mathcal{L}_\mathrm{ELBO} = \sum_{i=1}^{n} \mathds{E}_{q(f_i)}\left[ \log p(N_i|f_i) \right] -  \mathcal{KL}[q(\vect{u})||p(\vect{u})] \, ,
\end{equation}
where the first term consists of a sum of one dimensional integrals
which can be computed easily using Gauss-Hermite quadrature, and the
second term is the KL divergence between two multivariate Gaussian
distributions, since $p(\vect{u})=\mathcal{N}(\vect{0}, \K_{\Z\Z})$
due to the Gaussian process assumption. Optimizing the ELBO is
equivalent to maximizing the marginal log-likelihood in
Eq.~\eqref{eq:gp_mll}.

We use the
\texttt{ApproximateGP}\footnote{\url{https://docs.gpytorch.ai/en/latest/examples/04_Variational_and_Approximate_GPs/Non_Gaussian_Likelihoods.html}}
implementation of \texttt{GPyTorch} to model and optimize
$f(\vect{x}_i)$ with a custom implementation of the negative binomial
likelihood between $N(\vect{x}_i)$ and the measurements number counts
from the simulations, $N_i$.

\section{Emulator performance}\label{app:emulator_performance}
\begin{figure*}
  \centering
  \includegraphics[width=\textwidth]{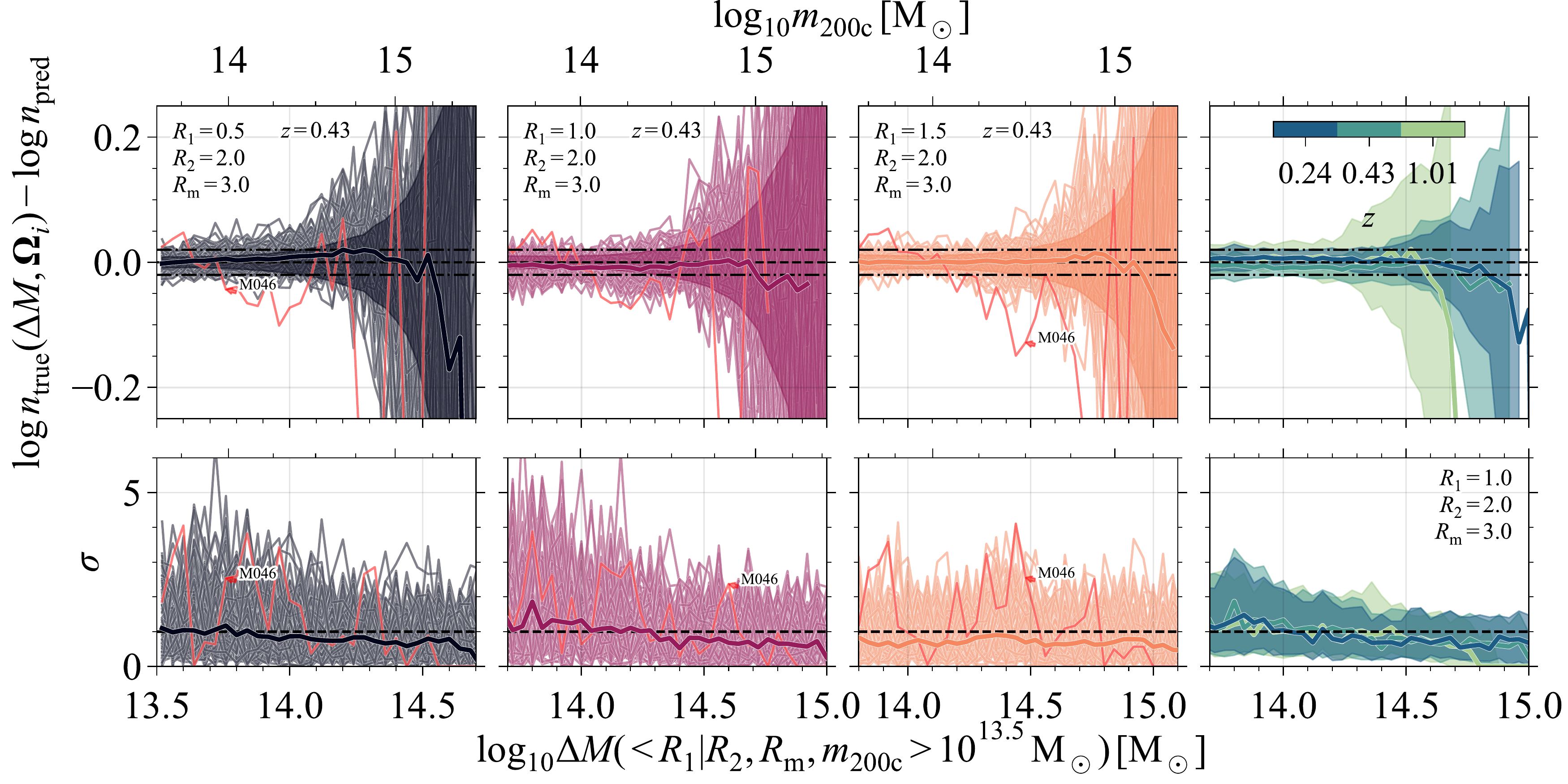}
  \caption{\emph{Top row:} The residuals between the best-fit emulator
    and the individual simulations in the hypercube for different
    apertures at fixed redshift $z=0.43$ (\emph{first three columns})
    and at different redshifts for a fixed aperture
    $(\Rin, \Rout, \Rmax) = (1, 2, 3) \, \cmpc$ (\emph{rightmost
      column}). All halo samples were selected with
    $m_\mathrm{200c} > 10^{13.5} \, \msun$. For the aperture
    variations, we show the individual simulations and for the
    redshift variations the 16th to 84th percentile scatter to avoid
    clutter. The thick, coloured line indicates the median emulator
    residual and the shaded regions in the aperture plots indicate the
    median of the bootstrapped simulation uncertainties. The black,
    dash-dotted lines indicate the $\pm 2 \, \percent$ deviation. The
    median emulator residual is within the simulation uncertainty, and
    unbiased for low aperture masses and $\Rin = 0.5, 1.5 \, \cmpc$,
    while biased low at the $\approx 1 \, \percent$ level for
    $\Rin=1.0 \, \cmpc$. For abundant, low-aperture mass haloes the
    emulator reaches an accuracy of $\approx 2 \, \percent$ for the
    bulk of the simulations, rarely exceeding the $5 \, \percent$
    level. The most significant outlier, M046, which was run with a
    smaller box size, is indicated with a red line in the different
    panels. \emph{Bottom row:} The Gaussian-equivalent significance of
    the likelihood ratio between the simulated data and the emulator.
    The median significance is mostly $\approx 1 \, \sigma$. The
    significance for the individual simulations can vary wildly, but,
    as the top row shows, the fractional uncertainty remains
    reasonable.}
  \label{fig:n_M_perf}
\end{figure*}
The approximate Gaussian process does not sample the simulation
inputs, but instead optimizes the inducing point locations to
accurately reproduce the posterior of the full Gaussian process, i.e.
Eq.~\eqref{eq:gp_inducing_kl}. Hence, we will not trivially reproduce
the simulation aperture mass function. In our set-up, we first
normalize the input parameters, $\X$, so that all parameters lie
between $0$ and $1$. We use 500 inducing points in the
\texttt{ApproximateGP} variational distribution and minimize the
marginal likelihood, approximated by the
\texttt{gpytorch.mlls.VariationalELBO}, with the Adam optimizer with a
learning rate of 0.01 and mini-batches of 512 observations each. These
settings resulted in the fastest loss function minimization in a
coarse, manual search for the optimal parameter settings. We resample
the initial hyperparameters 5 times from their uniform priors to avoid
local minima in the optimization. The emulator parameters are
specified by the inducing point locations, $\vect{u}$, from
Eq.~\eqref{eq:gp_q_dist} in Appendix~\ref{app:emulation}, the Gaussian
process mean, $\mu$, from Eq.~\eqref{eq:gp_mean}, and the kernel
lengthscales and normalization, $\bm{\ell}$ and $\sigma$,
respectively, from Eq.~\eqref{eq:gp_rbf}. We choose uniform priors
$\vect{u} \sim \mathcal{U}(0, 1)$, $\mu \sim \mathcal{U}(-1, 1)$,
$\bm{\ell} \sim \mathcal{U}(0.05, 2.0)$, and
$\sigma^2 \sim \mathcal{U}(0.05, 2.0)$.

In the top row of Fig.~\ref{fig:n_M_perf}, we show the resulting
absolute deviation between the emulated latent function,
Eq.~\eqref{eq:gp_latent}, and the normalized number density from the
simulation for different apertures and all cosmologies. The first
three columns correspond to the different aperture sizes at $z=0.43$,
and the final column shows the median and 16th to 84th percentile
scatter for the emulator at different redshifts. For low aperture
masses, the emulator error rarely exceeds the $5 \, \percent$
difference level, and the bulk of the simulations have residuals
within $\pm 2 \, \percent$ for the high abundance aperture mass
regime. The median deviation is biased slightly low for
$\Rin=1.0 \, \cmpc$, but it is within $\pm 2 \, \percent$ for all
aperture sizes and all but the most massive haloes. The bulk of the
simulations lack haloes at the highest aperture masses resulting in
the noticeable downturn. While the fractional deviation becomes large,
it is still within the variance of the simulations, which is shown as
the shaded region.

To quantify the quality of the fit in the high-mass tail, we show the
equivalent Gaussian significance of the deviation between the
simulated data and the emulator. We compute the significance by
calculating the difference between the log-likelihood of the measured
number counts in the simulation given the predicted aperture mass
function of the emulator,
$\ln \mathcal{L}(N_\mathrm{true}|N_\mathrm{pred})$, and the
log-likelihood of the emulated number counts,
$\ln \mathcal{L}(N_\mathrm{pred}|N_\mathrm{pred})$, and converting
this probability ratio into the equivalent Gaussian confidence
interval $n \sigma$ around the mean expectation, $\mu$, given by
$\ln P(\mu + n\sigma) - \ln P(\mu)$. We show the significance of the
deviation between the emulator and the simulation data in the bottom
row of Fig.~\ref{fig:n_M_perf}. Individual simulations behave
erratically for low aperture masses, rapidly oscillating between large
and small significance, but rarely exceeding $3 \, \sigma$. The median
significance of all cosmologies, on the other hand, is consistently
$\approx 1 \, \sigma$. For the high-aperture mass tail, with large
fractional deviations between the emulator and the simulations, the
significance of the deviation $\lesssim 1 \, \sigma$, indicating that
the emulator captures the trends in the data to within the shot-noise.

\begin{figure*}
  \centering
  \includegraphics[width=\textwidth]{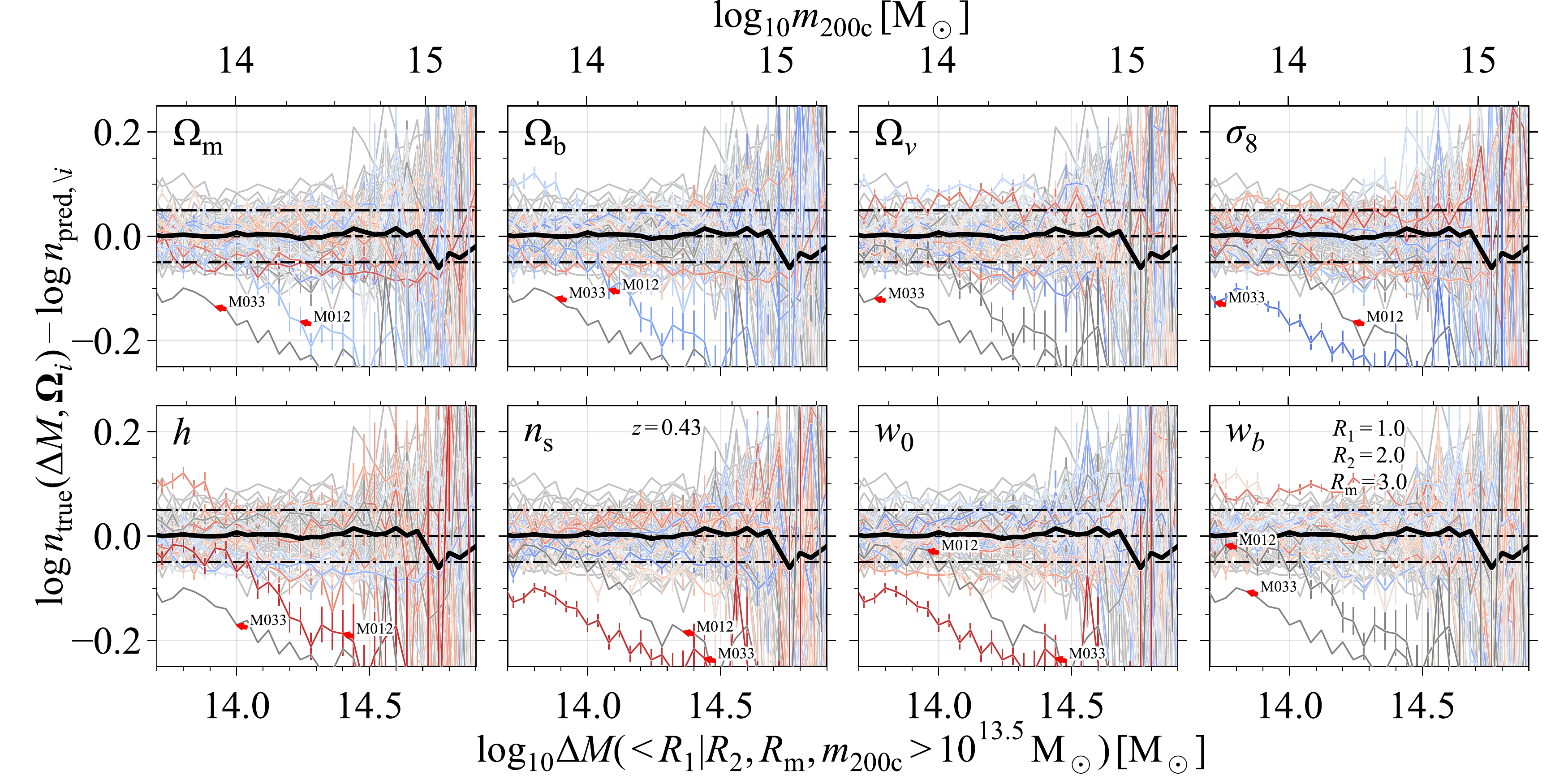}
  \caption{The residuals between the leave-one-out emulator and its
    held out simulation for all simulations that are not at the edge
    of the parameter space. The cosmological parameters are indicated
    in the top-left corner of each panel. All emulators have apertures
    $(\Rin, \Rout, \Rmax) = (1, 2, 3) \, \cmpc$ and have been
    calibrated at $z=0.43$. All halo samples were selected with
    $m_\mathrm{200c} > 10^{13.5} \, \msun$. Coloured lines indicate
    the simulations that are within the first (blue shades) or last
    (red shades) decile for the plotted cosmological parameter, with
    the saturation indicating the ordering. The black line indicates
    the median result for all simulations and the shaded region the
    16th to 84th percentile scatter. The black, dash-dotted lines
    indicate the $\pm 5 \, \percent$ region. Most emulators are able
    to reproduce the held out simulation prediction within
    $\approx 5 \, \percent$. The most significant outliers, M012 and
    M033, are indicated with red arrows and are simulations that are
    close to the edge of the parameter space for several cosmological
    parameters.}
  \label{fig:n_M_loo}
\end{figure*}
Finally, we also perform a leave-one-out test on all simulations that
are not at the edge of the parameter space for any of the cosmological
parameters. In Fig.~\ref{fig:n_M_loo}, we show how accurately the
emulator predicts the aperture mass function for all left-out
simulations. We colour the lines for simulations with cosmological
parameters that are within the first or the last decile of the
hypercube with different shades of blue and red, respectively, with
darker shades indicating more significant outliers. The emulator can
predict the outcome of most simulations to within
$\approx 5 \, \percent$ up to the tail of the mass function. The most
significant deviations are found for simulations that are close to the
edge of the cosmological parameter space in one or more dimensions.
The accuracy achieved by the emulator in the leave-one-out test
indicates that the emulator generalizes well beyond the trained
simulation inputs.

%%%%%%%%%%%%%%%%%%%%%%%%%%%%%%%%%%%%%%%%%%%%%%%%%%
% Don't change these lines
\bsp	% typesetting comment
\label{lastpage}
\end{document}